\documentclass[prx,aps,amssymb,superscriptaddress,notitlepage,twocolumn]{revtex4-2}
\usepackage{changes}
\usepackage{booktabs}%\usepackage[final]{changes}
\usepackage{amsmath}
\usepackage{amssymb}
\usepackage{amsthm}
\usepackage{amsfonts}
\usepackage{listings}
\usepackage{physics}
\usepackage{enumerate}
\usepackage{latexsym}
\usepackage{psfrag}
\usepackage{bm}
\usepackage{graphicx}
\usepackage[caption=false]{subfig}
\usepackage{blkarray}
\usepackage{array}
\usepackage{color}
\usepackage[normalem]{ulem}
\usepackage{hyperref}
\usepackage{framed}
\hypersetup{
     colorlinks = true,
     linkcolor = magenta,
     citecolor = magenta
}

\def\beq{\begin{equation}}
\def\eeq{\end{equation}}
\def\bal{\begin{aligned}}
\def\eal{\end{aligned}}

\newcommand{\kk}{\mathbf{k}}
\newcommand{\bq}{\mathbf{q}}

\begin{document}
\title{Exploring Many-Body Quantum Geometry Beyond the Quantum Metric with Correlation Functions: A  Time-Dependent Perspective}
\author{Yuntao Guan}
\affiliation{Department of Physics, University of Illinois Urbana-Champaign, Urbana IL 61801, USA}
\affiliation{Anthony J.~Leggett Institute for Condensed Matter Theory, University of Illinois Urbana-Champaign, Urbana IL 61801, USA}

\author{Barry Bradlyn}
\email{bbradlyn@illinois.edu}
\affiliation{Department of Physics, University of Illinois Urbana-Champaign, Urbana IL 61801, USA}
\affiliation{Anthony J.~Leggett Institute for Condensed Matter Theory, University of Illinois Urbana-Champaign, Urbana IL 61801, USA}

\date{\today}

\begin{abstract}
The quantum geometric tensor and quantum Fisher information have recently been shown to provide a unified geometric description of the linear response of many-body systems. 
However, a similar geometric description of higher-order perturbative phenomena including nonlinear response in generic quantum systems is lacking. 
In this work, we develop a general framework for the time-dependent quantum geometry of many-body systems by treating external perturbing fields as coordinates on the space of density matrices. 
We use the Bures distance between the initial and time-evolved density matrix to define geometric quantities through a perturbative expansion. 
To lowest order, we derive a time-dependent generalization of the Bures metric related to the spectral density of linear response functions, unifying previous results for the quantum metric in various limits and providing a geometric interpretation of Fermi's golden rule. 
At next order in the expansion, we define a time-dependent Bures-Levi-Civita connection for general many-body systems. 
We show that the connection is the sum of one contribution that is related to a second-order nonlinear response function, and a second contribution that captures the higher geometric structure of first-order perturbation theory. 
We show that in the quasistatic, zero-temperature limit for noninteracting fermions, this Bures connection reduces to the known expression for band-theoretic Christoffel symbols. 
Our work provides a systematic framework to explore many-body quantum geometry beyond the quantum metric and highlights how higher-order correlation functions can probe this geometry.
\end{abstract}
\maketitle

\section{Introduction} % (fold)
\label{sec:introduction}

Recent work has shown a deep connection between topological properties of electronic wavefunctions and geometric properties related to their localizability. 
For noninteracting systems, the Berry curvature of ground state wavefunctions can be viewed as the imaginary part of a quantum geometric tensor~\cite{provost1980riemannian,ma2010abelian}. 
The real part of this tensor---known as the single-particle quantum metric---is directly related to the gauge-invariant contribution to the minimum localization length of Wannier functions in an insulator~\cite{marzari1997maximally}. 
The close relation between Berry curvature and the single-particle quantum metric has been exploited to study bounds on localization and energy gap in topological materials~\cite{ozawa2021relations,onishi2024fundamental}. 
The relationship between localization, geometry, and topology has also been investigated as a criteria for designing interacting topological insulators by studying the dependence of projected form factors for interactions on the single-particle quantum metric~\cite{roy2014band,neupert2015fractional,wang2021exact,ledwith2023vortexability}. 
The discovery of engineered flat band materials has ushered in a renaissance in the study of the single-particle quantum metric. 
The quenching of kinetic energy in these systems allow for the geometric properties of Bloch wavefunctions (viewed as functions of crystal momentum) to play an major role in determining properties like projected interaction strengths and superfluid stiffness~\cite{torma2023essay,torma2022superconductivity,yu2024quantum,herzog2022superfluid,huhtinen2022revisiting,liang2017band}. 
Additional geometric quantities arising from derivatives of the quantum geometric tensor have also been introduced for noninteracting systems~\cite{jiang2025revealing,ahn2020lowfrequency,ahn2022riemannian,mitscherling2024gauge,avdoshkin2023extrinsic}. 
Both the single-particle quantum metric and these additional geometric quantities appear in the optical response of quantum materials, due to the connection between optical response, position operator matrix elements, and localization~\cite{verma2025quantum,ahn2022riemannian,morimoto2016topological,morimoto2023geometric,jankowski2024quantized,sodemann2015quantum,avdoshkin2024multi,jankowski2025optical,shinada2025quantum,kang2025measurements,dejuan2017quantized,flicker2018chiral,rees2020helicitydependent,kaushik2021magnetic,alexandradinata2024quantization,zhu2024anomalous}.

Beyond single-particle systems, pioneering work by Souza, Wilkens, and Martin~\cite{souza2000polarization} showed how a many-body generalization of the single-particle metric first introduced in Ref.~\cite{provost1980riemannian} can be related to macroscopic polarization fluctuations in insulators. 
Like its single-particle counterpart, this many-body quantum metric is the real part of a Hermitian quantum geometric tensor whose imaginary part gives the Berry curvature on the space of twisted boundary conditions related to the Hall conductance~\cite{niu1985quantized,resta2005electron,resta2011insulating}. Recently, Ref.~\cite{hetenyi2023fluctuations} showed that at zero temperature, higher-order geometric quantities could be defined from many-body wavefunction overlaps.

At nonzero temperatures and for generic many-body systems, a unifying picture of the connection between the many-body quantum metric and response to perturbations is given through the quantum Fisher information. 
The quantum Fisher information gives the essentially quantum contribution to the infinitesimal distance between density matrices, and was originally introduced to quantify the distinguishability of states in a quantum information context~\cite{Bengtsson_Zyczkowski_2006,braunstein1994statistical}. 
In pioneering work, it was shown in Ref.~\cite{hauke2016measuring} that the quantum Fisher information can be expressed in terms of moments of linear response functions. 
This has allowed for the study of many-body quantum geometry and quantum entanglement in condensed matter systems through the study of density response~\cite{balut2025quantum,balut2025quantuma,kruchkov2025topologicalcontrolquantumspeed,onishi2025quantum,mendez2024low,mao2025lowenergy} and spin response~\cite{mazza2024quantum,fang2025amplified,wang2025local}. 
Additionally, for deformations corresponding to twisted boundary conditions, the Fisher information reproduces the Souza-Wilkens-Martin sum rule relating the many-body metric to the optical conductivity~\cite{ji2025density,onishi2025quantum}. 
However, for generic many-body systems the influence of quantum geometry beyond linear response remains relatively unexplored. 
Generalizing the geometric framework for nonlinear optical response for noninteracting fermion systems to finite-temperature, disordered, and many-body systems remains an open problem. 

In this work, we take a first step to address this problem by investigating how to formulate many-body quantum geometry beyond the quantum metric. 
We take as our starting point the Bures distance, which gives a very general geometric structure to the space of density matrices~\cite{bures1969extension,Bengtsson_Zyczkowski_2006,nielsen2010quantum}. 
Generalizing from previous works, we consider the geometry of trajectories in the space of density matrices given by time-dependent perturbation theory, where the full spatiotemporal profile of external perturbing fields serve as our coordinates. 
Expanding the distance to lowest order in perturbation theory yields a many-body quantum (Bures) metric. 
However, the Bures metric we derive in this way is intrinsically time-dependent due to the time-dependence of the external perturbations, and is related to recently introduced time-dependent metric tensors associated to twisted boundary conditions~\cite{verma2024instantaneous,ji2025density,resta2025nonadiabatic}. 
Like other formulations of the many-body quantum metric, we show that our time-dependent Bures metric can be expressed as a weighted Fourier transform of a linear response function.

Going beyond the quantum metric, we are able to expand the Bures distance to higher order, yielding a definition of a many-body quantum geometric connection (Christoffel symbol). 
Our result generalizes known formulations of the  Christoffel symbols for noninteracting fermions to finite temperature and many-body correlated systems. 
Furthermore, the Christoffel symbols we introduce are explicitly time-dependent, exposing the geometric structures associated to higher-order time-dependent perturbation theory. 
We show that this Bures connection can be expressed as a weighted Fourier transform of a three-operator correlation function associated to the perturbation. 
Unlike for the metric, only part of the connection is expressible in terms of a response function. 
Our work opens up the door for a systematic exploration of many-body quantum geometry beyond clean noninteracting fermion systems at zero temperature, and highlights the importance of the geometry of time-dependent perturbations.

The remainder of this work is organized as follows. 
First, in Sec.~\ref{sec:geometry_of_time_dependent_perturbation_theory} we introduce a geometrical description of time-dependent perturbation theory. 
We show how the time-dependent amplitudes of external driving fields can be viewed as coordinates on the space of density matrices for a generic quantum system. 
We focus on the Bures distance expressed in terms of these coordinates, expanding it order-by-order in the perturbation to obtain geometric quantities. 
In Sec.~\ref{sec:bures_metric_fisher_information}, we focus on the lowest order expansion which gives the Bures metric, or alternatively the quantum Fisher information. 
Contrasting with previous works~\cite{balut2025quantum,hauke2016measuring,carollo2020geometry,lambert2023classical}, our formalism gives a time-dependent Bures metric on a space parametrized by the entire temporal history of the external perturbations. 
We show that our formalism recovers known relationships between linear response functions and the quantum Fisher information in the special cases of instantaneous and time-independent (quasistatic) perturbations. 
We apply our results to the special cases of density and current response. 
Next, in Sec.~\ref{sec:bures_connection_and_second_order_response} we expand the Bures distance to next order in the perturbation to define Levi-Civita connection associated to time-dependent perturbations of the density matrix. 
We show that the connection generically has two contributions. 
One contribution arises from the change in the Fisher information to next order in the perturbation, and can be related to the second-order spectral density for nonlinear response. 
The second contribution is independent of the second-order response function. 
We show that the connection can be expressed in terms of a single frequency-dependent correlation function. 
We examine the instantaneous and quasistatic limits of the connection, as well as the application to density and current response. 
In the special case of noninteracting fermions, we show that the quasistatic limit of the Levi-Civita connection for current response is proportional to the band-theoretic quantum geometric Christoffel symbols. 
We conclude in Sec.~\ref{sec:outlook} by discussing the outlook for experimental probes of many-body quantum geometry.

% section introduction (end)

\section{Geometry of Time-Dependent Perturbation Theory} % (fold)
\label{sec:geometry_of_time_dependent_perturbation_theory}

We consider a quantum system with Hamiltonian
\begin{equation}\label{eq:ham}
     H=H_0 + \kappa f^\mu(t)B_\mu\eta(t),
\end{equation}
where the unperturbed Hamiltonian $H_0$ is time independent, $f^\mu(t)$ are a set of external probe perturbing fields that couple to operators $B_\mu$, where $\mu=1,\dots,N$ indexes the different perturbations; here and throughout this work we will use the convention that repeated indices are implicitly summed. 
We have introduced a parameter $\kappa$ that we will use to organize terms in our perturbative expansion; we will set $\kappa=1$ at the end of our calculations. 
Furthermore, $\eta(t)$ is a function that ensures the perturbation vanishes at $t\rightarrow-\infty$. 
Common choices for $\eta(t)$ are
\begin{equation}
     \eta(t) = \begin{cases}
     e^{\epsilon t},\,\,\, \epsilon\rightarrow 0 \\
     \Theta(t)
     \end{cases}
\end{equation}
where $\Theta(t)$ is the Heaviside step function. 
We assume that our system starts in an initial thermal state with inverse temperature $\beta$, such that the density matrix $\rho(t)$ satisfies the initial condition
\begin{equation}
     \rho(t\rightarrow-\infty) = \rho_0 = e^{-\beta H_0}/Z_0,
\end{equation}
with $Z_0 = \mathrm{tr}{e^{-\beta H_0}}$. 
We will be interested in the case where the external probe fields $\kappa f^\mu(t)$ are small perturbations. 
In this case we can expand the density matrix $\rho(t)$ in a series order-by-order in $\kappa$,
\begin{equation}\label{eq:rho_expansion}
\rho(t)=\rho_0+\kappa \rho_1(t)+\kappa^2 \rho_2(t)+\dots
\end{equation}
where time-dependent perturbation theory gives (we work in units with $\hbar=1$ throughout)
\begin{align}\label{eq:rho_n}
&\rho_n(t)= (-i)^n\int_{-\infty}^tdt_1\int_{-\infty}^{t_1}dt_2\dots\int_{-\infty}^{t_{n-1}}dt_n\prod_{i=1}^{n}f^{\mu_i}(t_i)\eta(t_i)\nonumber \\ 
&\times\underbrace{\left[B_{\mu_1}(t_1-t),\left[B_{\mu_2}(t_2-t),\left[\dots \left[B_{\mu_n}(t_n-t),\rho_0\right]\right]\right]\right]}_{\text{$n$ times}}.
\end{align}
We will be interested in measuring by how much $\rho(t)$ deviates from $\rho_0$. To do so, we will first in Sec.~\ref{sub:expanding_the_time_dependent_bures_metric} introduce the Bures metric and its time-dependent expansion. Then, in Sec.~\ref{sub:interpretation_of_the_geometry} we will discuss how the different terms in the expansion can be interpreted in terms of quantum geometry.

\subsection{Expanding the Time-Dependent Bures Metric}\label{sub:expanding_the_time_dependent_bures_metric}
To measure the distinguishability of $\rho(t)$ from $\rho_0$, we can make use of the Bures distance $d_B$, which quantifies the distance between density matrices. 
For two density matrices $\rho_1$ and $\rho_2$, the Bures distance is defined as~\cite{bures1969extension,uhlmann1992metric,uhlmann1995geometric}
\begin{equation}\label{eq:buresdistance}
d_B(\rho_1,\rho_2) = \sqrt{2-2\mathrm{tr}\left[\left(\rho_1^{\frac{1}{2}}\rho_2\rho_1^{\frac{1}{2}}\right)^{\frac{1}{2}}\right]}.
\end{equation}

Our goal will be to compute the Bures distance $d_B(\rho_0,\rho(t))$ between the initial density matrix $\rho_0$ and the perturbed density matrix $\rho(t)$, order by order in $\kappa$ [equivalently, order by order in $f^\mu(t)$]. 
This will allow us to define the many-body quantum geometry of the perturbed density matrix. 
Note that our approach differs from that of the recent Ref.~\cite{Wang_2025} in that we work directly with the many-body density matrix, and we focus on many-body systems. 
We also work directly with the perturbed density matrix relevant to condensed matter experiments. 

We start by viewing the functions $\kappa f^\mu(t)$ as coordinates on an infinite-dimensional manifold. 
We can write the Bures distance as the integral of a line element
\begin{equation}
d_B(\rho_0,\rho(t)) = \int_0^\kappa d\kappa'\frac{ds(\kappa'\vec{f}(t))}{d\kappa'}.
\end{equation}
Next, we can write the infinitesimal line element in terms of a metric tensor on the manifold,
\begin{equation}\label{eq:buresmetricdef}
\left(\frac{ds}{d\kappa}\right)^2 = \int dt_1 dt_2 g_{\mu\nu}(\kappa \vec{f},t,t_1,t_2)f^\mu(t_1)f^\nu(t_2),
\end{equation} 
where we have made use of the fact that our coordinates on the manifold depend linearly on $\kappa$, $d(\kappa f^\mu)/d\kappa = f^\mu$, since we have chosen a straight line path in parameter space. 
The functional $g_{\mu\nu}(\kappa \vec{f},t,t_1,t_2)$ is the Bures metric on the space of density matrices parametrized by $\kappa\vec{f}(t')$, and depends in principle on the perturbations at all times $t'<t$. 
Viewing $t_1$ and $t_2$ as additional ``indices'' for the external fields $f^\mu(t)$, the time integrations in Eq.~\eqref{eq:buresmetricdef} play the same role as the implicit sum over repeated indices.

We can combine Eqs.~\eqref{eq:buresdistance} and \eqref{eq:buresmetricdef} to get
\begin{align}\label{eq:buresimplicit}
&\left(\int_0^\kappa d\kappa'\sqrt{ \int dt_1 dt_2 g_{\mu\nu}(\kappa' \vec{f},t,t_1,t_2)f^\mu(t_1)f^\nu(t_2) }\right)^2 \nonumber \\ 
& = 2-2\mathrm{tr}\left[\left(\rho_0^{\frac{1}{2}}\rho(t)\rho_0^{\frac{1}{2}}\right)^{\frac{1}{2}}\right],
\end{align}
which will let us solve for the Bures metric order-by-order in $\kappa$. 
In particular, we will expand both sides of Eq.~\eqref{eq:buresimplicit} in $\kappa$. 
Expanding the left hand side will give us a power series whose coefficients are expressible in terms of the metric $g_{\mu\nu}$ and its (functional) derivatives, evaluated at the origin for the unperturbed system with $\kappa=0$; Expanding the right hand side will give us a power series whose coefficients are expressible in terms of matrix elements of the perturbations $B_\mu$ and the unperturbed density matrix $\rho_0$ in the basis of eigenstates of $H_0$. 
Equating terms order-by-order will let us define the Bures metric \emph{and its derivatives}, giving us access to the geometry of time-dependent perturbation theory.

Concretely, we can expand the left hand side of Eq.~\eqref{eq:buresimplicit} order-by-order in $\kappa$ to find (see Appendix~\ref{sec:power_series_expansion_of_the_bures_distance} for details)
\begin{align}\label{eq:metric_and_connection_def}
d_B&(\rho_0,\rho(t))^2\approx \kappa^2\int dt_1 dt_2 g_{\mu\nu}(t,t_1,t_2) f^\mu(t_1)f^\nu(t_2) \nonumber \\ 
&+ \kappa^3\int dt_1dt_2dt_3 \Gamma_{\mu\lambda\nu}(0,t,t_1,t_3,t_2)f^\mu(t_1)f^\nu(t_2)f^\lambda(t_3),
\end{align}
which defines the Bures metric
\begin{equation}\label{eq:metric_at_orgin}
g_{\mu\nu}(t,t_1,t_2) \equiv g_{\mu\nu}(\kappa\vec{f}=0,t,t_1,t_2)
\end{equation}
and the symmetrized Christoffel symbol
\begin{equation}\label{eq:connection_as_var_deriv}
\Gamma_{\mu\lambda\nu}(t,t_1,t_3,t_2) = \left.\frac{1}{2}\frac{\delta g_{\mu\nu}(\kappa\vec{f},t,t_1,t_2)}{\delta(\kappa f^\lambda(t_3))}\right|_{\kappa=0}
\end{equation}
evaluated at the origin $\kappa=0$. 
Note from Eq.~\eqref{eq:metric_and_connection_def} that only the part of $\Gamma_{\mu\lambda\nu}(t,t_1,t_3,t_2)$ totally symmetric under exchanges $(\mu_i,t_i)\leftrightarrow (\mu_j,t_j)$ enters into the expansion of the Bures distance. 

Also note that from now on, we adopt a strict criterion for ``geometric'' data: a physical quantity is ``geometric'' only if it constitutes the full contribution to the Bures distance at a specific order $\mathcal{O}(\kappa^n)$, or if the full contribution can be expressed entirely in terms of this quantity. Under this definition, for example, the Bures metric [Eq.~\eqref{eq:metric_at_orgin}], the full Bures connection [Eq.~\eqref{eq:connection_as_var_deriv}], the linear-order spectral density $\chi''_{\mu \nu}(\omega)$ [Eq.~\eqref{eq:chipp_def}], and the three-operator correlation function $S_{\mu_1 \mu_2 \mu_3}(\omega_1,\omega_2)$ [Eq.~\eqref{eq:S_function_def}] we will introduce later are geometric.

\subsection{Interpretation of the Geometry}\label{sub:interpretation_of_the_geometry}

Let us now discuss how to interpret these geometric structures. In our framework, the Bures metric gives a Riemannian structure on the space of density matrices $\mathcal{D}$, which consists of all Hermitian, positive semi-definite, and unit-trace operators $\rho$. To study the Bures distance between the thermal and perturbed density matrices, we construct a map from an infinite-dimensional function space of external fields $\mathcal{F}$, parameterized by coordinates $\{\kappa f^{\mu}(\tau) \}$ for $\mu\in\{1,2,\dots,N\}, \tau \in (-\infty, t]$, to the space of density matrices $\mathcal{D}$. This mapping corresponds to time-evolving an initial thermal density matrix under the perturbations $\{\kappa f^\mu(\tau)\}$. Consequently, the time-dependent Bures metric and connection derived in Eq.~\eqref{eq:metric_and_connection_def} are the \textit{pullbacks} of the geometric structure on $\mathcal{D}$ onto the parameter space of external fields $\mathcal{F}$.

This construction generalizes the band-theoretic interpretation of the Fubini-Study metric and Berry curvature in fermion systems at zero temperature. As summarized in Table~\ref{table0: band_vs_rho}, for noninteracting crystalline systems, the analogous map is from the Brillouin zone $\mathbf{k} \in T^d$ to the projective Hilbert space of Bloch wavefunctions $\ket{u_{\kk}} \in \mathbb{C}P^{N-1}$. For interacting systems with periodic boundary conditions, one instead threads gauge fluxes $\phi_\mu$ through the system~\cite{niu1985quantized}. The relevant parameter space then becomes the flux torus $(\phi_1,\dots,\phi_d)\in T^d$, which maps to the many-body ground states subject to twisted boundary conditions. Our formalism goes beyond these two specific parameter spaces by considering the infinite-dimensional function space $\mathcal{F}$ of external driving fields.

In our time-dependent framework, the infinitesimal variations of the external fields $\delta f^\mu(\tau)$ define the components of tangent vectors in the parameter space $\mathcal{F}$. The pushforward of this vector to the tangent space of $\mathcal{D}$ yields the infinitesimal change of the density matrix $\delta \rho(t)$. Moreover, if we use the eigenbasis of $H_0$ to define the basis vectors $\{\ket{n}\bra{m}\}$ for the tangent space of $\mathcal{D}$, then the components of the tangent vector $\delta \rho(t)$ are given by $(\delta \rho)_{nm} = \bra{n} \delta \rho \ket{m}$. Therefore, the Bures distance defined on $\mathcal{D}$ is $d_B^2(\rho_0, \rho(t)) = g_{nm,k\ell}  (\delta \rho)_{nm} (\delta \rho)_{k\ell}$, and applying the pullback from $\mathcal{D}$ to $\mathcal{F}$ yields the Bures metric and connection in Eq.~\eqref{eq:metric_and_connection_def}.
As such, we can think of the pairs $(\mu,t)$ as an infinite-dimensional set of indices for the parameter space $\mathcal{F}$, where the sum over repeated indices in Eq.~\eqref{eq:metric_and_connection_def} is given by the implied summation over repeated Greek indices and the explicit integration over repeated time arguments. 
Viewed through this lens, the definition Eq.~\eqref{eq:connection_as_var_deriv} is a variational generalization of the standard covariance condition for the metric,
\begin{equation}
\partial_{a}g_{bc} - \Gamma_{cab}-\Gamma_{bac}=0,
\end{equation}
where now the indices $a,b,c$ represent both discrete Greek and continuous time indices. 
It is also important to note that the Christoffel symbols are not tensors, so that their expressions are not covariant under changes of coordinate system. 
In our context, the dynamics Eq.~\eqref{eq:ham} defines for us a preferred coordinate system in terms of the experimentally-accessible external fields. 
This is distinct from, e.g., general relativity, but reminiscent of the connection between geometric response and elasticity where the ``lab frame'' takes special precedence~\cite{kleinert1989gauge,zaanen2004duality,katanaev1992theory,bradlyn2015lowenergy}. 
As we do not expect any experimentalist to change the parameterization of their driving fields or time coordinate as a function of field amplitude, we neglect the coordinate-system dependence of Eq.~\eqref{eq:connection_as_var_deriv} in the remainder of this work.

\begin{table*}[t] 
\centering
\renewcommand{\arraystretch}{1.8} 
\begin{tabular}{l c c}
\toprule
\textbf{Geometric Framework} & \textbf{Band Geometry} & \textbf{Time-Dependent Bures Geometry} \\
\midrule
\textbf{State} & $\ket{u_{n}(\mathbf{k})}$  & $\rho[f^\mu](t)$ \\
\textbf{State Space} & Complex projective space $\mathbb{C}P^{N-1}$ & Space of density matrices $\mathcal{D}$ \\
\textbf{Parameter Space} & Brillouin zone $T^d$ & Function space of external fields $\mathcal{F}$ \\
\textbf{Coordinates on Parameter Space} & $\mathbf{k} = (k^1,...,k^d)$ & $\{f^\mu(\tau)\}$ for $\tau\in(-\infty,t]$ \\
\textbf{Tangent Vector in Parameter Space} & $\Delta k^\mu \frac{\partial}{\partial k^\mu}$ & $\displaystyle \int_{-\infty}^{t}  d\tau \ \delta f^\mu(\tau) \frac{\delta}{\delta f^\mu(\tau)}$ \\
\textbf{Tangent Vector in State Space} & $\ket{\delta u_n} =  \frac{\partial}{\partial k^\mu}\ket{u_{n}(\mathbf{k})} \Delta k^\mu$ & $\displaystyle \delta \rho(t) = \int_{-\infty}^{t} d\tau \ \delta f^\mu(\tau) \frac{\delta \rho(t)}{\delta f^\mu(\tau)} $  \\
\textbf{Metric Line Element $ds^2$} & $g_{\mu \nu}(\mathbf{k}) \Delta k^\mu \Delta k^\nu$ & $\displaystyle \int  dt_1 dt_2 \ g_{\mu\nu}(\vec{f},t,t_1, t_2)\delta f^\mu(t_1)\delta f^\nu(t_2)$\\
\bottomrule
\end{tabular}
\caption{Comparison of the underlying mathematical structures between standard noninteracting band geometry and our time-dependent geometric framework for density matrices.} 
\label{table0: band_vs_rho}
\end{table*}
\renewcommand{\arraystretch}{1}

Finally, we comment on the regime of validity of this geometric expansion. The perturbative approach here relies strictly on the analyticity of the Bures metric $g^{\mu \nu}(\kappa \vec{f},t,t_1,t_2)$ at the origin $\kappa = 0$. If $g^{\mu \nu}$ is non-analytic at the origin, the Taylor expansion used to derive the Bures metric and connection breaks down. This naturally occurs when a system approaches a quantum phase transition, where key information-theoretic quantities such as fidelity, quantum Fisher information, and the Bures metric itself, are known to exhibit critical divergent behavior~\cite{quan2006decay,zanardi2007ground,campos2007quantum,hauke2016measuring,carollo2020geometry}.

The geometric quantities defined through Eq.~\eqref{eq:metric_and_connection_def} depend explicitly on time $t$, as does the Bures distance itself. 
We thus see that the time-dependence of the density matrix defines a time-dependent geometry. 
Eq.~\eqref{eq:buresimplicit} defines the time-dependent geometric structures in terms of matrix elements of the perturbed density matrix. 
In the remainder of this work, we will compute expressions for the Bures metric and Bures connection (at the origin of the parameter space $\kappa=0$) in terms of response functions and correlation functions for the unperturbed system. 
We begin in Sec.~\ref{sec:bures_metric_fisher_information} with the metric.

% section geometry_of_time_dependent_perturbation_theory (end)
\section{Bures Metric, Fisher Information, and Linear Response} % (fold)
\label{sec:bures_metric_fisher_information}

To compute the time-dependent Bures metric, we expand the right-hand side of Eq.~\eqref{eq:metric_and_connection_def} to quadratic order in $\kappa$. 
In terms of the perturbations \eqref{eq:rho_expansion} to the density matrix and the eigenstates $\ket{n}$ of the unperturbed Hamiltonian $H_0$, we find (See Appendix~\ref{sec:power_series_expansion_of_the_bures_distance} for details)
\begin{align}
\int dt_1dt_2g_{\mu\nu}&(t,t_1,t_2)f^\mu(t_1)f^\nu(t_2) \nonumber \\ 
&= \frac{1}{2}\sum_{nm}\frac{|\bra{n}\rho_1\ket{m}|^2}{p_n+p_m}\nonumber \\ 
&=\frac{1}{4}F_Q[\rho_1],\label{eq:fisher_info_main}
\end{align}
where $p_n=\bra{n}\rho_0\ket{n}$ is the occupation probability of the eigenstate $\ket{n}$. 
The quantity $F_Q[\rho_1]$ is the quantum Fisher information associated to the leading order correction $\rho_1(t)$~\cite{braunstein1994statistical}. 
While the relation between the Fisher information and the Bures metric is well-known in general, we can now apply Eq.~\eqref{eq:fisher_info_main} to our case of interest by inserting the perturbation theory expression Eq.~\eqref{eq:rho_n} for $\rho_1(t)$. 
Taking the convergence factors to be exponentials $\eta(t)=e^{\epsilon t}$,
we find that the quantum Fisher information is given by
\begin{align}\label{eq:fq_sum_over_states}
&F_Q[\rho_1] = 2\int dt_1dt_2f^\mu(t_1)f^\nu(t_2)\left[\prod_{i=1}^2 e^{\epsilon t_i}\Theta(t-t_i)\right. \nonumber \\ 
&\left.\times\sum_{nm}\frac{(p_n-p_m)^2}{p_n+p_m}e^{i(E_n-E_m)(t_1-t_2)}\bra{n}B_\mu\ket{m}\bra{m}B_\nu\ket{n}\right],
\end{align}  
where $E_n$ are the energy eigenvalues of the states $\ket{n}$, and $\Theta(t)$ is the Heaviside step function. 
Following the logic of Ref.~\cite{hauke2016measuring}, we can use the identity 
\begin{equation}\label{eq:tanhidentity}
\frac{p_n-p_m}{p_n+p_m} = \int d\omega\tanh\frac{\beta\omega}{2}\delta(\omega+E_n-E_m)
\end{equation}
to re-express Eq.~\eqref{eq:fq_sum_over_states} as
\begin{align}\label{eq:fq_in_terms_of_chi}
F_Q&[\rho_1]= -\frac{2}{\pi}\int dt_1dt_2f^\mu(t_1)f^\nu(t_2)\left[\prod_{i=1}^2 e^{\epsilon t_i}\Theta(t-t_i)\right. \nonumber \\ 
&\times\left. \int d\omega e^{i\omega(t_1-t_2)}\tanh\frac{\beta\omega}{2}\chi''_{\nu\mu}(\omega)\right],
\end{align}
where
\begin{equation}\label{eq:chipp_def}
\chi''_{\nu\mu}(\omega) = -\frac{1}{2}\int dt e^{i\omega t}\langle\left[B_\nu(t),B_\mu(0)\right]\rangle_0
\end{equation}
is the spectral density for the linear response of $\langle B_\nu\rangle$ to the perturbation $f^\mu(t)B_\mu$. In particular, $\chi''_{\nu\mu}(\omega)$ is the anti-Hermitian part of the linear response function~\cite{forster1975hydrodynamic}. 
Note that the term in brackets in Eq.~\eqref{eq:fq_in_terms_of_chi} is symmetric under the interchange $(\mu,t_1)\leftrightarrow(\nu,t_2)$ due to the antisymmetry of $\chi''_{\nu\mu}(\omega)$ under $\omega\leftrightarrow -\omega$ in Eq.~\eqref{eq:chipp_def}. 
Furthermore, since the perturbation $f^\mu(t)B_\mu$ is a Hermitian operator, the right hand side of Eq.~\eqref{eq:fq_in_terms_of_chi} is explicitly real. 
Using these properties, we can take two variational derivatives of Eq.~\eqref{eq:fisher_info_main} and use Eq.~\eqref{eq:fq_in_terms_of_chi} to find the time-dependent Bures metric. 
Multiplying by a factor of $1=\lim_{\epsilon\rightarrow 0}e^{-2\epsilon t}$, we find that
\begin{align} 
g_{\mu\nu}(t,t_1,t_2)&\equiv\frac{1}{8}\frac{\delta^2 F_Q[\rho_1]}{\delta f^\mu(t_1) \delta f^\nu(t_2)}e^{-2\epsilon t} \nonumber \\ 
&=-\frac{1}{2\pi}\prod_{i=1}^2 e^{\epsilon(t_i-t)}\Theta(t-t_i)\nonumber \\ 
&\times\mathrm{Re}\left[\int d\omega e^{i\omega(t_1-t_2)}\tanh\frac{\beta\omega}{2}\chi''_{\nu\mu}(\omega)\right].\label{eq:bures_metric_main_result}
\end{align}
Note that time-translation invariance of the unperturbed Hamiltonian ensures that the Bures metric is a function of time differences only,
\begin{equation}
g_{\mu\nu}(t,t_1,t_2)\equiv g_{\mu\nu}(t-t_1,t-t_2),
\end{equation}
which is manifest in Eq.~\eqref{eq:bures_metric_main_result}.

Eq.~\eqref{eq:bures_metric_main_result} is our first main result. 
It generalizes the known expressions for the Bures metric for instantaneous perturbations at short times~Ref.~\cite{hauke2016measuring,balut2025quantum}, as well as the expression for the metric associated to quasistatic, time-independent perturbations~\cite{lambert2023classical}. 
For positive arguments, Eq.~\eqref{eq:bures_metric_main_result} coincides with the real part of the time-dependent quantum geometric tensor introduced recently in Refs.~\cite{ji2025density,scandi2023quantum}. 
Here, however, we have derived our expression by considering the Bures distance on the trajectory of the density matrix as it evolved in time. 
As such, our results go beyond Ref.~\cite{ji2025density} by demonstrating that this time-dependent metric tensor quantifies the geometry of time-dependent evolution of the density matrix in perturbation theory. 
Furthermore, we will see that by keeping track of the convergence factors and Heaviside functions, we can constrain the geometric quantities via causality and thereby examine several experimentally relevant limits of the time-dependent Bures metric.

To see concretely how Eq.~\eqref{eq:bures_metric_main_result} generalizes existing expressions for the Fisher information, we can use it to compute the infinitesimal Bures distance $F_Q[\rho_1]$ for restricted classes of perturbations. 
Geometrically, this amounts to considering the induced metric on a subspace of the space of density matrices, parametrized by the restricted set of perturbations. 
The two classes of perturbations we will consider are ultrashort instantaneous perturbations, and slow, quasistatic (approximately time-independent) perturbations.

\subsection{The Instantaneous Limit}\label{sub:the_instantaneous_limit}
Let us consider the restricted class of instantaneous perturbations
\begin{equation}\label{eq:inst_pert}
f^\mu_{\mathrm{inst}}(t) = f_0^\mu\delta(t),
\end{equation}
where $f^\mu_0$ are time independent and serve as the infinitesimal tangent vectors to a subspace of the space of density matrices. 
Inserting Eq.~\eqref{eq:inst_pert} into Eq.~\eqref{eq:fq_in_terms_of_chi} and taking two derivatives with respect to $f^\mu_0$, we can define the instantaneous Bures metric
\begin{align}\label{eq:inst_metric}
g^\mathrm{inst}_{\mu\nu}(t) &= \frac{1}{8}\frac{\partial^2 F_Q[\rho_1]}{\partial f_0^\mu\partial f_0^\nu} = g_{\mu\nu}(t,t) \nonumber \\ 
&=-\frac{1}{2\pi}\Theta(t)\int d\omega \tanh\frac{\beta\omega}{2} \mathrm{Re}\chi''_{\nu\mu}(\omega).
\end{align}
For $t<0$, $\rho(t<0)=\rho_0$ for an instantaneous perturbation and hence the Bures metric (and in fact the entire Bures distance) is trivial. 
For $t>0$, we see that the instantaneous Bures metric coincides with (one quarter of) the quantum Fisher information as defined in Refs.~\cite{hauke2016measuring,balut2025quantum}. 
Furthermore, the instantaneous metric $g^\mathrm{inst}_{\mu\nu}(t)$ is closely related to the time-dependent metric after a quench. 
Consider a quench perturbation of the form $f^\mu(t)=\Theta(t)g^\mu(t)$ satisfying $g^\mu(0)=f^\mu_0$. 
Inserting this into Eq.~\eqref{eq:fq_in_terms_of_chi} and expanding about $t=0$, we find that Fisher information $F^\mathrm{quench}_Q[\rho_1]$ after the quench is given by
\begin{equation}
F^\mathrm{quench}_Q[\rho_1] = 4 t^2f_0^\mu f_0^\nu \lim_{\tau \rightarrow 0^+} g^\mathrm{inst}_{\mu\nu}(\tau) + \mathcal{O}(t^3),
\end{equation}
in agreement with the definition of the quantum Fisher information given in Ref.~\cite{balut2025quantum}. 
\subsubsection{Density-Density Response in the Instantaneous Limit}
\label{subsec: lineardensity}

As an example, we will consider the instantaneous Bures metric associated to density perturbations, to make contact with the results of Refs.~\cite{balut2025quantum,balut2025quantuma,kruchkov2025topologicalcontrolquantumspeed,wang2025local,onishi2025quantum}. 
We consider a scalar density perturbation (i.e. spatially-varying electric potential) to the system Hamiltonian $H_0$,
\begin{equation}
H = H_0 + \frac{1}{V}\sum_{\bq} \phi_{\bq}(t) \rho_{-\bq},
\end{equation}
where $V$ is the volume of the system. 
In the instantaneous limit, $\phi_{\bq}(t) = \phi_{\bq} \delta(t)$ is a pulse that only perturbs the system at $t=0$. 
From Eq.~\eqref{eq:inst_metric}, we find the instantaneous Bures metric is given by
\begin{equation}\label{eq:inst_metric_for_density_general}
g_{\rho_{\bq'}\rho_\bq }(t,t) = -\frac{1}{2\pi V^2} \Theta(t) \int d\omega \tanh\frac{\beta \omega}{2} \mathrm{Re} \chi''_{ \rho_{\bq}\rho_{\bq'}}(\omega).
\end{equation}
For systems with translation invariance, the spectral density $\chi''_{ \rho_{\bq}\rho_{\bq'}}(\omega)$ can be rewritten in terms of the standard density-density response function as  
\begin{align} \label{eq: linear spectral density for density}
\chi''_{ \rho_{\bq}\rho_{\bq'}}(\omega)  &= \delta_{\bq+\bq',0} \pi \sum_{mn} (p_m-p_n) |\bra{n} \rho_{\bq} \ket{m}|^2\nonumber\\ 
&\times  \delta(\omega+E_n-E_m) \nonumber \\ 
&=V\delta_{\bq+\bq',0}\chi''(\bq,\omega),
\end{align}
where 
\begin{equation}
\chi''(\bq,\omega) = -\frac{1}{2V}\int dt e^{i\omega t} \langle \left[\rho_{\bq}(t),\rho_{-\bq}(0)\right]\rangle_0.
\end{equation}
We see that $\chi''_{\rho_\bq\rho_{\bq'}}(\omega)$ is purely real. 
Thus, we can rewrite the instantaneous metric in Eq.~\eqref{eq:inst_metric_for_density_general} as
\begin{align}\label{eq:simplified_density_metric}
g_{\rho_{\bq{}'}\rho_\bq }(t,t) &= -\frac{1}{V}\delta_{\bq+\bq',0}\frac{1}{2\pi} \Theta(t) \int_{-\infty}^{\infty} d\omega \tanh\frac{\beta \omega}{2} \chi''(\bq,\omega) \nonumber \\ 
&\equiv\frac{1}{V}\delta_{\bq+\bq',0} g_{\bq}(t).
\end{align}
We see that $g_{\bq}(t)$ gives the only nonzero contribution to the instantaneous Bures metric for density perturbations at wavevector $\bq$. 
Eq.~\eqref{eq:simplified_density_metric} is one quarter of the quantum Fisher information associated to density perturbations as defined in Refs.~\cite{balut2025quantum,balut2025quantuma,kruchkov2025topologicalcontrolquantumspeed}. 

Going further, we can use the fluctuation-dissipation theorem 
\begin{equation}
\chi''(\bq,\omega) = -\frac{1}{2} \left(1-e^{-\beta \omega}\right) S(\bq,\omega),
\end{equation}
where $S(\bq,\omega) = \frac{1}{V}\int d t e^{i\omega t} \left\langle \rho_{\bq}(t) \rho_{-\bq}(0) \right\rangle_c$ is the dynamic structure factor (the $c$ subscript denotes the connected correlation function), to rewrite $g_{\bq}(t)$ as
\begin{equation}\label{eq:g_in_terms_of_S}
\begin{aligned}
g_{\bq }(t) = &-\frac{1}{2\pi} \Theta(t) \int_{-\infty}^{\infty} d\omega \tanh\frac{\beta \omega}{2} \chi''(\bq,\omega)\\
=&-\frac{1}{2\pi} \Theta(t)  \int_{0}^{\infty} d\omega \tanh \frac{\beta \omega}{2} \left[\chi''(\bq, \omega)-\chi''(\bq ,-\omega) \right]\\
=&-\frac{1}{2\pi} \Theta(t)  \int_{0}^{\infty} d\omega \tanh \frac{\beta \omega}{2} \left[\chi''(\bq, \omega)+\chi''(-\bq ,\omega) \right]\\
=& \frac{\Theta(t)}{4\pi}  \int_{0}^{\infty} d\omega \tanh \frac{\beta \omega}{2} \left(1-e^{-\beta \omega}\right) \\ &\times\left[S(\bq,\omega)+S(-\bq,\omega)\right].
\end{aligned}
\end{equation}
Note that we have not assumed that our system has inversion or time-reversal symmetry; in the presence of either of these symmetries, $S(\bq,\omega)=S(-\bq,\omega)$.
At zero temperature, Eq.~\eqref{eq:g_in_terms_of_S} for $g_{\bq}(t)$ reduces to
\begin{align}\label{eq: quantum metric for density example at T=0}
g_{\bq }(t) &\underset{T\rightarrow 0}{=}\frac{1}{4\pi} \Theta(t)  \int_{0}^{\infty} d\omega\left[S(\bq,\omega)+S(-\bq,\omega)\right] \nonumber\\ 
&= \frac{1}{4\pi} \Theta(t)  \int_{-\infty}^{\infty} d\omega\left[S(\bq,\omega)+S(-\bq,\omega)\right] \nonumber \\ 
&= \frac{1}{2}\Theta(t)[s_2(\bq)+s_2(-\bq)],
\end{align}
where we used $S(\bq,\omega) = 0$ for $\omega <0$ at zero temperature, and the static structure factor $s_2(\bq)$ is defined as
\begin{equation}\label{eq: integral of S(q,omega)}
s_2(\bq) = \frac{1}{V}\langle\rho_\bq\rho_{-\bq}\rangle_0-N^2/V\delta_{0\bq}.
\end{equation}

We note that Ref.~\cite{tam2024topological} showed that for a one-dimensional noninteracting Fermi gas at small $|q|$, $s_2(q) \sim \frac{|q|}{2\pi} \chi_F$, where $\chi_F$ counts the number of components of the Fermi sea, i.e. half the number of Fermi points. 
Therefore, at small $|q|$ $g_{q}(t) \sim \Theta(t) \frac{|q|}{2\pi} \chi_F$ in one-dimensional Fermi gas. 
This relates the instantaneous Bures metric at small $|q|$ to a topological invariant of the Fermi surface in one dimension.

\subsection{The Quasistatic Limit}\label{sub:the_quasistatic_limit}

In the opposite limit, we can consider a restricted class of time-independent perturbations 
\begin{equation}\label{eq:qs_pert}
f_{\mathrm{q.s.}}^\mu(t) = f_\mathrm{i}^\mu.
\end{equation}
In this case, we expect the time-dependence of the density matrix $\rho(t)$ to arise entirely from the fact that the eigenstates $\ket{n}$ of $H_0$ do not coincide with the instantaneous eigenstate $\ket{n(\vec{f}_\mathrm{i})}$ of $H$. 
This is the setup that is routinely considered in the quantum information literature, where the Fisher information is used to quantify the distinguishability of states that depend parametrically on some external fields $f^\mu_\mathrm{i}$~\cite{lambert2023classical}.  

As in Sec.~\ref{sub:the_instantaneous_limit}, We can insert the quasistatic form Eq.~\eqref{eq:qs_pert} of the perturbation into Eq.~\eqref{eq:fisher_info_main} and take two derivatives with respect to the perturbing fields to define the Bures metric restricted to the subspace of quasi-static perturbations. 
Concretely, we find
\begin{align} \label{eq: time_int_of_g}
g^{\mathrm{q.s.}}_{\mu\nu} &= \frac{1}{8}\frac{\partial^2 F_Q[\rho_1]}{\partial f_\mathrm{i}^\mu\partial f_\mathrm{i}^\nu}e^{-2\epsilon t} \nonumber \\ 
&=\int dt_1\int dt_2 g_{\mu\nu}(t-t_1,t-t_2),
\end{align}
which shows that the quasistatic Bures metric is the integral of the time-dependent Bures metric over both its time arguments. Mathematically, the time integral appears since the Bures distance in this limit is a function of $\{f^\mu \}$ since the external fields are independent of time in this limit. Thus, the sum over $\mu$ and the integral over $t$ are decoupled, and we can perform the time integral independent of the external fields. Physically, Eq.~\eqref{eq: time_int_of_g} can be interpreted as the accumulation of the quantum metric contributions over the time domain, constrained by the causal factors in Eq.~\eqref{eq:bures_metric_main_result}.

We can use our explicit expression Eq.~\eqref{eq:bures_metric_main_result} for the time-dependent Bures metric in terms of the spectral density to carry out the time integrals, yielding
\begin{align}\label{eq:g_qs_intermediate_step}
g^{\mathrm{q.s.}}_{\mu\nu}=-\lim_{\epsilon\rightarrow 0}\frac{1}{2\pi}\int d\omega \frac{e^{2\epsilon t}}{\omega^2+\epsilon^2}\tanh\frac{\beta\omega}{2}\mathrm{Re}\chi''_{\nu\mu}(\omega),
\end{align}
where we have made the limit $\epsilon\rightarrow 0$ explicit. 
We can carefully take the limit as $\epsilon\rightarrow 0$ by noting that the real part of the spectral density is an odd function of $\omega$, hence near $\omega=0$ the integrand in Eq.~\eqref{eq:g_qs_intermediate_step} is regular in the limit of $\epsilon\rightarrow 0$. 
Thus, the Bures metric on the manifold of quasistatic perturbations is time-independent and given by
\begin{equation} \label{quantum metric at q.s. limit}
g_{\mu\nu}^\mathrm{q.s.}= -\frac{1}{2\pi}\int d\omega \frac{1}{\omega^2}\tanh\frac{\beta\omega}{2}\mathrm{Re}\chi''_{\nu\mu}(\omega).
\end{equation}
This coincides with the quantum Fisher information computed in time-independent perturbation theory for time-independent perturbations~\cite{lambert2023classical}. 

\subsubsection{Current-Current Response in the Quasistatic Limit}\label{ssub:current_current_response_in_the_quasistatic_limit}

As a specific example, we can consider the quasistatic metric associated to perturbations of the system by a constant vector potential. 
For systems with periodic boundary conditions, this is equivalent to studying the change in the density matrix under twists of boundary conditions. 
A constant vector potential enters the Hamiltonian through minimal coupling, where the momentum operator for each particle is shifted by a constant $p_\mu\rightarrow p_\mu-A_\mu$. 
Here $\mu$ ranges over spatial directions of the system, and for simplicity we have set the charge of the particles $q=1$. 
The perturbed Hamiltonian can be expanded order-by-order in the vector potential as
\begin{equation}\label{eq:vec_potential_ham}
H_A=H_0 -\sum_{n=1}^{\infty}\frac{1}{n!}j^{\mu_1\dots\mu_n}A_{\mu_1}\dots A_{\mu_n},
\end{equation}
with
\begin{equation}\label{eq:current_vertex_defs}
 j^{\mu_1 \mu_2 ... \mu_n} =-\frac{\partial^n H_A}{\partial A_{\mu_1} \partial A_{\mu_2}...\partial A_{\mu_n}}\Bigg|_{\mathbf{A}= 0}.
 \end{equation} 
Since the first-order perturbed density matrix $\rho_1(t)$ depends only on
\begin{equation}
j^\mu = -\left.\frac{\partial H_A}{\partial A_\mu}\right|_{\mathbf{A}=0},
\end{equation}
the spectral density $[\chi'']^{\mu\nu}(\omega)$ that enters the metric Eq.~\eqref{eq:bures_metric_main_result} is
\begin{align}\label{eq:current_spectral_density}
[\chi'']^{\mu\nu}(\omega) &= -\frac{1}{2}\int dt e^{i\omega t}\langle\left[j^\mu(t),j^\nu(0)\right]\rangle_0 \nonumber\\
&=\pi \sum_{m\neq n} (p_m-p_n) \bra{n} j^\nu \ket{m} \bra{m} j^{\mu} \ket{n}\nonumber \\ 
&\times \delta(\omega+E_n-E_m).
\end{align}
We have made explicit for later convenience that the diagonal terms $m=n$ do not contribute to the spectral density and so can be excluded from the sum. 
For systems with normalizable eigenstates (such as those with periodic boundary conditions) this follows straightforwardly from the definition of the spectral density in terms of a commutator; however when we consider the current response in extended systems, it is important to explicitly exclude the possibly divergent diagonal terms. 
Note that this exclusion is also manifest in Eq.~\eqref{eq:fq_sum_over_states}, which shows the diagonal terms should not contribute to the metric since they do not involve changes to the density matrix.

The spectral density $[\chi'']^{\mu\nu}(\omega)$ is closely related to the Hermitian part of the conductivity tensor
\begin{equation}\label{eq:sigma_kubo}
\begin{aligned}
\sigma^{\mu\nu}(\omega) &= -\frac{i}{V\omega^+}\langle j^{\mu\nu}\rangle_0 \\ &+ \frac{1}{V\omega^+}\int_0^\infty dt e^{i\omega^+t}\langle\left[j^\mu(t),j^\nu(0)\right]\rangle_0.
\end{aligned}
\end{equation} 
Focusing on the Hermitian part 
\begin{equation}\label{eq:sigma_hermitian_def}
\sigma_{\mathrm{abs}}^{\mu\nu}(\omega)=\frac{1}{2}\left[\sigma^{\mu\nu}(\omega)+\sigma^{\nu\mu}(\omega)^*\right]
\end{equation}
of the conductivity and isolating the terms proportional to $\delta(\omega)$, we find~\cite{resta2018drude}
\begin{equation}\label{eq:sigma_spectral_density_reln}
\sigma_{\mathrm{abs}}^{\mu\nu}(\omega)=\pi\delta(\omega)D^{\mu\nu} - \mathrm{P}\frac{1}{V\omega}[\chi'']^{\mu\nu}(\omega),
\end{equation}
where 
\begin{equation}\label{eq:drude_weight}
D^{\mu\nu} = -\frac{1}{V}\langle j^{\mu\nu}\rangle_0 -\frac{i}{2V}\int dt e^{-\epsilon|t|}\mathrm{sign}(t)\langle\left[j^\mu(t),j^\nu(0)\right]\rangle_0
\end{equation}
is the Drude weight, and $\mathrm{P}$ denotes the principal value~\footnote{Note that the integral weighted by $\mathrm{sign}(t)$ in Eq.~\eqref{eq:drude_weight} is the time-domain representation of the Hilbert transform, such that this contribution of the Drude weight comes from evaluating $[\chi']^{\mu\nu}(0)$, where $[\chi']^{\mu\nu}$ is the Kramers-Kronig partner of $[\chi'']^{\mu\nu}$}. 
We can then insert Eqs.~\eqref{eq:sigma_spectral_density_reln} and \eqref{eq:current_spectral_density} into our general expression Eq.~\eqref{quantum metric at q.s. limit} for the quasistatic Bures metric to find
\begin{align}\label{eq:g_qs_current_general}
g_{\mathrm{q.s.}}^{\mu\nu} &= \frac{V}{2\pi}\int d\omega \frac{1}{\omega}\tanh\frac{\beta\omega}{2}\left[\mathrm{Re}\sigma^{\mu\nu}_\mathrm{abs}(\omega)-\pi D^{\mu\nu}\delta(\omega)\right] \nonumber\\ 
&=\frac{V}{\pi}\int_0^\infty d\omega \frac{1}{\omega}\tanh\frac{\beta\omega}{2}\left[\mathrm{Re}\sigma^{\mu\nu}_\mathrm{abs}(\omega)-\pi D^{\mu\nu}\delta(\omega)\right] \nonumber \\ 
&=\frac{V}{\pi}\int_{0^+}^\infty  d\omega \frac{1}{\omega}\tanh\frac{\beta\omega}{2}\mathrm{Re}\sigma^{\mu\nu}_{\mathrm{abs}}(\omega),
\end{align}
where we used the fact that $\sigma^{\mu\nu}_{\mathrm{abs}}(\omega)$ is an even function of $\omega$, and $0^+$ indicates that the lower limit of integration should be taken to zero from above \emph{after} performing the integration to exclude the Drude weight. 

Note also that we write Eq.~\eqref{eq:g_qs_current_general} with upper indices, since the current operator and conductivity tensor naturally have raised spatial indices. 
We emphasize that unlike the case in general relativity, $g^{\mu\nu}_\mathrm{q.s.}$ is a metric, not an inverse metric. 
It originates as the leading-order contribution to the infinitesimal Bures distance Eq.~\eqref{eq:buresdistance}, where the naturally-lower-indexed vector potentials $A_\mu$ are the infinitesimal line elements in parameter space (components of tangent vectors).

Eq.~\eqref{eq:g_qs_current_general} shows that the quasistatic Bures metric associated with constant vector potential deformations (twisted boundary conditions) 
is given by a finite-temperature generalization of the Souza-Wilkens-Martin sum rule; indeed at zero temperature we recover the known result~\cite{souza2000polarization,souza2025optical,verma2024instantaneous,verma2025framework}
\begin{align} \label{eq: SWM sum rule}
g_{\mathrm{q.s.}}^{\mu\nu} &\underset{T\rightarrow 0}{=} \frac{V}{\pi}\int_{0^+}^\infty  d\omega \frac{1}{\omega}\mathrm{Re}\sigma^{\mu\nu}_{\mathrm{abs}}(\omega).
\end{align}
To understand this, first note that at $T=0$, the density matrix reduces to a pure (ground) state. 
Second, a constant vector potential $\mathbf{A}$ does not generate any electromagnetic field but creates a gauge flux. 
Physically, this is equivalent to imposing twisted boundary conditions parameterized by twist angles $\bm{\theta}$ across the system boundaries, as we discussed in Sec.~\ref{sub:interpretation_of_the_geometry}. Thus the quasistatic Bures metric becomes equal to the twist-angle quantum metric.
Furthermore, we note that in the noninteracting limit, the results of Ref.~\cite{souza2000polarization} show that the twist-angle quantum metric of the many-body ground state reduces to the integral of the single-particle Fubini-Study metric over the first Brillouin zone.

Additionally, in extended systems where off-diagonal matrix elements of the position operator can be defined, we can use the definition
\begin{equation}
\bra{n} j^\mu \ket{m} = i (E_n-E_m) \bra{n} X^\mu \ket{m}
\end{equation}
for $n\neq m$ to rewrite the spectral density as
\begin{align}\label{eq:jj_spectral_density_in_terms_of_x}
[\chi'']^{\mu\nu}(\omega) = \pi \omega^2 \sum_{m\neq n} (p_m-p_n)& \bra{n} X^\nu \ket{m} \bra{m} X^{\mu} \ket{n}\nonumber\\ 
&\times \delta(\omega+E_n-E_m).
\end{align}
Inserting this into Eq.~\eqref{quantum metric at q.s. limit} and taking the zero-temperature limit yields
\begin{equation}\label{eq:g_qs_in_terms_of_X}
g_{\mathrm{q.s.}}^{\mu\nu}\underset{T\rightarrow 0}{=}\mathrm{Re} \left[\sum_{n>0} \bra{0}X^\mu \ket{n} \bra{n} X^\nu \ket{0} \right].
\end{equation}
As shown in Ref.~\cite{souza2000polarization}, the second cumulant of the many-body position operator on the right-hand side of Eq.~\eqref{eq:g_qs_in_terms_of_X} at zero temperature is the many-body Fubini-Study metric (quantum metric tensor) associated with twisting boundary conditions in the ground state. 
Thus, we see that the quasistatic metric Eq.~\eqref{eq:g_qs_current_general} generalizes the many-body quantum metric tensor of the ground state to nonzero temperature. 
Additionally, our analysis shows that at zero temperature the many-body quantum metric tensor coincides with the quasistatic Bures metric quantifying the Fisher information associated with twisting boundary conditions.

To conclude this subsection, we recall for later convenience that in the case of crystalline noninteracting fermion systems, we can re-express the spectral density Eq.~\eqref{eq:jj_spectral_density_in_terms_of_x} in terms of single-particle dipole transition matrix elements. 
In particular, introducing single-particle Bloch eigenstates $\ket{\psi_{\alpha\kk}}=e^{i\kk\cdot\mathbf{x}}\ket{u_{\alpha\kk}}$ with single-particle energies $E_{\alpha\kk}$, we have
\begin{equation}\label{eq:jj_spectral_density_in_terms_of_x_single_particle}
[\chi'']^{\mu\nu}(\omega) = \pi \omega^2 \sum_{\substack{\alpha\neq \beta \\ \kk}} \Delta f_{\beta\alpha\kk} r^{\nu}_{\alpha \beta \kk} r^{\mu}_{\beta \alpha\kk}\delta(\omega+\Delta E_{\alpha \beta\kk}),
\end{equation}
where $\Delta f_{\beta\alpha\kk}=f_{\beta\kk}-f_{\alpha\kk}$ is a difference of Fermi distributions (occupation probabilities), $\Delta E_{\alpha \beta\kk}=E_{\alpha\kk}-E_{\beta\kk}$ and the dipole matrix elements are defined as the off-diagonal components of the Berry connection,
\begin{equation}\label{eq:rab_def}
r_{\alpha \beta \kk}^{\mu} = \bra{u_{\alpha \kk}} i \partial^{\mu} \ket{u_{\beta \kk}},
\end{equation}
where $\partial^{\mu} \equiv \partial/\partial{k_{\mu}}$, and the inner product of cell-periodic functions is normalized to one unit cell. 
Inserting Eq.~\eqref{eq:jj_spectral_density_in_terms_of_x_single_particle} into the definition Eq.~\eqref{quantum metric at q.s. limit} and taking the zero temperature limit shows that the quasistatic Bures metric for noninteracting fermions gives
\begin{equation} \label{eq: free-fermion many-body quantum metric}
g^{\mu \nu}_{\mathrm{q.s.}} \underset{T\rightarrow 0}{=}\sum_{\substack{\alpha \in \mathrm{unocc}\\ \beta \in \mathrm{occ}, \, \kk }} \mathrm{Re} \left(r^{\mu}_{\alpha \beta \kk} r^{\nu}_{\beta \alpha \kk} \right) = \sum_{\substack{\alpha \in \mathrm{unocc}\\ \beta \in \mathrm{occ}, \, \kk }} g^{\mu \nu}_{\alpha \beta}(\mathbf{k}) ,
\end{equation}
where we recognize the right-hand side is the Brillouin-zone average of the single-particle quantum metric tensor $g^{\mu \nu}(\mathbf{k}) =\sum_{\substack{\alpha \in \mathrm{unocc}\\ \beta \in \mathrm{occ} }} \mathrm{Re} \left(r^{\mu}_{\alpha \beta \kk} r^{\nu}_{\beta \alpha \kk} \right)$~\cite{souza2000polarization,ahn2022riemannian,resta2025nonadiabatic}.

\subsection{Relation between Quasistatic and Instantaneous Metrics}

In Secs.~\ref{sub:the_instantaneous_limit} and \ref{sub:the_quasistatic_limit} we examined the time-dependent Bures metric restricted to two seemingly opposite classes of perturbations, namely instantaneous pulses and quasistatic, time-independent perturbations. 
Here we show that for pairs of observables related by a conservation law, we can relate the instantaneous and quasistatic Bures metrics. 
In particular, let us consider two perturbations
\begin{align}
H_1 &= H_0 + \sum_{\mu} f^{\mu}(t) B_\mu,\nonumber \\ 
 H_2 &= H_0+\sum_{\mu}h^\mu(t) C_\mu,
\end{align}
where the pair of observables $B_\mu$ and $C_\mu$ satisfy the equations of motion
\begin{equation}\label{eq:C_B_relation}
C_\mu = \partial_t B_\mu = i\left[H_0,B_\mu\right].
\end{equation}
Using Eq.~\eqref{eq:inst_metric} the Bures metric corresponding to the Hamiltonian $H_1$ in the instantaneous limit is
\begin{equation}\label{eq:Bmu_inst_metric}
g^{\mathrm{inst}}_{B_\mu B_\nu}(t) = - \frac{1}{2\pi} \Theta (t)\int d\omega  \tanh \frac{\beta \omega}{2}\mathrm{Re}\chi''_{B_\nu B_\mu}(\omega),
\end{equation}
where
\begin{equation}
\chi''_{B_\nu B_\mu}(\omega) = -\frac{1}{2} \int dt e^{i \omega t} \left\langle \left[B_{\nu}(t), B_{\mu}(0)\right]\right\rangle_0. 
\end{equation}
Using Eq.~\eqref{eq:C_B_relation}, we can integrate by parts twice to relate $\chi''_{B_\nu B_\mu}(\omega)$ to the spectral density $\chi''_{C_\nu C_\mu}(\omega)$ for the Hamiltonian $H_2$,
\begin{equation}
\begin{aligned}
\omega^2 \chi''_{B_\nu B_\mu}(\omega) &= \frac{1}{2} \int dt \frac{d^2}{dt^2}\left(e^{i \omega t} \right) \left\langle \left[B_{\nu}(0), B_{\mu}(-t)\right]\right\rangle_0\\
&=-\frac{1}{2} \int dt e^{i \omega t}  \left\langle \left[  C_{\nu}(t), C_{\mu}(0)\right]\right\rangle_0\\
&=\chi''_{C_\nu C_\mu}(\omega).
\end{aligned}
\end{equation}
Substituting this into Eq.~\eqref{eq:Bmu_inst_metric} and comparing with our definition Eq.~\eqref{quantum metric at q.s. limit}, we find that
\begin{align}
g^{\mathrm{inst}}_{B_\mu B_\nu}(t) &= - \frac{1}{2\pi} \Theta (t)\int d\omega  \tanh \frac{\beta \omega}{2}\frac{\mathrm{Re}\chi''_{C_\nu C_\mu}(\omega)}{\omega^2} \nonumber \\
&=\Theta (t) g^{\mathrm{q.s.}}_{C_\mu C_\nu}. 
\end{align}
Thus, at any positive time, the instantaneous metric associated with the perturbation $B_\mu$ coincides with the quasistatic metric associated with the perturbation $C_\mu$.

We can apply this to the case of instantaneous density response as in Sec.~\ref{subsec: lineardensity}. 
Using the charge conservation equation 
\begin{equation}\label{eq:continuity}
\partial_t \rho_{\bq}(t) =- i\bq \cdot \mathbf{j}_{\bq}(t),
\end{equation}
 we find in the limit of small $|\bf{q}|$
\begin{equation}
\begin{aligned}
\omega^2 \chi''(\bq,\omega) &= \frac{1}{V} \omega^2 \chi''_{\rho_{\bq} \rho_{-\bq}}(\omega) = \frac{1}{V} \chi''_{\partial_t \rho_{\bq}, \partial_t \rho_{-\bq}} (\omega) \\
&=-\frac{1}{2V} q_\mu q_\nu \int dt e^{i\omega t} \left\langle \left[  j^{\nu}_{\bq}(t), j^{\mu}_{-\bq}(0)\right]\right\rangle_0\\
 & \underset{\bq \rightarrow 0}{=}\frac{q_\mu q_\nu}{V}\left(-\frac{1}{2}\right)\int dt e^{i\omega t} \left\langle \left[  j^{\nu}(t), j^{\mu}(0)\right]\right\rangle_0\\
&\underset{\bq \rightarrow 0}{=}\frac{q_\mu q_\nu}{V} [\chi'']^{\nu \mu}(\omega ),
\end{aligned}
\end{equation}
where $[\chi'']^{\nu \mu}(\omega )$ is the spectral density for the uniform current operator defined in Eq.~\eqref{eq:current_spectral_density}. 
Applying this to our expression Eq.~\eqref{eq:simplified_density_metric} for the instantaneous metric associated to density perturbations, we find
\begin{align}
g_{\bq}(t) &= -\frac{1}{2\pi} \Theta(t) \int d\omega \tanh\frac{\beta \omega}{2} \chi''(\bq,\omega) \nonumber \\
&=-\frac{1}{2\pi} \Theta(t) \frac{q^\mu q^\nu}{V}\int d\omega \frac{1}{\omega^2} \tanh\frac{\beta \omega}{2}  [\chi'']^{\nu \mu}(\omega ) \nonumber \\ 
&= \Theta(t) q_\mu q_\nu \frac{1}{V} g^{\mu \nu}_{\mathrm{q.s.}},
\end{align}
where $g^{\mu \nu}_{\mathrm{q.s.}}$ is the quasistatic metric associated to perturbations by a uniform electric field given in Eq.~\eqref{eq:g_qs_current_general}. 
Thus, we see that the instantaneous Bures metric measured via density-density response is proportional to the longitudinal part of the quasistatic Bures metric associated to electric field perturbations, as discussed in Refs.~\cite{balut2025quantum,resta2006polarization}. We emphasize that the quasistatic Bures metric associated with electric field perturbations measures the strength of polarization fluctuations in materials. For insulators, this means that the Bures metric on this parameter space allows for the experimental determination of electron localization from density-density or conductivity measurements. This approach has recently been used to experimentally probe electron localization and entanglement in ionic and covalent insulators~\cite{balut2025quantum,balut2026fundamental}. It can also be shown using the relation between the Bures metric and linear response theory that there are electrostatic bounds on these two Bures metrics that are satisfied by any material and constrain the  degree of entanglement present in the bonds of insulators~\cite{onishi2025quantum}. Turning to metals, one crucial point about the longitudinal part of the quasistatic Bures metric is that it remains well-defined even when there is no energy gap. This is because (in contrast to the transverse part) it incorporates electron-electron correlations and screening due to the Coulomb interaction. As such, this Bures metric provides a witness to entanglement and information contained in correlated metals. This has recently been used to experimentally probe the differences between Fermi liquids and strange metals~\cite{balut2025quantuma,fang2025amplified,chowdhury2026information}.

\subsection{Bures Metric in the Infinite Time Limit}
Using our time-dependent formalism, we can now examine the Bures metric and quantum geometry beyond the instantaneous and quasistatic limits. 
As an experimentally-relevant example, we can consider the infinite-time limit of the Bures metric.  We imagine a scenario in which we slowly ramp up an external probe of a system, allow the system to interact with the external field for some time, and then slowly turn off the external field. 
At the end of this process, we can ask by what distance the density matrix has moved, as measured by the Bures metric. 
This corresponds to evaluating the infinitesimal distance in Eq.~\eqref{eq:fisher_info_main} and Eq.~\eqref{eq:fq_in_terms_of_chi} on a restricted submanifold of parameter space spanned by perturbations $\bar{f}^\mu(t)$ that decay sufficiently fast as $|t|\rightarrow\infty$; in particular, we will assume that $\bar{f}^{\mu}(t)$ decays exponentially fast as $|t|\rightarrow\infty$. 
For $\bar{f}^\mu(t)$ satisfying this property, we can safely ignore the convergence factors $e^{\epsilon t}$ in our expression Eq.~\eqref{eq:fq_in_terms_of_chi} for the Fisher information. 

We can now use Eqs.~\eqref{eq:fisher_info_main} and \eqref{eq:fq_in_terms_of_chi} to find the infinitesimal Bures metric after the perturbation has turned off. 
Taking the $t\rightarrow\infty$ limit, we find
\begin{align}\label{eq:fq_infinite_time}
\bar{F}_Q &\equiv \lim_{t\rightarrow\infty} F_Q[\rho_1]  \nonumber \\ 
&= -\frac{2}{\pi}\int dt_1dt_2d\omega\Big[ \bar{f}^\mu(t_1)e^{i\omega t_1} \bar{f}^\nu(t_2)e^{-i\omega t_2} \nonumber \\ 
&\left.\times \tanh\frac{\beta\omega}{2}\chi''_{\nu\mu}(\omega)\right] \nonumber \\ 
&=-\frac{2}{\pi}\int d\omega \bar{f}^\nu(-\omega)\left[\tanh\frac{\beta\omega}{2}\chi''_{\nu\mu}(\omega)\right]\bar{f}^\mu(\omega),
\end{align}
We see that we can view the Fourier components $\bar{f}^\mu(\omega)$ of the perturbation as the infinitesimal tangent vectors to the space of density matrices. 
Taking two variational derivatives to define the frequency-dependent Bures metric in the infinite time limit, we find
\begin{align}\label{eq:infinite_time_g}
\bar{g}_{\mu\nu}(\omega_1,\omega_2) &= \frac{1}{8}\frac{\delta \bar{F}_Q}{\delta\bar{f}^\mu(\omega_1)\delta\bar f^\nu(\omega_2)} \nonumber \\ 
&=-\frac{1}{2\pi}\delta(\omega_1+\omega_2)\tanh\frac{\beta\omega_1}{2}\chi''_{\mu\nu}(\omega_1).
\end{align}
This shows that each Fourier component of the pulse $\bar{f}(t)$ contributes an amount $-\tanh\frac{\beta\omega}{2}\chi''_{\mu\nu}(\omega)/2\pi$ to the distance between $\rho(t\rightarrow\infty )$ and $\rho_0$. 
This result has a very simple physical interpretation. 
Using Eq.~\eqref{eq:chipp_def}, we can rewrite the product $f^\nu(-\omega)\chi''_{\nu\mu}(\omega)f^\mu(\omega)$ in Eq.~\eqref{eq:fq_infinite_time} as
\begin{align}\label{eq:fgr}
-\bar{f}^\nu(-\omega)&\chi''_{\nu\mu}(\omega)\bar{f}^\mu(\omega) \nonumber \\ &= \pi\sum_{n\neq m}\Big[(p_n-p_m)\left|\bra{m}\bar{f}^\mu(\omega)B_\mu\ket{n}\right|^2\nonumber \\ 
&\times\delta(\omega+E_n-E_m)\Big],
\end{align}
where we have used the fact that $f^\mu(t)$ is real to write $\bar{f}^\mu(-\omega) = [\bar{f}^\mu(\omega)]^*$. 
We recognize the summand in Eq.~\eqref{eq:fgr} as (half) the Fermi's golden rule transition rate from state $\ket{n}$ to state $\ket{m}$ induced by the perturbation at frequency $\omega$, per unit frequency squared. 
Additionally, the $\tanh\beta\omega/2$ factor in Eq.~\eqref{eq:infinite_time_g} measures by how much a transition at frequency $\omega$ can change the density matrix, by virtue of Eq.~\eqref{eq:tanhidentity}: it gives the difference in occupation probabilities between the two states in the transition, weighted by their average occupation probability, interpolating from $0$ when both states have the same occupation probability, to a maximum of $1$ when one state is occupied and the other unoccupied. 

Viewed in this light, Eq.~\eqref{eq:infinite_time_g} for the long-time Bures metric counts the number of transitions induced by the perturbation at frequency $\omega$, weighted by a thermal factor that accounts for how distinguishable the initial and final state are relative to the equilibrium thermal distribution, as quantified by their relative difference in probability.
Intuitively, this shows that after long times, the infinitesimal Bures distance between the equilibrium and perturbed density matrix is given by counting how many transitions to thermally inaccessible states were induced by the perturbation. 
This gives a geometric interpretation to Fermi's golden rule. 
At low temperatures, $\tanh\beta\omega/2\rightarrow \mathrm{sign}(\omega)$, yielding
\begin{equation}\label{eq:bar_g_t_0}
\bar{g}_{\mu\nu}(\omega_1,\omega_2)\underset{T\rightarrow0}{=}-\frac{1}{2\pi}\delta(\omega_1+\omega_2)\mathrm{sign}(\omega_1)\chi''_{\mu\nu}(\omega_1).
\end{equation}
 Turning to the Fisher information Eq.~\eqref{eq:fq_infinite_time}, we can use Eqs.~\eqref{eq:infinite_time_g} and \eqref{eq:bar_g_t_0}, and exploit the symmetry of $\chi''_{\mu\nu}$ to find
\begin{equation}
\bar{F}_Q\underset{T\rightarrow0}{=}-\frac{4}{\pi}\int_0^\infty d\omega \bar{f}^\nu(-\omega)\chi''_{\nu\mu}(\omega)\bar{f}^\mu(\omega).
\end{equation}
At $T=0$, we then see that the Fisher information (and hence the Bures distance) is proportional to the total number of transitions out of the ground state induced by the perturbation.

On the opposite extreme, we can consider the classical limit $T\rightarrow\infty$, in which case we can Taylor expand 
\begin{equation}
\tanh\frac{\beta\omega}{2}\approx \frac{\beta\omega}{2}+\mathcal{O}(T^{-2}).
\end{equation}
Inserting this into Eq.~\eqref{eq:infinite_time_g}, we find that the classical limit of the Bures metric at infinite times is given by
\begin{align}
\bar{g}_{\mu\nu}(\omega_1,\omega_2)&\underset{T\rightarrow\infty}{=}\frac{1}{2 T}\delta(\omega_1+\omega_2)\left[-\frac{\omega_1}{2\pi}\chi''_{\mu\nu}(\omega_1)\right] \nonumber \\ 
&\underset{T\rightarrow\infty}{=}\frac{1}{2 T}\delta(\omega_1+\omega_2)W_{\mu \nu}(\omega),
\end{align}
where we recognize $W_{\mu\nu}(\omega)$ as the total work done on the system per unit frequency per unit intensity of the perturbation~\cite{forster1975hydrodynamic}. 
This gives a geometric interpretation to power dissipation: for frequencies much smaller than the temperature, the dissipated power at that frequency is proportional to the Bures distance between the initial equilibrium density matrix and the density matrix after applying the perturbation.

Our results, particularly Eq.~\eqref{eq:infinite_time_g}, highlight the geometric interpretation of the spectral density $\chi''_{\mu\nu}(\omega)$ for any perturbation. 
This shows that the anti-Hermitian part of the response function $\chi_{\mu\nu}(\omega)$ is proportional to the infinite time limit of the Bures metric associated to the perturbation $f^\mu(\omega)B_\mu$. 
In the particular case of a uniform vector potential, this implies via Eq.~\eqref{eq:sigma_spectral_density_reln} that the Bures metric in the long time limit is proportional to the regular part $\sigma^{\mu\nu}_\mathrm{abs}(\omega)-D^{\mu\nu}\delta(\omega)$ of the Hermitian component of the conductivity tensor. 
This generalizes to all temperatures the recent findings of Ref.~\cite{resta2025nonadiabatic} where it was pointed out that the regular part of the zero-temperature conductivity tensor could be given a geometric interpretation.

We have thus seen that the time-dependent Bures metric introduced in Eq.~\eqref{eq:bures_metric_main_result} provides a unifying picture for quantum geometry in linear response. 
In the instantaneous limit we recover the geometric interpretation of the density-density response function as well as the relation of Ref.~\cite{hauke2016measuring} between spectral density and short-time Fisher information. 
On the other hand, the quasistatic limit of the time-dependent metric allowed us to derive a finite-temperature generalization of the Souza-Wilkens-Martin sum rule relating conductivity to the quantum metric tensor on the space of twisted boundary conditions. 
Finally, in the infinite time limit we saw that the time-dependent Bures metric reveals the quantum geometric meaning of Fermi's golden rule and energy dissipation. 
Next, we will move to next order in the perturbation and study the connection between second-order time-dependent perturbation theory and quantum geometry via the Bures metric compatible Christoffel symbols.

\subsection{Summary}
In this section, we have studied three specific limits of the Bures metric and showed that, we not only unifies geometric results into a single framework valid for finite-temperature and interacting systems, but also provides a novel geometric interpretation of Fermi's Golden rule. To summarize:
\begin{enumerate}
\item By considering current perturbations in the quasistatic limit, we recover the finite-temperature generalization of the twist-angle quantum metric and the Souza-Wilkens-Martin sum rule.

\item By considering density perturbations in the instantaneous limit, we establishes a relationship between the metric/quantum Fisher information and the dynamic structure factor, connecting the geometric quantity to experimentally measurable observables.

\item We identify the infinite-time metric as the transition rate in Fermi's Golden Rule.
\end{enumerate}

For clarity, the explicit expressions for the Bures distance and the zero-temperature metrics under specific perturbations are listed in Table~\ref{table1: Metric_and_limits}. 

\renewcommand{\arraystretch}{1.8}  % 1.5× default line spacing in table
\begin{table*}[t] 
\centering
\begin{tabular}{cccc}
\toprule 
\textbf{Limits} & \textbf{Bures distance at $\mathcal{O}(\kappa^2)$} & \textbf{Perturbations} & \textbf{Bures metric at $T=0$} \\
\midrule
Instantaneous   & $\frac{-1}{2\pi}\Theta(t) f^\mu f^\nu \int d\omega \tanh\frac{\beta\omega}{2} \mathrm{Re}\chi''_{\nu\mu}(\omega)$ \eqref{eq:inst_metric}  & $\ \frac{1}{V}\sum_{\mathbf{q}} \phi_{\mathbf{q}} \delta(t) \rho_{-\mathbf{q}}\ $  & $\frac{\Theta(t)}{4\pi V}   \int d\omega\left[S(\mathbf{q},\omega)+S(-\mathbf{q},\omega)\right]$  \eqref{eq: quantum metric for density example at T=0} \\
Quasistatic     & $\frac{-1}{2\pi}f^\mu f^\nu \int d\omega \frac{1}{\omega^2}\tanh\frac{\beta\omega}{2}\mathrm{Re}\chi''_{\nu\mu}(\omega)$ \eqref{eq:g_qs_intermediate_step}  & $-j^\mu A_{\mu}+\dots$  & $\frac{V}{\pi}\int_{0^+}^\infty  d\omega \frac{1}{\omega}\mathrm{Re}\sigma^{\mu\nu}_{\mathrm{abs}}(\omega)$ \eqref{eq: SWM sum rule} \\
Infinite Time   & $\frac{-1}{2\pi}\int d \omega \bar{f}^{\nu}(-\omega) \tanh\frac{\beta \omega }{2} \chi''_{\mu \nu}(\omega) \bar{f}^\mu(\omega)$ \eqref{eq:fq_infinite_time} & any & $\frac{-1}{2\pi}\delta(\omega_1+\omega_2)\mathrm{sign}(\omega_1) \chi''_{\mu\nu}(\omega_1)$ \eqref{eq:bar_g_t_0}  \\
\bottomrule
\end{tabular}
\caption{Summary of limits, leading-order Bures distance formulas, perturbations, and zero-temperature Bures metrics.} \label{table1: Metric_and_limits}
\end{table*}  
\renewcommand{\arraystretch}{1}

% section bures_metric_fisher_information (end)
\section{Bures Connection and Second-Order Response} % (fold)
\label{sec:bures_connection_and_second_order_response}

We now turn back to Eqs.~\eqref{eq:metric_and_connection_def} and \eqref{eq:connection_as_var_deriv} to compute the time-dependent Bures metric. 
As shown in Appendix~\ref{sec:power_series_expansion_of_the_bures_distance}, we can Taylor expand the right-hand side of Eq.~\eqref{eq:buresimplicit} in terms of matrix elements of the perturbed density matrix to find
\begin{equation}
\begin{aligned}
\int dt_1dt_2dt_3 \Gamma_{\lambda\mu\nu}(t,t_1,t_2,t_3)f^\mu(t_1)f^\nu(t_2)f^\lambda(t_3)\\
\equiv \Gamma^{\mathrm{f}}(t)+\Gamma^\mathrm{in}(t),
\end{aligned}
\end{equation}
where the two contributions $\Gamma^{\mathrm{in}}(t)$ and $\Gamma^\mathrm{f}(t)$ are defined as
\begin{equation} \label{eq: intrinsic part of Bures connection}
\Gamma^{\mathrm{in}}(t)=-\frac{1}{2}\sum_{n\ell\ell'}\frac{\bra{n}\rho_1\ket{\ell'}\bra{\ell'}\rho_1\ket{\ell}\bra{\ell}\rho_1\ket{n}}{(p_\ell+p_n)(p_n+p_{\ell'})},
\end{equation}
and
\begin{align}
\label{eq: response part of Bures connection}
\Gamma^{\mathrm{f}}(t)=\frac{1}{2}\sum_{n\ell}\frac{\bra{n}\rho_2\ket{\ell}\bra{\ell}\rho_1\ket{n}+\bra{n}\rho_1\ket{\ell}\bra{\ell}\rho_2\ket{n}}{p_n+p_\ell}.
\end{align}

We see that $\Gamma^{\mathrm{f}}(t)$ depends on both the first-order and second-order perturbations of the density matrix, while $\Gamma^\mathrm{in}(t)$ depends only on the first-order perturbation. 
As such, they have distinct origins. $\Gamma^{\mathrm{f}}(t)$ arises from replacing $ \rho_1$ of $F_Q[\rho_1]$ in Eq.~\eqref{eq:fisher_info_main} by the total correction $\delta \rho/\kappa = \rho_1 +\kappa \rho_2 +...$, and then Taylor expanding this quantum Fisher information $F_{Q}[\delta \rho /\kappa]$ to next-order in $\kappa$.
Hence, $\Gamma^{\mathrm{f}}(t)$ can be interpreted as the ``Fisher information'' contribution to the connection. 
On the other hand,  $\Gamma^{\mathrm{in}}(t)$ arises from expanding the trace in Eq.~\eqref{eq:buresimplicit} to third order in $\rho_1$. In fact, for all $n>1$, a term depending on the product of $n$ matrix elements of $\rho_1$ appears at order $\mathcal{O}(\kappa^n)$  in the Bures distance expansion, reflecting the intrinsic structure of the expansion.
Eq.~\eqref{eq: intrinsic part of Bures connection} can thus be viewed as an ``intrinsic" contribution to the connection arising from the third-order change to the Bures distance induced by $\rho_1(t)$. This term arises in the expansion of the Bures distance even when only linear perturbations to the density matrix are considered (as in, e.g. Ref.~\cite{braunstein1994statistical})
We can use the expression Eq.~\eqref{eq:rho_n} for the density matrix in time-dependent perturbation theory to evaluate both Eqs.~\eqref{eq: intrinsic part of Bures connection} and \eqref{eq: response part of Bures connection}.

The intrinsic contribution $\Gamma^\mathrm{in}(t)$ can be written in terms of a three-operator correlation function
\begin{equation}
\begin{aligned}\label{eq:S_function_def}
S_{\mu_1\mu_3\mu_2}(\omega_1,\omega_2)=& \frac{1}{4\pi^2} 
\int dt_1 dt_2 e^{i\omega_1 t_1} e^{i \omega_2 t_2} \\ 
&\times
\langle \left[B_{\mu_3}(-t_1)B_{\mu_1}(0),B_{\mu_2}(-t_1-t_2)\right] \rangle_0,
\end{aligned}
\end{equation}
which involves a single commutator. 
Since $S_{\mu_1\mu_3\mu_2}(\omega_1,\omega_2)$ does not involve a nested commutator, it is distinct from a second-order response function. 
Rather, $S_{\mu_1\mu_3\mu_2}(\omega_1,\omega_2)$ can be thought of as a generalized second-order dynamic structure factor.

In terms of $S_{\mu_1\mu_3\mu_2}(\omega_1,\omega_2)$, we show in Appendix~\ref{sub:intrinsic_contribution} that
\begin{equation}\label{eq:gamma_in_integral_def}
\Gamma^{\mathrm{in}}(t)\\
=\int\prod_i dt_if^{\mu_i}(t_i) \Gamma^\mathrm{in}_{\mu_1\mu_2\mu_3}(t,t_1,t_2,t_3)
\end{equation}
with 
\begin{equation}\label{eq:intrinsic_gamma_after_var_deriv}
\begin{aligned}
&\Gamma^\mathrm{in}_{\mu_1\mu_2\mu_3}(t,t_1,t_2,t_3) = -\frac{1}{12}\prod_i\Theta(t-t_i)e^{\epsilon t_i} \\ &\times\mathrm{Im}\Big[\int d\omega_1d\omega_2\tanh\frac{\beta\omega_1}{2}\tanh\frac{\beta(\omega_1-\omega_2)}{2} \\ &\times e^{i[(\omega_1-\omega_2)t_3+\omega_2t_2-\omega_1t_1]}S_{\mu_1\mu_3\mu_2}(\omega_1,\omega_2)\Big] \\
&+\{(\mu_i,t_i)\leftrightarrow(\mu_j,t_j)\},
\end{aligned}
\end{equation}
where we have explicitly symmetrized under all exchanges $(\mu_i,t_i)\leftrightarrow(\mu_j,t_j)$. 
Comparing with the definition Eq.~\eqref{eq:connection_as_var_deriv}, we see that Eq.~\eqref{eq:intrinsic_gamma_after_var_deriv} gives the intrinsic contribution to the time-dependent Bures connection arising from $\Gamma^\mathrm{in}(t)$. 
Thus, the intrinsic contribution to the Bures connection is related to second-order fluctuations of the perturbation $B_\mu$ as defined in the correlation function $S_{\mu_1\mu_3\mu_2}(\omega_1,\omega_2)$.

On the other hand, the Fisher contribution to the Bures connection in Eq.~\eqref{eq: response part of Bures connection} can be directly related to second-order response. 
Following Refs.~\cite{bradlyn2024spectral,sinha2025imaginary,kugler2021multipoint}, we can introduce the spectral density
\begin{equation}\label{eq:2nd_order_chipp_def}
\begin{aligned}
\chi''_{\mu_1\mu_3\mu_2}(\omega_1,\omega_2)&=\frac{1}{4}\int dt_1 dt_2 e^{i\omega_1 t_1} e^{i \omega_2 t_2} \\ &\times\left\langle \left[ 
\left[ B_{{\mu_1}}(0),B_{{\mu_3}}(-t_1)\right], B_{{\mu_2}}(-t_1-t_2) \right] \right\rangle_0.
\end{aligned}
\end{equation}
The second-order spectral density Eq.~\eqref{eq:2nd_order_chipp_def} determines the second-order causal response function
\begin{equation}
\begin{aligned}
\langle B_{\mu_1}\rangle_0(\omega) = \int d\omega_1 d\omega_2& f^{\mu_2}(\omega_1)f^{\mu_3}(\omega_2)\delta(\omega-\omega_1-\omega_2) \\ 
&\times\chi_{\mu_1\mu_2\mu_3}(\omega_1,\omega_2) 
\end{aligned}
\end{equation}
via a generalized Kramers-Kronig relation
\begin{equation}\begin{aligned}
\chi_{\mu_1\mu_2\mu_3}(\omega_1,\omega_2) = \frac{1}{2\pi^2}\int&\frac{d\alpha_1d\alpha_2}{(\alpha_1-\omega_1^+-\omega_2^+)(\alpha_2-\omega_2^+)} \\ &\times \chi''_{\mu_1\mu_3\mu_2}(\omega_1,\omega_2) \\ 
&+\{(\omega_1,\mu_2)\leftrightarrow (\omega_2,\mu_3)\}.
\end{aligned}
\end{equation}
Similar to how the time-dependent Bures metric can be expressed via Eq.~\eqref{eq:bures_metric_main_result} in terms of the spectral density for linear response, we find that the Fisher contribution to the Bures connection takes the form (see Appendix~\ref{sub:fisher_contribution} for details)
\begin{equation} \label{eq:gamma f integral def}
\Gamma^\mathrm{f}(t) =\int\prod_i dt_if^{\mu_i}(t_i) \Gamma^\mathrm{f}_{\mu_1\mu_2\mu_3}(t,t_1,t_2,t_3)
\end{equation}
with
\begin{equation}
\begin{aligned}\label{eq:fisher_gamma_after_var_derivative}
&\Gamma^\mathrm{f}_{\mu_1\mu_2\mu_3}(t,t_1,t_2,t_3)= \frac{\Theta(t_1-t_2)}{6\pi^2} \prod_i\Theta(t-t_i)e^{\epsilon t_i} \\ &\times \mathrm{Im} \Big[ \int  d\omega_1 d\omega_2 \tanh\frac{\beta (\omega_1-\omega_2)}{2} \\ &\times e^{i[(\omega_1-\omega_2)t_3+\omega_2t_2-\omega_1 t_1]} \chi''_{\mu_1\mu_3\mu_2}(\omega_1,\omega_2) \Big] \\ 
&+\{(\mu_i,t_i)\leftrightarrow(\mu_j,t_j)\},
\end{aligned}
\end{equation}
where we have explicitly symmetrized $\Gamma^\mathrm{f}_{\mu_1\mu_2\mu_3}(t,t_1,t_2,t_3)$ under all permutations of pairs $(\mu_i,t_i)\leftrightarrow(\mu_j,t_j)$. 
% section bures_connection_and_second_order_response (end)

As we have emphasized, the two contributions $\Gamma^\mathrm{in}(t)$ and $\Gamma^\mathrm{f}(t)$ to the Bures connection have distinct physical origins, and are defined in terms of different properties of the unperturbed system: $\Gamma^\mathrm{in}(t)$ is defined in terms of the correlation function $S_{\mu_1\mu_3\mu_2}(\omega_1,\omega_2)$, while $\Gamma^\mathrm{f}(t)$ is directly related to the spectral density for second-order response functions, $ \chi''_{\mu_1\mu_3\mu_2}(\omega_1,\omega_2)$. 
Going further, we note that the second-order spectral density satisfies a generalized fluctuation-dissipation theorem,
\begin{equation}
\begin{aligned} \label{eq: relation between chi'' and S}
\chi''_{\mu_1\mu_3\mu_2}(\omega_1,\omega_2) = -\pi^2 \Big[& S_{\mu_1 \mu_3 \mu_2}(\omega_1,\omega_2) \\ &-S_{\mu_3 \mu_1 \mu_2} (\omega_2-\omega_1,\omega_2) \Big],
\end{aligned}
\end{equation}
which follows directly from the definitions Eqs.~\eqref{eq:2nd_order_chipp_def} and \eqref{eq:S_function_def}. 
Because the intrinsic contribution to the Bures connection $\Gamma^\mathrm{in}_{\mu_1\mu_2\mu_3}(t,t_1,t_2,t_3)$ as defined in Eq.~\eqref{eq:intrinsic_gamma_after_var_deriv} is explicitly symmetric under the relabeling and change of integration variables $(\mu_1,t_1) \leftrightarrow (\mu_3,t_3), \omega_1 \rightarrow \omega_2-\omega_1, \omega_2\rightarrow \omega_2$, the generalized fluctuation-dissipation relation shows that $\Gamma^\mathrm{in}(t)$ is independent of the second-order spectral density. 

Thus, unlike for the Bures metric, the Bures connection is not entirely determined by a response function. 
The generalized fluctuation-dissipation relation Eq.~\eqref{eq: relation between chi'' and S} shows that the fundamental quantity that determines the time-dependent Bures connection is the correlation function $S_{\mu_1 \mu_3 \mu_2}(\omega_1,\omega_2)$; both $\Gamma^\mathrm{in}(t)$ and $\Gamma^\mathrm{f}(t)$ can be expressed as a double frequency integral of $S_{\mu_1 \mu_3 \mu_2}(\omega_1,\omega_2)$. 
Nevertheless, we see that the perturbative corrections $\Gamma^\mathrm{f}(t)$ to the Fisher information (and hence to the Bures metric) \emph{are} determined by second-order response functions. 
However, $\Gamma^\mathrm{f}(t)$ does not appear alone in quantum geometric quantities like Eq.~\eqref{eq:connection_as_var_deriv}. 
As such, we expect the second-order response function and its moments to contain non-geometric data in general. Note that we call it non-geometric since unlike $S_{\mu_1 \mu_3 \mu_2}(\omega_1,\omega_2)$, it does not capture the full contribution to the Bures distance at a specific order.
This is consistent with prior work examining the geometric interpretations of the second-order conductivity tensor~\cite{ahn2020lowfrequency,ahn2022riemannian,cook2017design,avdoshkin2024multi}, as we will discuss in more detail in Sec.~\ref{sub:current_perturbations_connection} below.

Having derived our main results Eqs.~\eqref{eq:intrinsic_gamma_after_var_deriv} and \eqref{eq:fisher_gamma_after_var_derivative}, we can now compute the Bures connection restricted to the case of instantaneous and quasistatic perturbations. 
Doing so will yield the induced connection on a subspace of the space of density matrices, parametrized by the restricted set of perturbations, and in the particular coordinate system determined by the amplitude of the external fields.

\subsection{The Instantaneous Limit}\label{sub:bures_instantaneous}
In the instantaneous limit, we again consider the restricted class of perturbations 
\begin{equation}
f^\mu_\mathrm{inst}(t)=f^\mu_0\delta(t),
\end{equation}
or equivalently the short time expansion of a perturbation $f^\mu(t) = \Theta(t)[f^\mu_0+\mathcal{O}(t)]$. 
In this limit, we can carry out the time integrals in Eqs.~\eqref{eq:gamma_in_integral_def} and \eqref{eq:gamma f integral def} trivially. 
Furthermore, we can use the generalized fluctuation-dissipation relation \eqref{eq: relation between chi'' and S} to rewrite the total connection $\Gamma^\mathrm{in}(t)+\Gamma^\mathrm{f}(t)$ in terms of the correlation function $S_{\mu_1\mu_3\mu_2}(\omega_1,\omega_2)$. 
Doing so, and using the fact that as a distribution $\Theta(0)=1/2$, we find
\begin{equation}
\Gamma^{\mathrm{inst}}(t)=\Gamma^\mathrm{in}(t)+\Gamma^\mathrm{f}(t)=f^{\mu_1}_0f^{\mu_2}_0f^{\mu_3}_0 \Gamma^\mathrm{inst}_{\mu_1\mu_2\mu_3}(t)
\end{equation}
with
\begin{equation}
\begin{aligned}\label{eq:inst_connection}
\Gamma^\mathrm{inst}_{\mu_1\mu_2\mu_3}(t)&=-\frac{\Theta(t)}{2}\mathrm{Im}\int d\omega_1d\omega_2\left( \tanh\frac{\beta \omega_1}{2}+1 \right)\\ &\times\left(\tanh \frac{\beta (\omega_1-\omega_2)}{2}+1 \right) S_{(\mu_1 \mu_2 \mu_3)}(\omega_1,\omega_2),
\end{aligned}
\end{equation}
where 
\begin{equation}
S_{(\mu_1 \mu_2 \mu_3)}(\omega_1,\omega_2)
\end{equation}
denotes the totally symmetric part of $S_{\mu_1 \mu_2 \mu_3}(\omega_1,\omega_2)$ under $\mu_i\leftrightarrow\mu_j$, and we have added
\begin{equation}
0=\int d\omega_1d\omega_2 S_{\mu_1\mu_2\mu_3}(\omega_1,\omega_2)
\end{equation}
to the right hand side of Eq.~\eqref{eq:inst_connection}. 
Eq.~\eqref{eq:inst_connection} gives the general expression for the Bures connection in the instantaneous limit. 

Using the definition Eq.~\eqref{eq:S_function_def} of the correlation function $S_{\mu_1 \mu_2 \mu_3}(\omega_1,\omega_2)$, we can examine the universal behavior of $\Gamma^\mathrm{inst}_{\mu_1\mu_2\mu_3}(t)$ in the zero- and high- temperature limits First, taking $T=0$, we have
\begin{equation}
\begin{aligned}
\Gamma^\mathrm{inst}_{\mu_1\mu_2\mu_3}(t)\underset{T=0}{=}-2\Theta(t)\mathrm{Im}\int &d\omega_1d\omega_2\Theta(\omega_1)\Theta(\omega_1-\omega_2) \\ &\times S_{(\mu_1 \mu_2 \mu_3)}(\omega_1,\omega_2).
\end{aligned}
\end{equation}
We can carry out the integral using the spectral representation of $S_{\mu_1 \mu_2 \mu_3}(\omega_1,\omega_2)$ given in Eq.~\eqref{eq:s_appendix_def} to find
\begin{equation}
\begin{aligned}
\Gamma^\mathrm{inst}_{\mu_1\mu_2\mu_3}(t)&\underset{T=0}{=}
-\frac{1}{6}\Theta(t) \mathrm{Im}\sum_{\underset{m\neq 0}{nm}}\Big[\Theta(E_m-E_n) \Theta(E_0-E_n) \\ &\times\big[\bra{0} B_{\mu_1} \ket{n} \bra{n} B_{\mu_2}\ket{m} \bra{m} B_{\mu_3} \ket{0} \\
&-\bra{0}B_{\mu_1}\ket{m} \bra{m}B_{\mu_2} \ket{n} \bra{n} B_{\mu_3} \ket{0} \big] \\ 
& +(\mu_i\leftrightarrow \mu_j)\Big] \\ 
&=0,
\end{aligned}
\end{equation}
which follows from evaluating the step function $\Theta(E_0-E_n)=1/2\delta_{n0}$.
Note that in our framework the Bures distance is expanded about the origin of the parameter space where the perturbation strength $\kappa=0$. Geometrically, the vanishing of the Bures connection here indicates that at zero temperature, the lab frame we choose naturally coincides with a normal coordinate system~\cite{nakahara2018geometry} for the parameter space at the origin, where the lab frame is defined by the physical driving fields $\{\kappa f^\mu(\tau)\}$.

While the low-temperature limit of $\Gamma^\mathrm{inst}_{\mu_1\mu_2\mu_3}(t)$ is trivial, the high-temperature limit is more interesting. 
Expanding the $\tanh$ functions in Eq.~\eqref{eq:inst_connection} and retaining only the leading order terms in $1/T$, we find
\begin{equation}\label{eq:high_t_sum}
\begin{aligned}
\Gamma^\mathrm{inst}_{\mu_1\mu_2\mu_3}(t)\underset{T\rightarrow \infty}{=}-\frac{\Theta(t)}{4T}\mathrm{Im}\int d\omega_1d\omega_2 (2\omega_1-\omega_2)\times \\
S_{(\mu_1 \mu_2 \mu_3)}(\omega_1,\omega_2).
\end{aligned}
\end{equation}
Inserting the definition Eq.~\eqref{eq:S_function_def} and integrating by parts, we then find
\begin{equation}\label{eq:inst_connection_high_T}
\begin{aligned}
\Gamma^\mathrm{inst}_{\mu_1\mu_2\mu_3}(t)&\underset{T\rightarrow \infty}{=}\frac{\Theta(t)}{24T}\mathrm{Re}\Big\langle \left[\left[\dot{B}_{\mu_1},B_{\mu_2}\right],B_{\mu_3}\right]\Big\rangle_0 \\ &+ \{(\mu_i\leftrightarrow \mu_j)\}.
\end{aligned}
\end{equation}
We thus see that the high-temperature behavior of the instantaneous connection is given by an equal-time average, arising from the generalized sum rule Eq.~\eqref{eq:high_t_sum}. 

\subsubsection{Density Perturbations in the Instantaneous Limit}\label{sub:density_perturbations_connection}

The sum rule Eq.~\eqref{eq:inst_connection_high_T} for the high temperature asymptotic behavior of the instantaneous Bures connection has important implications for density perturbations. 
As in Sec.~\ref{subsec: lineardensity}, we consider the instantaneous connection associated to a perturbation of the form
\begin{equation}
H=H_0+\frac{1}{V}\sum_\mathbf{q}\phi_\bq\delta(t)\rho_{-\bq}.
\end{equation}
The instantaneous Bures connection associated to the perturbation $\phi_{\bq}$ is then given by Eq.~\eqref{eq:inst_connection} with
\begin{equation}
\begin{aligned}
S_{\bq_1,\bq_3,\bq_2}(\omega_1,\omega_2) &= \frac{\delta_{\bq_3,-\bq_1-\bq_2}}{4\pi^2 V^3}\int dt_1dt_2e^{i\omega_1 t_1} e^{i \omega_2 t_2} \\ 
&\times
\langle \left[\rho_{\bq_3}(-t_1)\rho_{\bq_1}(0),\rho_{\bq_2}(-t_1-t_2)\right] \rangle_0.
\end{aligned}
\end{equation}
Taking the high-temperature limit, we find from Eq.~\eqref{eq:inst_connection_high_T} that the instantaneous connection associated to density response becomes
\begin{equation}
\Gamma^\mathrm{inst}(t) = \sum_{\mathbf{q}_1\mathbf{q}_2\mathbf{q}_3}\mathrm{Re}\left(\Gamma^\mathrm{inst}_{\bq_1\bq_2\bq_3}(t)\phi_{-\bq_1}\phi_{-\bq_2}\phi_{-\bq_3}\right)
\end{equation}
with
\begin{equation}\label{eq:rho_connection_inst_1}
\begin{aligned}
\Gamma^\mathrm{inst}_{\bq_1\bq_2\bq_3}(t)&\underset{T\rightarrow \infty}{=}\frac{\Theta(t)\delta_{\bq_3,-\bq_1-\bq_2}}{24V^3T}\Big\langle \left[\left[\dot{\rho}_{\bq_1},\rho_{\bq_2}\right],\rho_{\bq_3}\right]\Big\rangle_0 \\ &+ (\bq_i\leftrightarrow \bq_j).
\end{aligned}
\end{equation}
The nested commutator in Eq.~\eqref{eq:rho_connection_inst_1} can be simplified by means of the continuity equation \eqref{eq:continuity}, along with the fact that $\rho_q$ is the generator of gauge transformations. 
In particular, for operators minimally coupled to the vector potential, we have~\cite{mckay2024charge,bradlyn2024spectral}
\begin{equation}
[O(A),\rho_\bq] = \bq_\mu\frac{\delta O(A)}{\delta A_\mu(-\bq)}.
\end{equation}
We thus find that Eq.~\eqref{eq:rho_connection_inst_1} simplifies to
\begin{equation}\label{eq:inst_density_connection_fsum}
\begin{aligned}
\Gamma^\mathrm{inst}_{\bq_1\bq_2\bq_3}(t)&\underset{T\rightarrow \infty}{=}\frac{\Theta(t)\delta_{\bq_3,-\bq_1-\bq_2}}{4V^3T}\bq_1^\mu\bq_2^\nu\bq_3^\lambda\\ &\times\Big\langle \frac{\delta H_0(A)}{\delta A_\mu(-\bq_1)\delta A_\nu(-\bq_2)\delta A_\lambda(-\bq_3)}\Big\rangle_0\Bigg|_{\mathbf{A} \rightarrow 0}.
\end{aligned}
\end{equation}
We recognize the third variational derivative of the Hamiltonian with respect to the vector potential as the generalized diamagnetic current $-j^{\mu \nu \lambda}_{\bq_1,\bq_2,\bq_3}$ defined in Ref.~\cite{mckay2024charge}, which is a momentum-dependent generalization of Eq.~\eqref{eq:current_vertex_defs}. 
This same average appears in the generalized $f$-sum rule for nonlinear density and current response~\cite{bradlyn2024spectral,watanabe2020generalized}. 
Thus Eq.~\eqref{eq:inst_density_connection_fsum} shows that these nonlinear sum rules capture geometric properties of the Bures connection. 
We note that for non-relativistic systems with quadratic dispersion, Eq.~\eqref{eq:inst_density_connection_fsum} vanishes identically.

\subsection{The Quasistatic Limit}\label{sub:bures_quasistatic}

Next, we turn our attention to the quasistatic limit, where the perturbations $f^\mu(t)\rightarrow f_\mathrm{i}^\mu$ independent of time except for the exponential factors which ensure the perturbation is turned on slowly. 
Analogous to Sec.~\ref{sub:the_quasistatic_limit}, we can insert the quasistatic form of the metric into Eqs.~\eqref{eq:gamma_in_integral_def} and \eqref{eq:gamma f integral def} to find
\begin{equation}
\begin{aligned}\label{eq:qs_connections_as_time_integrals}
\Gamma^\mathrm{f}_\mathrm{q.s.}(t) &= \prod_i f^{\mu_i}_\mathrm{i}\int dt_i \Gamma^{\mathrm{f}}_{\mu_1\mu_2\mu_3}(t,t_1,t_2,t_3),\\
\Gamma^\mathrm{in}_\mathrm{q.s.}(t) &= \prod_i f^{\mu_i}_\mathrm{i}\int dt_i \Gamma^{\mathrm{in}}_{\mu_1\mu_2\mu_3}(t,t_1,t_2,t_3),
\end{aligned}
\end{equation}
which shows that the quasistatic connection is given by the time integral of the general Bures connection. 

Let us now evaluate each integral in Eq.~\eqref{eq:qs_connections_as_time_integrals} explicitly. 
We begin with the Fisher contribution. 
Taking care of the convergence factors in Eq.~\eqref{eq:fisher_gamma_after_var_derivative}, we can carry out the time integral to find
\begin{equation} \label{eq: star 1}
\begin{aligned}
&\Gamma^{\mathrm{f}}_{\mathrm{q.s.}} =-\frac{1}{\pi^2} \mathrm{Re} \biggl[ e^{3\epsilon t} \prod_i f^{\mu_i} \int d\omega_1 d\omega_2 \tanh\frac{\beta(\omega_1-\omega_2)}{2}  \\
& \times\frac{\chi''_{\mu_1\mu_3 \mu_2}(\omega_1,\omega_2)}{(\omega_1-\omega_2-i\epsilon)(\omega_2 -i \epsilon) (\omega_1-\omega_2+i\epsilon)}   \biggr]\\
=&-\frac{1}{\pi^2} \mathrm{Re} \Bigg[ \prod_i f^{\mu_i}\int d\omega_1 d\omega_2 \tanh\frac{\beta(\omega_1-\omega_2)}{2} \\ &\times \mathrm{P} \frac{\chi''_{\mu_1\mu_3 \mu_2}(\omega_1,\omega_2) }{(\omega_1-\omega_2)^2\omega_2} \Bigg]\\
=& \mathrm{Re} \biggl[ \prod_i f^{\mu_i} \int d\omega_1 d\omega_2 \bigl(\tanh\frac{\beta(\omega_1-\omega_2)}{2} \mathrm{P} \frac{S_{\mu_1\mu_3 \mu_2}(\omega_1,\omega_2)}{(\omega_1-\omega_2)^2\omega_2}\\
& \quad \quad \quad \quad \quad \quad \quad \quad \quad +\tanh\frac{\beta \omega_1}{2} \mathrm{P} \frac{S_{\mu_1\mu_3 \mu_2}(\omega_1,\omega_2)}{\omega_1^2\omega_2} \bigr)\biggr],
\end{aligned}
\end{equation}
where we used the facts that $\tanh\frac{\beta (\omega_1-\omega_2)}{2}=0$ for $\omega_1-\omega_2=0$ and $\chi''_{\mu \lambda \nu}(\omega_1,\omega_2=0)=0$ due to Eq.~\eqref{result 2 reused in appendix B}. These imply that the integrand is zero for $\omega_1-\omega_2=0$ or $\omega_2=0$ for any finite $\epsilon$, and so only the principal value contributes. 

Using analogous logic, we can carry out the integral for  $\Gamma^{\mathrm{in}}_\mathrm{q.s}$ to find
\begin{equation} \label{eq: star 3}
\begin{aligned}
\Gamma^{\mathrm{in}}_\mathrm{q.s.} = &\frac{1}{2} \mathrm{Re} \biggl[ \prod_i f^{\mu_i} \int  d\omega_1 d\omega_2 \tanh \frac{\beta \omega_1}{2} \tanh \frac{\beta (\omega_1-\omega_2)}{2}\\
& \times \mathrm{P} \frac{S_{\mu_1\mu_3 \mu_2}(\omega_1,\omega_2)}{\omega_1 \omega_2 (\omega_1-\omega_2)}  \biggr].
\end{aligned}
\end{equation}
Taken together, Eqs.~\eqref{eq: star 1} and\eqref{eq: star 3} show that in the quasistatic limit, the Bures connection can be expressed in terms of negative frequency moments of the correlation function $S_{\mu_1\mu_3 \mu_2}(\omega_1,\omega_2)$. 
This is analogous to the case of the quasistatic metric considered in Sec.~\ref{sub:the_quasistatic_limit}, where we saw that the quasistatic Bures metric was defined in terms of negative frequency moments of the linear spectral density (which is related to an equilibrium correlation function by the fluctuation dissipation theorem). 
Eqs.~\eqref{eq: star 1} and \eqref{eq: star 3} are general, giving the quasistatic Bures connection for arbitrary perturbations to any many-body systems. 
We will now apply these results to compute the quasistatic Bures connection associated to constant vector potential perturbations (twisted boundary conditions).

\subsubsection{Current Perturbations in the Quasistatic Limit}\label{sub:current_perturbations_connection}
Let us again consider the Hamiltonian $H_0$ perturbed by a constant vector potential as defined in Eqs.~\eqref{eq:vec_potential_ham} and \eqref{eq:current_vertex_defs}. 
Note that unlike the general case we have considered so far, the current perturbation  in Eq.~\eqref{eq:vec_potential_ham} includes \emph{diamagnetic} contributions that are nonlinear in the external vector potential $A_\mu$. 
While this posed no problem for computing the metric in Sec.~\ref{ssub:current_current_response_in_the_quasistatic_limit} since the diamagnetic correction does not change $\rho_1$, we must now take into account the contributions of the diamagnetic current to the perturbed density matrix $\rho_2(t)$. 
Concretely, we have for the perturbed density matrices 
\begin{equation}
\rho_1(t) = i \int_{-\infty}
^t dt_1 A_\mu(t_1) [j^\mu(t_1-t),\rho_0],
\end{equation}
and
\begin{equation}
\begin{aligned}\label{eq:rho2diadef}
\rho_2(t)  &= - \int_{-\infty}
^t dt_1 \int_{-\infty}^{t_1}dt_2\Big(  A_\mu(t_1) A_\nu(t_2) \\ 
&\times [j^\mu(t_1-t),[j^{\nu}(t_2-t),\rho_0]]\Big)\\
+&\frac{i}{2} \int_{-\infty}^{t} dt_1 A_\mu(t_1) A_\nu(t_1) [j^{\mu \nu}(t_1-t),\rho_0], \\ 
&\equiv \rho_2^\mathrm{para}(t)+\rho_2^\mathrm{dia}(t).
\end{aligned}
\end{equation}
We can see that the $\rho_2^{\mathrm{para}}(t)$ gives Eq.~\eqref{eq: star 1}, and $\rho_1(t)$ gives $\Gamma^{\mathrm{in}}$ in the Bures connection Eq.~\eqref{eq: star 3}. 
However, based on the most general expression for the Fisher part of the Bures connection in Eq.~\eqref{eq: response part of Bures connection}, there is an additional term appearing in the Fisher part of the Bures connection that comes from the diamagnetic density matrix correction $\rho^{\mathrm{dia}}_2(t)$. 
Therefore, $\Gamma = \Gamma^{\mathrm{in}}+\Gamma^{\mathrm{f}}$ still holds, but $\Gamma^{\mathrm{f}}$ can be further divided into the sum of $\Gamma^{\mathrm{para}}$ and $\Gamma^{\mathrm{dia}}$. We can use Eqs.~\eqref{eq: star 3} to compute $\Gamma^{\mathrm{in}}$, and use Eq.~\eqref{eq: star 1} to compute the paramagnetic part $\Gamma^{\mathrm{para}}$, taking into account that the current perturbation enters into Eq.~\eqref{eq:vec_potential_ham} with a minus sign.

Using $\rho_2^\mathrm{dia}(t)$, we can compute the diamagnetic part $\Gamma^{\mathrm{dia}}$ of the Fisher contribution $\Gamma^{\mathrm{f}}$  (see Appendix~\ref{subappendix: derivation of diamagnetic part of Bures connection} for details)
\begin{equation} \label{eq: final result for the generic gamma dia}
\begin{aligned}
\Gamma^{\mathrm{dia}}(t) = &\sum_{n\ell} \frac{\bra{n}\rho_2^{\mathrm{dia}} \ket{\ell} \bra{\ell} \rho_1 \ket{n}}{p_n+p_\ell}\\
=&-\frac{1}{2\pi}\int_{-\infty}^{t}dt_1 A_\mu(t_1) A_\nu(t_1) \int_{-\infty}^{t}dt_2 A_{\lambda}(t_2) \times\\
&\int d\omega  \tanh\frac{\beta \omega }{2}e^{i\omega (t_2-t_1)}  \chi''_{j^{\mu \nu} j^{\lambda}}(\omega),
\end{aligned}
\end{equation}
where $ \chi''_{j^{\mu \nu} j^{\lambda}}(\omega)$ is the spectral density 
\begin{equation}\label{eq:diamag_spectral_density}
 \chi''_{j^{\mu \nu} j^{\lambda}}(\omega) = -\frac{1}{2} \int dt e^{i\omega t}\left\langle \left[j^{\mu \nu}(t), j^\lambda(0)\right]\right\rangle_0,
\end{equation}
which can be used to measure the response of $j^{\mu \nu}$ to the perturbation from $j^\lambda$, and appears as the spectral density for one of the generalized diamagnetic terms in Kubo formulas for the nonlinear optical conductivity~\cite{parker2019diagrammatic}

Using the fact that $\Gamma^\mathrm{dia}(t)$ is explicitly real, we find
\begin{equation}
\begin{aligned}
\Gamma^{\mathrm{dia}}(t) = -\frac{1}{2\pi}\int_{-\infty}^{t}dt_1 A_\mu(t_1) A_\nu(t_1) \int_{-\infty}^{t}dt_2 A_{\lambda}(t_2) \\
\times \int   d\omega  \tanh\frac{\beta \omega }{2}\mathrm{Re} \left[e^{i\omega (t_2-t_1)}  \chi''_{j^{\mu \nu} j^{\lambda}}(\omega) \right].
\end{aligned}
\end{equation}
We can now take the quasistatic limit. 
Since Eq.~\eqref{eq:diamag_spectral_density} is a linear spectral density, we can use the results of Sec.~\ref{sub:the_quasistatic_limit} to carry out the time integrals in  the quasistatic limit to find
\begin{equation} \label{eq: star 2}
\Gamma^{\mathrm{dia}}_\mathrm{q.s.} = -\frac{1}{2\pi}A_\mu A_\nu A_\lambda  \int   d\omega \tanh \frac{\beta \omega}{2} \mathrm{P} \left[\frac{{\mathrm{Re} \chi''_{j^{\mu \nu}j^\lambda}(\omega )}}{\omega^2 }\right].
\end{equation}

We now apply these results to study the quasistatic Bures connection in the case of noninteracting crystalline systems at zero temperature, where there has been much interest in quantum geometric contributions to nonlinear conductivities~\cite{cook2017design,avdoshkin2024multi,ahn2022riemannian,ahn2020lowfrequency,morimoto2016topological}. 
For noninteracting systems we can, as in Sec.~\ref{ssub:current_current_response_in_the_quasistatic_limit}, express the matrix elements of the paramagnetic and diamagnetic current in terms of the single-particle dipole transition matrix elements, which themselves can be expressed in terms of the Berry connection and derivatives of the single-particle energies. 
The details of the derivation have been included in App.~\ref{Free-fermion calculations of the second-order spectral density} and \ref{Computing the Bures connection for free-fermion crystalline systems at zero temperature}.  
Three parts of the Bures connection at zero temperature are
\begin{equation} \label{eq: computed response part of Bures connection for current example}
\begin{aligned}
\Gamma^{\mathrm{para}}_{\mathrm{q.s.}}& = A_\mu A_\nu A_\lambda \times\\
\mathrm{Re} \biggl[&+2i\sum_{\substack{\alpha \neq \beta \\ \alpha \neq  \gamma,\kk}} \Delta f_{\beta \gamma\kk } \frac{\Delta E_{\beta \alpha\kk}}{\Delta E_{\gamma \alpha\kk}}\mathrm{sign}(\Delta E_{\gamma \alpha\kk})r^\mu_{\alpha\beta\kk} r^\nu_{\beta \gamma\kk} r^\lambda_{\gamma \alpha\kk}\\
&-\sum_{\alpha \beta \kk} \Delta f_{\beta \alpha\kk } \frac{\partial^\mu \Delta E_{\beta \alpha\kk}}{\Delta E_{\beta \alpha\kk}}\mathrm{sign}({\Delta E_{ \beta \alpha\kk}})   r^\nu_{\beta \alpha\kk} r^\lambda_{\alpha \beta\kk} 
\biggr],
\end{aligned}
\end{equation}
\begin{equation} \label{eq: computed diamagnetic part of Bures connection for current example}
\begin{aligned}
\Gamma^{\mathrm{dia}}_\mathrm{q.s.}& = A_\mu A_\nu A_\lambda  \mathrm{Re} \Biggl[\sum_{\alpha\beta \kk}\Delta f_{ \beta \alpha \kk}\mathrm{sign}(\Delta E_{ \beta \alpha \kk})\times \\
\bigg(&\frac{1}{2} (\partial^\mu r^\nu_{\alpha \beta \kk}) r^\lambda_{ \beta \alpha \kk} +  r^\nu_{\alpha \beta \kk}r^\lambda_{ \beta \alpha \kk} \frac{\partial^\mu\Delta E_{ \beta \alpha \kk}}{\Delta E_{ \beta \alpha \kk}}  \\
&-i r^\nu_{\alpha \beta \kk }r^\mu_{\alpha \alpha \kk} r^\lambda_{ \beta \alpha \kk} +i\sum_{\substack{\gamma\neq \alpha  \\ \gamma\neq\beta  }} r^\mu_{\alpha \gamma \kk} r^\nu_{\gamma \beta \kk} r^\lambda_{ \beta \alpha \kk} \frac{\Delta E_{\gamma \beta \kk}}{\Delta E_{ \beta \alpha \kk}}  \bigg) \Biggr],
\end{aligned}
\end{equation}
and
\begin{equation} \label{eq: computed intrinsic part of Bures connection for current example}
\begin{aligned}
\Gamma^{\mathrm{in}}_\mathrm{q.s.} =  A_\mu A_\nu A_\lambda &\mathrm{Re} \Bigg[ \frac{1}{2} i\sum_{\substack{\alpha \neq \beta \\ \alpha \neq \gamma,\mathbf{k}}} \mathrm{sign}(\Delta E_{ \beta \alpha \kk}) \mathrm{sign}(\Delta E_{\gamma \alpha \kk}) \times \\
&\Delta f_{ \gamma \beta \kk} \big[f_{\alpha \kk}-(1-f_{\alpha \kk})  \big] r^\lambda_{\alpha\beta \kk} r^\nu_{\beta \gamma \kk} r^\mu_{\gamma \alpha \kk} \Bigg].
\end{aligned}
\end{equation}
Individually, we see that the contributions $\Gamma^{\mathrm{para}},\Gamma^\mathrm{dia}$ and $\Gamma^\mathrm{in}$ contain both energy-independent terms (depending only on occupation factors and off-diagonal dipole matrix elements), as well as energy-dependent terms that depend explicitly on the energy differences and their derivatives. 
However, when we add together Eqs.~\eqref{eq: computed response part of Bures connection for current example}--\eqref{eq: computed intrinsic part of Bures connection for current example}, we find that all the energy-dependent terms cancel. 
As shown in Appendix~\ref{Cancellation of terms in Bures connection}, the Bures connection $\Gamma$ is given by
\begin{equation}\label{eq:ff_connection_final}
\Gamma^{\mathrm{in}}_{\mathrm{q.s.}}+\Gamma^\mathrm{para}_{\mathrm{q.s.}}+\Gamma^\mathrm{dia}_{\mathrm{q.s.}} =-\frac{1}{2}  A_\mu A_\nu A_\lambda \sum_{\substack{\alpha    \in \mathrm{unocc} \\  \beta \in \mathrm{occ},\mathbf{k}}} \partial^\mu (r^\nu_{\alpha \beta \kk} r^\lambda_{ \beta \alpha \kk}).
\end{equation}
We can compare Eq.~\eqref{eq:ff_connection_final} with the definition of the Christoffel symbols computed from the noninteracting $\kk$-dependent metric tensor in Ref.~\cite{ahn2020lowfrequency}. These were defined, using the $\kk$-dependent quantum metric for noninteracting systems in Eq.~\eqref{eq: free-fermion many-body quantum metric} (and recalling that the sign convention for $\Gamma$ is set by noting that $g^{\mu\nu}_\mathrm{q.s.}$ is a metric and not an inverse metric) as
\begin{equation}  \label{eq: sum_christoffel}
\Gamma^{\mu \nu \lambda}= \frac{1}{2} \sum_{\mathbf{k}} \left[ \partial^\nu g^{\mu \lambda} (\mathbf{k})+ \partial^{\lambda} g^{\mu \nu} (\mathbf{k}) -\partial^{\mu} g^{\nu \lambda} (\mathbf{k}) \right].
\end{equation}
We see then that in the zero temperature, noninteracting limit, the Bures connection Eq.~\eqref{eq:ff_connection_final} can be written as $-A_\mu A_\nu A_\lambda \Gamma^{\mu \nu \lambda}$ which coincides with the \textit{symmetric} part of the noninteracting Christoffel symbols. Thus, in the noninteracting limit our results are consistent with the interpretation of the symmetrized Bures connection as the quantum metric dipole~\cite{mitscherling2024gauge}.

As such, our general expressions Eqs.~\eqref{eq: star 1}, \eqref{eq: star 3}, and \eqref{eq: star 2} for the quasistatic Bures connection generalize the noninteracting Christoffel symbols to finite temperature, and to general interacting systems. 
The sum over $\mathbf{k}$ in Eq.~\eqref{eq: sum_christoffel} becomes the integral over $\mathbf{k}$ in the thermodynamic limit. Since the integrand consists of total derivatives with respect to $\kk$, the quasistatic Bures connection $\Gamma^{\mu \nu \lambda}$ can be expressed as a boundary term, i.e., the Fermi surface integral for metals or semimetals, due to partially filled bands. However, for band insulators, there is no Fermi surface and all bands are either fully occupied or unoccupied, so the range of the $\kk$ integral is the full first Brillouin zone. 
In this case, using the periodicity of $r^\mu_\kk$ in the first Brillouin zone, the sum in Eq.~\eqref{eq: sum_christoffel} vanishes and the Bures connection is zero.

Finally, we comment on the relation of our symmetrized quasistatic Bures connection and recently-introduced approximate geometric sum rules for the nonlinear shift current in noninteracting systems~\cite{cook2017design,ahn2020lowfrequency,avdoshkin2024multi}. 
For noninteracting systems, the steady-state DC current generated by an oscillating electric field, and can be expressed as ~\cite{ahn2020lowfrequency}
\begin{equation}
\sigma_{\mathrm{shift}}^{\mu;\nu \lambda} (\omega) = \frac{\pi}{V} \sum_{\alpha  \beta \kk} \Delta f_{\alpha \beta \kk} I_{\alpha \beta \kk}^{\mu;\nu \lambda} \delta(\omega -\Delta E_{\beta\alpha \kk }),
\end{equation}
with
\begin{equation}
I_{\alpha \beta \kk}^{\mu;\nu \lambda} = -i (r_{\alpha \beta \kk}^\lambda r_{ \beta \alpha \kk}^{\nu,\mu} - r_{\alpha \beta \kk}^{\lambda,\mu} r_{ \beta \alpha \kk }^\nu)
\end{equation}
and the covariant derivative
\begin{equation}
r^{\nu,\mu}_{\beta \alpha \kk} \equiv \partial^\mu r_{\beta \alpha \kk}^{\nu} -i(r_{\beta \beta \kk}^{\mu}-r_{\alpha \alpha \kk}^{\mu})r_{\beta \alpha \kk}^{\nu}.
\end{equation}
It has recently been shown that when virtual transitions to highly-excited bands are suppressed or can otherwise be ignored, the shift current satisfies an approximate geometric sum rule~\cite{ahn2020lowfrequency,avdoshkin2024multi}
\begin{equation} \label{eq: approx_sum_rule_Im}
\int_{0}^{+\infty} d\omega \ \mathrm{Im}\sigma_{\mathrm{shift}}^{\mu;\nu \lambda} (\omega)\approx \frac{\pi}{V} \left(\Gamma^{\lambda \mu \nu} - \Gamma^{\nu \mu \lambda}\right).
\end{equation}
involving the \emph{antisymmetric} part of the noninteracting connection. 
Thus, the symmetrized part of the quasistatic Bures connection for noninteracting fermions that we derived here is \emph{independent} of the shift current sum rule, and provides access to a complementary aspect of quantum geometry in those systems. 
We note also that it has recently been appreciated that the contributions to the sum rule arising from the virtual transitions have an interpretation in terms of a multi-state torsion tensor~\cite{jankowski2024quantized,avdoshkin2024multi}. 
However, the symmetrized quasistatic Bures connection that we computed is independent of the torsion, as we defined it as a metric compatible Levi-Civita connection.

\section{Outlook} 
\label{sec:outlook}
In this work, we have presented a general framework for studying the time-dependent quantum geometry of many-body systems, where we view time-dependent external fields as coordinates on the space of accessible density matrices. 
By focusing on the Bures distance Eq.~\eqref{eq:buresdistance} between the equilibrium density matrix $\rho_0$ and the time-evolved state $\rho(t)$, we were able to expand to find expressions for the Bures metric $g_{\mu\nu}(t)$ [Eq.~\eqref{eq:bures_metric_main_result}] and (the symmetric part of) the Levi-Civita connection for the Bures metric [Eqs.~\eqref{eq:intrinsic_gamma_after_var_deriv} and \eqref{eq:fisher_gamma_after_var_derivative}]. 
For the Bures metric, our results unify several threads in the literature studying the quantum geometry associated to instantaneous density response~\cite{hauke2016measuring,balut2025quantum,balut2025quantuma,wang2025local,kruchkov2025topologicalcontrolquantumspeed} and quasistatic current response~\cite{souza2000polarization,verma2025framework,komissarov2024quantuma,onishi2024fundamental,torma2023essay,fang2024quantum}. 
Furthermore, our results provide a concrete geometric interpretation to recent time-dependent formulations of the quantum metric tensor, showing precisely how the time-dependent Bures metric relates to the evolution trajectory of the density matrix~\cite{ji2025density,resta2025nonadiabatic}. 
We clarified the precise relationship between the time-dependent Fisher information, Bures metric, and spectral density for linear response functions, and derived a relationship between the instantaneous and quasistatic limits of the Bures metric. 
We also provided a geometric interpretation of the spectral density and Fermi's golden rule in terms of the infinite-time limit of the Bures metric. 

Going beyond the commonly-studied lowest order, we showed that the Levi-Civita connection associated to the Bures metric can be expressed in terms of a particular three-operator correlation function $S_{\mu_1\mu_2\mu_3}(\omega_1,\omega_2)$ defined in Eq.~\eqref{eq:S_function_def}. 
We showed that the Bures connection contains two physically distinct contributions: the first originated from the correction to the Fisher information arising from the second-order perturbed density matrix. 
We showed that this Fisher contribution can be expressed in terms of the spectral density for second-order nonlinear response [Eq.~\eqref{eq:fisher_gamma_after_var_derivative}]. 
On the contrary, the second ``intrinsic'' contribution to the Bures connection arose from the third-order change of the Bures distance under the first-order perturbation, and hence we showed it is independent of the nonlinear response functions. 

We showed how to evaluate the time-dependent Bures connection in both the instantaneous and quasistatic limits, and applied our formalism to compute the connection associated to quasistatic current perturbations. 
For current perturbations, we showed that there is an inherent ``diamagnetic'' contribution to the Bures connection arising from the response of the diamagnetic current to the first-order current perturbation. 
We showed that for noninteracting fermions at zero temperature, our quasistatic Bures connection reduces to the known symmetric part of the band-theoretic Christoffel symbols~\cite{ahn2020lowfrequency,avdoshkin2024multi}, which are independent of the antisymmetric component that enters approximate sum rules for the shift current.

It is important to emphasize the differences between lowest-order quantum geometry---as quantified by the Bures metric---and the geometry of the Bures connection. 
We reviewed how the Bures metric could be entirely expressed in terms of an integral transform of a first-order spectral density. 
In particular, Eq.~\eqref{eq:infinite_time_g} shows that the infinite-time limit of the Bures metric, parametrized by the frequency components of the perturbation, is proportional to the linear spectral density as a function of frequency. 
Since the linear spectral density is itself the anti-Hermitian part of the causal response function, there is a direct link between linear response and the quantum metric. 
From this perspective, linear response theory has an essential geometric character and interpretation. 

On the other hand, the situation for the Bures connection is a bit less clear. 
One contribution to the connection, the Fisher contribution $\Gamma^\mathrm{f}_{\mu_1\mu_2\mu_3}(t,t_1,t_2,t_3)$ of Eq.~\eqref{eq:fisher_gamma_after_var_derivative}, is indeed the integral transform of a second-order spectral density. 
The second-order spectral density $\chi''_{\mu_1\mu_2\mu_3}(\omega_1,\omega_2)$ does determine the full second-order causal response function via a generalized Kramers-Kronig relation~\cite{bradlyn2024spectral}, but crucially it is not just the Hermitian or anti-Hermitian part of the response function. 
This makes $\chi''_{\mu_1\mu_2\mu_3}(\omega_1,\omega_2)$, and hence $\Gamma^\mathrm{f}_{\mu_1\mu_2\mu_3}(t,t_1,t_2,t_3)$, less directly experimentally accessible than the Bures metric.

On an even deeper level, our derivation of the Bures connection showed that $\Gamma^\mathrm{f}_{\mu_1\mu_2\mu_3}(t,t_1,t_2,t_3)$ is not a geometrical quantity when taken alone. 
This can be seen in the example of  current response for noninteracting fermions, where Eqs.~\eqref{eq: computed response part of Bures connection for current example} and \eqref{eq: computed diamagnetic part of Bures connection for current example} show that in the quasistatic limit $\Gamma^\mathrm{f}_\mathrm{q.s.} = \Gamma^\mathrm{para}_\mathrm{q.s.} + \Gamma^\mathrm{dia}_\mathrm{q.s.}$ depend explicitly on the single-particle eigenenergy differences.
Indeed, only when supplemented with the intrinsic contribution does the full connection become a geometric object.
When viewed in the context of second-order response theory, this goes part of the way to explain recent observations that sum rules for second-order response functions at zero temperature contain both ground state geometric contributions as well as ``multi-state'' contributions from virtual transitions that can dominate outside of certain special limits~\cite{jankowski2024quantized,jankowski2025optical,avdoshkin2024multi}. From another perspective, studying the relation between each contribution to the Bures connection and virtual transitions would
be interesting for further investigation.

Our results open up several avenues for future research. 
First, our general expressions Eqs.~\eqref{eq:intrinsic_gamma_after_var_deriv} and \eqref{eq:fisher_gamma_after_var_derivative}, along with the generalized fluctuation-dissipation relation Eq.~\eqref{eq: relation between chi'' and S} highlight the geometric significance of the correlation function $S_{\mu_1\mu_2\mu_3}(\omega_1,\omega_2)$. 
Designing experimental probes of these correlation functions for different perturbations, for example using density-coupled or optical fields as probes, would open up new routes to experimentally studying quantum geometry in condensed matter systems. 
This represents a promising route to measure robust signatures of quantum geometry beyond linear response. Additionally, because the correlation function $S_{\mu_1\mu_2\mu_3}(\omega_1,\omega_2)$ and our expressions for the Bures connection are valid at any temperature and regardless of system interactions, our formalism provides an experimentally-accessible probe of quantum geometry beyond the Bures metric for correlated quantum systems. Our work thus opens the door to experimentally studying higher-order quantum geometric effects beyond free- or nearly-free fermion systems.

Next, it is natural to expand the Bures distance to next order in time-dependent perturbations. 
Doing so would produce expressions for the time-dependent Riemann curvature tensor of the Bures metric in terms of four-operator correlation functions. 
Investigating the relationship between this Riemann curvature tensor and third-order response functions in specific cases is a fruitful area for future exploration, especially in light of results relating the noninteracting curvature tensor to components of the third-order optical conductivity~\cite{ahn2022riemannian}. 
Moreover, if we consider cases where the perturbations provide a normal coordinate system, such as the instantaneous connection at zero temperature (Sec.~\ref{sub:bures_instantaneous}) and the quasistatic connection at zero temperature for band insulators with current perturbations (Sec.~\ref{sub:current_perturbations_connection}), then expanding the Bures metric $g_{\mu\nu}(\kappa \vec{f},t,t_1,t_2)$ around the origin $\kappa = 0$ to order $\mathcal{O}(\kappa^2)$ provides a straightforward approach to derive the Riemann curvature tensor and its relation with response or correlation functions.

Additionally, recall that since we defined the time-dependent geometric objects by expanding the Bures distance, we only had access to the totally-symmetric component of the metric and Christoffel symbols. 
At lowest order, it is well-appreciated that the Bures metric can be viewed as the real part of a Hermitian tensor, whose imaginary part is given by the Uhlmann curvature~\cite{carollo2020geometry,ji2025density,leonforte2019haldane,leonforte2019uhlmann}. 
A fruitful area for future research would be to reformulate Eqs.~\eqref{eq: intrinsic part of Bures connection} and \eqref{eq: response part of Bures connection} in terms of a symmetrized trace of a generalized derivative, and then consider the antisymmetric parts. 
Doing so would be expected to yield additional quantum geometric information, using the techniques we outlined here. 
Additionally, it would allow for the exploration of ``multi-state geometry'' and torsion beyond the noninteracting limit~\cite{jankowski2024quantized,avdoshkin2024multi}. 
For instance, just as the approximate sum rule in Eq.~\eqref{eq: approx_sum_rule_Im} relates the \textit{imaginary} part of the shift conductivity to the antisymmetric part of the Christoffel symbol, the sum rule for the \textit{real} part yields the skewness of the polarization distribution~\cite{avdoshkin2024multi} when we neglect virtual transitions. 
Given that this skewness is governed by the geometric quantities distinct from the Christoffel symbols and derivatives of the Berry curvature, extending our time-dependent framework to capture the higher-order geometric quantities associated with the skewness and the real part of the shift conductivity is a promising direction for future research.

Finally, since our results are formulated in terms of density matrices, our approach extends naturally to open quantum systems with continuous time evolution. 
It would be interesting and experimentally relevant to generalize our approach to Lindbladian and non-Hermitian evolution. 
For instance, Refs.~\cite{albert2016geometry,carollo2020geometry} considered the metric tensor associated to perturbing the steady state density matrix for Lindbladian evolution with engineered steady state subspaces. 
More recently, Ref.~\cite{pan2020non,geier2022non} considered the general response theory for non-Hermitian evolution. 
Provided a time-dependent perturbation series for the density matrix evolution can be formulated, our techniques for defining time-dependent quantum geometry can be extended to open quantum systems as well. 
This would allow for the exploration of the connection between many-body quantum geometry and quantum information.

% section outlook (end)
\begin{acknowledgments}
This work is supported by the U.S. DOE, Office of Basic Energy Sciences, Grant No. DE-SC0026342.
\end{acknowledgments}

\appendix
\onecolumngrid

\section{Power Series Expansion of the Bures Distance}\label{sec:power_series_expansion_of_the_bures_distance}
In this Appendix, we show how to expand the Bures distance Eq.~\eqref{eq:buresimplicit} order-by-order in $\kappa$. 
Expanding the arc length integral on the left-hand side will define for us geometric structures at the origin $\kappa=0$: the Bures metric and the Bures connection. 
At the same time, expanding the expression for the Bures distance in terms of density matrices in the right-hand side will relate these geometric quantities to correlation functions in a quantum system.

\subsection{Expansion of the Arc Length Integral}

To begin, we expand the left-hand side of Eq.~\eqref{eq:buresimplicit} to third order in $\kappa$. 
To simplify notation, let us define
\begin{align}
D(t) & = \sqrt{\int dt_1dt_2 g_{\mu\nu}(0,t,t_1,t_2)f^\mu(t_1)f^\nu(t_2)} \\ 
x^\mu(\kappa,t) &= \kappa f^\mu(t).
\end{align}
Then we have 
\begin{align}
&\left(\int_0^\kappa d\kappa'\sqrt{ \int dt_1 dt_2 g_{\mu\nu}(\kappa' \vec{f},t,t_1,t_2)f^\mu(t_1)f^\nu(t_2) }\right)^2 \approx \left(\kappa D(t) + \frac{\kappa^2}{2}\frac{\int dt_1dt_2\left.\frac{\partial g_{\mu\nu}(x,t,t_1,t_2)}{\partial\kappa}\right|_{\kappa=0}f^\mu(t_1)f^\nu(t_2)}{2D(t)}\right)^2 \\ 
&\approx \kappa^2 D(t)^2 + \frac{\kappa^3}{2}\int dt_1dt_2\left.\frac{\partial g_{\mu\nu}(x,t,t_1,t_2)}{\partial\kappa}\right|_{\kappa=0}f^\mu(t_1)f^\nu(t_2) \\ 
&\approx \kappa^2 D(t)^2 + \frac{\kappa^3}{2}\int dt_1dt_2dt_3\left.\frac{\delta g_{\mu\nu}(x,t,t_1,t_2)}{\delta x^\lambda(t_3)}\right|_{\kappa=0}f^\mu(t_1)f^\nu(t_2)f^\lambda(t_3)
\end{align}

We can simplify this by introducing a variational generalization of the Christoffel symbols,
\begin{equation}
\frac{\delta g_{\mu\nu}(t,t_1,t_2)}{\delta x^\lambda(t_3)} \equiv \Gamma_{\mu\lambda \nu}(t,t_1,t_3,t_2) + \Gamma_{\nu\lambda\mu}(t,t_2,t_3,t_1),
\end{equation}
in which case we have
\begin{equation}\label{eq:buresmetricexpand}
\left(\int_0^\kappa d\kappa'\sqrt{ \int dt_1 dt_2 g_{\mu\nu}(\kappa' \vec{f},t,t_1,t_2)f^\mu(t_1)f^\nu(t_2) }\right)^2  \approx \kappa^2 D(t)^2 +\kappa^3 \int dt_1dt_2dt_3 \Gamma_{\lambda\mu\nu}(0,t,t_1,t_3,t_2)f^\mu(t_1)f^\nu(t_2)f^\lambda(t_3),
\end{equation}
which is equivalent to Eq.~\eqref{eq:metric_and_connection_def}. 
Thus we see that to second order in $\kappa$ the Bures distance is given by the Bures metric, as expected. 
The third order in $\kappa$ correction defines for us the Bures-Levi-Civita connection.

\subsection{Expansion of the Trace}
We will now expand the trace on right hand side of Eq.~\eqref{eq:buresimplicit} to determine explicit expressions for the metric and the connection in terms of matrix elements of the perturbation. 
To do so, we will follow the logic of Ref.~\cite{hubner1992explicit} and introduce
\begin{equation}
A(\kappa) = \left(\rho_0^{\frac{1}{2}}\rho(t)\rho_0^{\frac{1}{2}}\right)^{\frac{1}{2}}.
\end{equation}
We can make use of the identity
\begin{equation}\label{eq:Aidentity}
A(\kappa)^2 = \rho_0^{\frac{1}{2}}\rho(t)\rho_0^{\frac{1}{2}}
\end{equation}
to expand $A$ to third order in $\kappa$. 
In terms of $A$, we have
\begin{equation}\label{eq:buresAexpand}
d^2_B(\rho_0,\rho(t)) = 2-2\mathrm{tr}A(0) -2\kappa\mathrm{tr}{A'(0)} -\kappa^2\mathrm{tr}{A''(0)} -\frac{1}{3}\kappa^3\mathrm{tr}A'''(0)+\dots
\end{equation}
 Taking derivatives of $A(\kappa)$ and then setting $\kappa=0$, we have
\begin{equation}
\left.\frac{d^n}{d\kappa^n}(A(\kappa))^2\right|_{\kappa=0} = n!\rho_0^\frac{1}{2}\rho_n(t)\rho_0^\frac{1}{2}. 
\end{equation}
Explicitly, for $n=0,1,2,3$ we have, respectively
\begin{align}
A(0) &= \rho_0 \\ 
\left\{A'(0),A(0)\right\} &= \rho_0^\frac{1}{2}\rho_1(t)\rho_0^\frac{1}{2} \\ 
\left\{A''(0),A(0)\right\} + 2 [A'(0)]^2 & = 2\rho_0^\frac{1}{2}\rho_2(t)\rho_0^\frac{1}{2} \\ 
\left\{A'''(0),A(0)\right\} + 3\left\{A''(0),A'(0)\right\} & = 6\rho_0^\frac{1}{2}\rho_3(t)\rho_0^\frac{1}{2}
\end{align}
Taking matrix elements between eigenstates $\ket{n}$ of the unperturbed Hamiltonian with
\begin{equation}
H_0\ket{n}=E_n\ket{n},
\end{equation}
and defining 
\begin{equation}
p_n = \bra{n}\rho_0\ket{n} = e^{-\beta E_n}/Z_0,
\end{equation}
we find
\begin{align}
\bra{n}A(0)\ket{m} &= p_n\delta_{nm} \\ 
\bra{n}A'(0)\ket{m}(p_n+p_m) &= \sqrt{p_np_m}\bra{n}\rho_1(t)\ket{m} \\ 
\bra{n}A''(0)\ket{m}(p_n+p_m) + 2\sum_\ell \bra{n}A'(0)\ket{\ell}\bra{\ell}A'(0)\ket{m} & =2\sqrt{p_np_m}\bra{n}\rho_2(t)\ket{m} \\ 
\bra{n}A'''(0)\ket{m}(p_n+p_m) + 3\sum_\ell \bra{n}A''(0)\ket{\ell}\bra{\ell}A'(0)\ket{m} + \bra{n}A'(0)\ket{\ell}\bra{\ell}A''(0)\ket{m} & =6\sqrt{p_np_m}\bra{n}\rho_3(t)\ket{m} 
\end{align}
Using the fact that $p_n\neq 0$ for any finite inverse temperature $\beta$, we can divide each equation by $p_n+p_m$ to solve iteratively for the matrix elements of the derivatives of $A$. 
We find
\begin{align}
\bra{n}A'(0)\ket{m} & = \frac{\sqrt{p_np_m}\bra{n}\rho_1(t)\ket{m}}{p_n+p_m} \\ 
\bra{n}A''(0)\ket{m} &= \frac{2\sqrt{p_np_m}\bra{n}\rho_2(t)\ket{m}}{p_n+p_m} - 2\sum_\ell \frac{p_\ell\sqrt{p_np_m}\bra{n}\rho_1(t)\ket{\ell} \bra{\ell}\rho_1(t)\ket{m}}{(p_n+p_\ell)(p_\ell+p_m)(p_n+p_m)} \\ \bra{n}A'''(0)\ket{m} & =\frac{6\sqrt{p_np_m}\bra{n}\rho_3(t)\ket{m}}{p_n+p_m} -6\sum_\ell\frac{p_\ell\sqrt{p_np_m}\left(\bra{n}\rho_2\ket{\ell}\bra{\ell}\rho_1\ket{m}+\bra{n}\rho_1\ket{\ell}\bra{\ell}\rho_2\ket{m}\right)}{(p_n+p_m)(p_m+p_\ell)(p_\ell+p_n)} \nonumber \\ 
&+6\sum_{\ell\ell'}\Bigg[\frac{p_\ell p_{\ell'}\sqrt{p_np_m}}{(p_n+p_m)(p_\ell+p_{\ell'})(p_\ell+p_m)(p_n+p_\ell)} \nonumber \\ &\times\left(\frac{\bra{n}\rho_1\ket{\ell'}\bra{\ell'}\rho_1\ket{\ell}\bra{\ell}\rho_1\ket{m}}{p_n+p_{\ell'}} + \frac{\bra{n}\rho_1\ket{\ell}\bra{\ell}\rho_1\ket{\ell'}\bra{\ell'}\rho_1\ket{m}}{p_m+p_{\ell'}}\right)\Bigg]
\end{align}
Taking traces and using the fact that $\mathrm{tr}(\rho_n) = \delta_{n0}$, we have
\begin{align}
\mathrm{tr}{A(0)} &= 1\label{eq:A0} \\ 
\mathrm{tr}{A'(0)} &= 0\\ 
\mathrm{tr}{A''(0)} &= -\frac{1}{2}\sum_{nm}\frac{|\bra{n}\rho_1\ket{m}|^2}{p_n+p_m} \\ 
\mathrm{tr}{A'''(0)} &=\frac{3}{2}\left(\sum_{n\ell\ell'}\frac{\bra{n}\rho_1\ket{\ell'}\bra{\ell'}\rho_1\ket{\ell}\bra{\ell}\rho_1\ket{n}}{(p_\ell+p_n)(p_n+p_{\ell'})}-\sum_{n\ell}\frac{\bra{n}\rho_2\ket{\ell}\bra{\ell}\rho_1\ket{n}+\bra{n}\rho_1\ket{\ell}\bra{\ell}\rho_2\ket{n}}{p_n+p_\ell}\right)\label{eq:A3}
\end{align}

\section{Evaluation of the Bures Connection}\label{sec:evaluation_of_the_bures_connection}
In this Appendix, we give the detailed derivation of the contributions to the Bures connection in Eqs.~\eqref{eq:intrinsic_gamma_after_var_deriv} and \eqref{eq:fisher_gamma_after_var_derivative}.
\subsection{Intrinsic Contribution}\label{sub:intrinsic_contribution}
To evaluate the intrinsic contribution $\Gamma^{\mathrm{in}}(t)$ to the Bures connection, we can insert Eq.~\eqref{eq:rho_n} into Eq.~\eqref{eq: intrinsic part of Bures connection} to find
\begin{equation}
\begin{aligned}
&\Gamma^{\mathrm{in}}(t)\\
=&-\frac{1}{2} \sum_{n\ell m}\frac{\bra{n}\rho_1\ket{m}\bra{m}\rho_1\ket{\ell}\bra{\ell}\rho_1\ket{n}}{(p_\ell+p_n)(p_n+p_{m})}\\
=&-\frac{(-i)^3}{2} \int\prod_i dt_if^{\mu_i}(t_i)\Theta(t-t_i)e^{\epsilon t_i} \sum_{n\ell m}\frac{\bra{n}  [B_{\mu_1}(t_1-t),\rho_0]\ket{m}\bra{m}[B_{\mu_2}(t_2-t),\rho_0]\ket{\ell}\bra{\ell}[B_{\mu_3}(t_3-t),\rho_0]\ket{n}}{(p_\ell+p_n)(p_n+p_{m})}\\
=&\frac{-i}{2} \int\prod_i dt_if^{\mu_i}(t_i)\Theta(t-t_i)e^{\epsilon t_i} \sum_{nm\ell}\frac{p_m-p_n}{p_m+p_n} \frac{p_n -p_\ell}{p_n+p_\ell}(p_\ell - p_m) \times  \\
&\bra{n}B_{\mu_1}\ket{m}\bra{m}B_{\mu_2} \ket{\ell} \bra{\ell} B_{\mu_3} \ket{n} e^{-i(E_m-E_n)t_1}e^{-i(E_\ell-E_m)t_2}e^{-i(E_n-E_\ell)t_3}.
\end{aligned}
\end{equation}
By defining $\omega_1 = E_m-E_n,\omega_2 = E_m-E_\ell$ (so $\omega_1-\omega_2 = E_\ell-E_n$) and introducing two delta functions using Eq.~\eqref{eq:tanhidentity}, we get
\begin{equation} \label{*3 term}
\begin{aligned}
&\Gamma^{\mathrm{in}}(t)\\
=&\frac{i}{2}\int\prod_i dt_if^{\mu_i}(t_i)\Theta(t-t_i)e^{\epsilon t_i} \int d\omega_1 d\omega_2 \tanh\frac{\beta \omega_1}{2} \tanh \frac{\beta(\omega_1-\omega_2)}{2}e^{-i\omega_1 t_1}e^{i\omega_2 t_2}e^{i(\omega_1 -\omega_2)t_3}  \\
&\times \sum_{nm\ell}(p_\ell - p_m) \bra{n}B_{\mu_1}\ket{m}\bra{m}B_{\mu_2} \ket{\ell} \bra{\ell} B_{\mu_3} \ket{n} \delta(\omega_1 +E_n -E_m) \delta(\omega_2 + E_\ell -E_m),
\end{aligned}
\end{equation}
Focusing on the correlation function in Eq.~\eqref{*3 term}, we see that it is the Fourier transform of a three-operator correlation function,
\begin{equation}\label{eq:s_appendix_def}
\begin{aligned}
\sum_{nm\ell}(p_\ell - p_m) &\bra{n}B_{\mu_1}\ket{m}\bra{m}B_{\mu_2} \ket{\ell} \bra{\ell} B_{\mu_3} \ket{n} \delta(\omega_1 +E_n -E_m) \delta(\omega_2 + E_\ell -E_m) \\ 
&= \frac{1}{4\pi^2} \int dt_1 dt_2 e^{i\omega_1 t_1} e^{i \omega_2 t_2} \langle \left[B_{\mu_3}(-t_1)B_{\mu_1}(0),B_{\mu_2}(-t_1-t_2)\right] \rangle_0 \\ 
&\equiv S_{\mu_1 \mu_3 \mu_2}(\omega_1,\omega_2).
\end{aligned}
\end{equation}
Inserting Eq.~\eqref{eq:s_appendix_def} into Eq.~\eqref{*3 term}, we can write
\begin{equation}
\Gamma^{\mathrm{in}}(t)\\
=\int\prod_i dt_if^{\mu_i}(t_i) \Gamma^\mathrm{in}_{\mu_1\mu_2\mu_3}(t,t_1,t_2,t_3)
\end{equation}
with 
\begin{equation}
\begin{aligned}
\Gamma^\mathrm{in}_{\mu_1\mu_2\mu_3}(t,t_1,t_2,t_3) &= \prod_i\Theta(t-t_i)e^{\epsilon t_i}\frac{-1}{12}\mathrm{Im}\int d\omega_1d\omega_2\tanh\frac{\beta\omega_1}{2}\tanh\frac{\beta(\omega_1-\omega_2)}{2}e^{i[(\omega_1-\omega_2)t_3+\omega_2t_2-\omega_1t_1]}S_{\mu_1\mu_3\mu_2}(\omega_1,\omega_2) \\
&+(\mu_i,t_i)\leftrightarrow(\mu_j,t_j),
\end{aligned}
\end{equation}
where we have explicitly symmetrized $\Gamma_{\mu_1\mu_2\mu_3}(t,t_1,t_2,t_3)$ under all permutations of pairs $(\mu_i,t_i)\leftrightarrow(\mu_j,t_j)$, and used the fact that $\Gamma^{\mathrm{in}}(t)$ is explicitly real.

\subsection{Fisher Contribution}\label{sub:fisher_contribution}

We next turn to evaluate the Fisher contribution $\Gamma^\mathrm{f}(t)$ in Eq.~\eqref{eq: response part of Bures connection}. 
Inserting Eq.~\eqref{eq:rho_n}, we find
\begin{align}
\Gamma^{\mathrm{f}}(t) =
&\sum_{n\ell} \frac{\bra{n}\rho_2 \ket{\ell} \bra{\ell} \rho_1 \ket{n}}{p_n+p_\ell} \nonumber\\
=&\sum_{n\ell}(-i)^3\int_{-\infty}^{t} dt_1 \int_{-\infty}^{t_1}dt_2 \int_{-\infty}^{t}dt_3 f^{\mu_1}(t_1)  f^{\mu_2}(t_2) f^{\mu_3}(t_3)e^{\epsilon\sum_i t_i} \nonumber\\
&\times \frac{\bra{n}[B_{\mu_1}(t_1-t),[B_{\mu_2}(t_2-t),\rho_0]]\ket{\ell} \bra{\ell} [B_{\mu_3}(t_3-t),\rho_0] \ket{n}}{p_n+p_{\ell}} \nonumber \\
=&(-i)^3\int_{-\infty}^{t} dt_1 \int_{-\infty}^{t_1}dt_2 \int_{-\infty}^{t}dt_3 f^{\mu_1}(t_1)  f^{\mu_2}(t_2) f^{\mu_3}(t_3)e^{\epsilon\sum_i t_i}\times \label{part 1 of Bures connections}\\
&\Big[\sum_{nm\ell}(p_\ell-p_m)\tanh\frac{\beta(E_\ell-E_n)}{2} \bra{n}B_{\mu_1} \ket{m}\bra{m}B_{\mu_2} \ket{\ell} \bra{\ell} B_{\mu_3} \ket{n}e^{-i(E_m-E_n)t_1}e^{-i(E_\ell-E_m)t_2}e^{-i(E_n-E_\ell)t_3} \nonumber\\
&+\sum_{nm\ell}(p_n-p_m)\tanh\frac{\beta(E_\ell-E_n)}{2} \bra{n}B_{\mu_2} \ket{m}\bra{m}B_{\mu_1} \ket{\ell} \bra{\ell} B_{\mu_3} \ket{n} e^{-i(E_\ell-E_m)t_1}e^{-i(E_m-E_n)t_2}e^{-i(E_n-E_\ell)t_3}\Big]. \nonumber
\end{align}
Using Eq.~\eqref{eq:tanhidentity}, we can define $\omega_1 = E_m-E_n,\omega_2 = E_m-E_\ell$ in the terms multiplying $p_\ell-p_m$, and define $\omega_1 = E_\ell-E_m,\omega_2 = E_n-E_m$ in the terms multiplying $p_n-p_m$. 
We then arrive at 
\begin{equation} \label{result 1 reused in appendix B}
\begin{aligned}
&\Gamma^{\mathrm{f}}(t)\\
=&(-i)^3\int d\omega_1 d\omega_2 \tanh\frac{\beta (\omega_1-\omega_2)}{2} \int_{-\infty}^{t} dt_1 \int_{-\infty}^{t_1} dt_2 \int_{-\infty}^{t}dt_3f^{\mu_1}(t_1)  f^{\mu_2}(t_2) f^{\mu_3}(t_3)e^{\epsilon\sum_i t_i} e^{-i\omega_1 t_1}e^{i\omega_2 t_2}e^{i(\omega_1-\omega_2)t_3} \\
&\times \sum_{nm\ell}\big[+(p_\ell-p_m) \bra{n}B_{\mu_1} \ket{m}\bra{m}B_{\mu_2} \ket{\ell} \bra{\ell} B_{\mu_3} \ket{n} \delta(\omega_1 -E_m+E_n)\delta(\omega_2-E_m+E_\ell)\\
&\quad \quad \quad   \ - (p_m-p_n) \bra{n}B_{\mu_2} \ket{m}\bra{m}B_{\mu_1} \ket{\ell} \bra{\ell} B_{\mu_3} \ket{n} \delta(\omega_1 -E_\ell+E_m)\delta(\omega_2-E_n+E_m)\big].
\end{aligned}
\end{equation}
We can relabel intermediate states in the correlation function, $n \leftrightarrow \ell$ for the first term in brackets and $n \leftrightarrow m$ for the second term in brackets. 
Doing so, we have
\begin{equation} \label{result 2 reused in appendix B}
\begin{aligned}
-\sum_{n\ell m} (p_n-p_m) \Big[
&+\bra{n}B_{\mu_1} \ket{\ell} \bra{\ell} B_{\mu_3} \ket{m} \bra{m}B_{\mu_2} \ket{n}\delta(\omega_1 -E_\ell+E_n)\delta(\omega_2-E_m+E_n)\\
&-\bra{n} B_{\mu_3} \ket{\ell}\bra{\ell}B_{\mu_1} \ket{m}\bra{m}B_{\mu_2} \ket{n} \delta(\omega_1 -E_m+E_\ell)\delta(\omega_2-E_m+E_n)\Big] \\ 
&=- \frac{1}{4\pi^2} \int dt_1 dt_2 e^{i\omega_1 t_1} e^{i \omega_2 t_2} \left\langle \left[ 
\left[ B_{{\mu_1}}(0),B_{{\mu_3}}(-t_1)\right], B_{{\mu_2}}(-t_1-t_2) \right] \right\rangle_0 \\ 
&=-\frac{1}{\pi^2}\chi''_{\mu_1\mu_3\mu_2}(\omega_1,\omega_2),
\end{aligned}
\end{equation}
where $\chi''_{\mu_1\mu_3\mu_2}(\omega_1,\omega_2)$ is the spectral density for the second-order response function as defined in Refs.~\cite{bradlyn2024spectral,sinha2025imaginary}.  Combining this with the fact that $\Gamma^{\mathrm{f}}(t)$ is purely real, can rewrite Eq.\eqref{result 1 reused in appendix B} as
\begin{equation} \label{eq: final version of response term in Bures connection}
\Gamma^\mathrm{f}(t) =\int\prod_i dt_if^{\mu_i}(t_i) \Gamma^\mathrm{f}_{\mu_1\mu_2\mu_3}(t,t_1,t_2,t_3)
\end{equation}
with
\begin{equation}
\begin{aligned}
\Gamma^\mathrm{f}_{\mu_1\mu_2\mu_3}(t,t_1,t_2,t_3)&= \frac{1}{6\pi^2} \Theta(t_1-t_2)\prod_i\Theta(t-t_i)e^{\epsilon t_i} \mathrm{Im} \Big[ \int  d\omega_1 d\omega_2 \tanh\frac{\beta (\omega_1-\omega_2)}{2} e^{i[(\omega_1-\omega_2)t_3+\omega_2t_2-\omega_1 t_1]} \chi''_{\mu_1\mu_3\mu_2}(\omega_1,\omega_2) \Big]. \\ 
&+(\mu_i,t_i)\leftrightarrow(\mu_j,t_j),
\end{aligned}
\end{equation}
where we have explicitly symmetrized $\Gamma^\mathrm{f}_{\mu_1\mu_2\mu_3}(t,t_1,t_2,t_3)$ under all permutations of pairs$(\mu_i,t_i)\leftrightarrow(\mu_j,t_j)$, and used the fact that $\Gamma^{\mathrm{f}}(t)$ is explicitly real.

\section{Derivations of the Current Response in the Quasistatic Limit}
\label{sec: Bures connection of current example}
In this Appendix, we derive the Bures connection in the quasistatic limit corresponding to constant vector potential perturbations. 
First in Sec.~\ref{subappendix: derivation of diamagnetic part of Bures connection} we derive the diamagnetic contribution to the Bures connection. 
Then we specialize to a noninteracting fermion system and derive Eq.~\eqref{eq:ff_connection_final}. 

\subsection{Derivation of the Diamagnetic Part of the Bures Connection}
\label{subappendix: derivation of diamagnetic part of Bures connection}
In this section, we compute the diamagnetic contribution to the time-dependent Bures connection, $\Gamma^{\mathrm{dia}}$, as defined in Eq.~\eqref{eq: final result for the generic gamma dia}. 
Using Eq.~\eqref{eq:rho2diadef} for $\rho_2^{\mathrm{dia}}(t)$ gives
\begin{equation}
\begin{aligned}
&\Gamma^{\mathrm{dia}}(t) \\
= &\sum_{n\ell} \frac{\bra{n}\rho_2^{\mathrm{dia}} \ket{\ell} \bra{\ell} \rho_1 \ket{n}}{p_n+p_\ell}\\
=&\frac{(i)^2}{2}\int_{-\infty}^{t}dt_1\int_{-\infty}^{t}dt_2 A_\mu(t_1) A_\nu(t_1) A_\lambda(t_2) \sum_{n\ell} \frac{\bra{n} [j^{\mu \nu}(t_1-t),\rho_0] \ket{\ell} \bra{\ell} [j^{\lambda}(t_2-t),\rho_0]\ket{n}}{p_n+p_\ell}\\
=&\frac{1}{2}\int_{-\infty}^{t}dt_1\int_{-\infty}^{t}dt_2 A_\mu(t_1) A_\nu(t_1) A_\lambda(t_2) \sum_{n\ell} (p_n-p_\ell) \tanh\frac{\beta (E_\ell-E_n)}{2} \bra{n} j^{\mu \nu }\ket{\ell} \bra{\ell} j^{\lambda} \ket{n}  e^{-i (E_n-E_\ell)(t_2-t_1)}\\
=&\frac{1}{2}\int_{-\infty}^{t}dt_1 A_\mu(t_1) A_\nu(t_1) \int_{-\infty}^{t}dt_2 A_\lambda(t_2)  \int   d\omega  \tanh\frac{\beta \omega }{2}e^{i\omega (t_2-t_1)} \sum_{n\ell} (p_n-p_\ell)  \bra{n} j^{\mu \nu }\ket{\ell} \bra{\ell} j^{\lambda} \ket{n}   \delta(\omega-E_\ell+E_n)\\
=&-\frac{1}{2\pi}\int_{-\infty}^{t}dt_1 A_\mu(t_1) A_\nu(t_1) \int_{-\infty}^{t}dt_2 A_\lambda(t_2)  \int   d\omega  \tanh\frac{\beta \omega }{2}e^{i\omega (t_2-t_1)}  \chi''_{j^{\mu \nu} j^{\lambda}}(\omega),
\end{aligned}
\end{equation}
where in the last step we used
\begin{equation}
\chi''_{j^{\mu \nu} j^{\lambda}}(\omega) = -\frac{1}{2} \int dt e^{i\omega t}\left\langle \left[j^{\mu \nu}(t), j^\lambda(0)\right]\right\rangle_0 =  \pi \sum_{n \ell} (p_{\ell} -p_n) \bra{n} j^{\mu \nu} \ket{\ell} \bra{\ell} j^{\lambda} \ket{n} \delta (\omega -E_{\ell} +E_n).
\end{equation}
\subsection{Correlation Function and Spectral Density for Noninteracting Fermions}
\label{Free-fermion calculations of the second-order spectral density}
In this section we compute the correlation function $S^{\mu \lambda \nu} (\omega_1,\omega_2)$ and the second-order spectral density $[\chi'']^{\mu \lambda \nu}(\omega_1,\omega_2)$ for a noninteracting crystalline systems of fermions with a constant vector potential perturbation.

Using Eqs.~\eqref{eq:S_function_def}, \eqref{eq: relation between chi'' and S}, \eqref{result 1 reused in appendix B}, and \eqref{result 2 reused in appendix B}, we have
\begin{equation}
\begin{aligned}
[\chi'']^{\mu\lambda \nu}(\omega_1,\omega_2) = -\pi^2  \left[S^{\mu \lambda \nu}(\omega_1,\omega_2) -S^{\lambda \mu \nu} (\omega_2-\omega_1,\omega_2) \right] ,
\end{aligned}
\end{equation}
where
\begin{equation} \label{eq: Lehman reprsentation of S}
\begin{aligned}
S^{\mu \lambda \nu}(\omega_1,\omega_2) =& \sum_{nm\ell}(p_\ell - p_m) \bra{n}j^{\mu}\ket{m}\bra{m}j^\nu \ket{\ell} \bra{\ell} j^\lambda \ket{n} \delta(\omega_1  -E_m+E_n) \delta(\omega_2  -E_m+ E_\ell),\\
S^{\lambda \mu \nu}(\omega_2-\omega_1,\omega_2) =& \sum_{nm\ell} (p_m-p_n)\bra{n}j^{\nu}\ket{m}\bra{m}j^\mu \ket{\ell} \bra{\ell} j^\lambda \ket{n} \delta(\omega_1  -E_\ell+E_m) \delta(\omega_2  -E_n+ E_m).
\end{aligned}
\end{equation}

Let us first calculate $ S^{\mu \lambda \nu}(\omega_1,\omega_2)$. 
Since there is a factor $(p_\ell-p_m)$ in Eq.~\eqref{eq: Lehman reprsentation of S}, and the operator $j^\mu$ connecting $\ket{\ell}$ and $\ket{m}$ is a single-particle operator that is diagonal in $\kk$, we know that the fillings of states $\ket{\ell}$ and $\ket{m}$ differ by exactly by one electron at the same $\kk$ in the band structure. 
Therefore, the inner products between all other filled states gives $1$ and we only need to consider one single-particle state for $\ket{\ell}$ and $\ket{m}$. 
By using $j^\nu = \sum_{\beta \gamma \kk} \bra{\psi_{\beta \kk}}j^{\nu} \ket{\psi_{ \gamma\kk}} c^\dagger_{\beta \kk} c_{ \gamma\kk}$, denoting the many-body state $\ket{\ell}$, $\ket{m}$ in the Fock basis as $\ket{\{n^{\ell}_{\alpha}\}}$, $\ket{\{n^{m}_{\alpha}\}}$ and requiring that the two states must be different, we have
\begin{equation}
\begin{aligned}
&\bra{\ell} j^\nu \ket{n} =\sum_{\beta \gamma \kk} \bra{\psi_{\beta \kk}}j^{\nu} \ket{\psi_{ \gamma\kk}} \bra{\{n^{\ell}_{\alpha}\}}c^\dagger_{\beta \kk} c_{ \gamma\kk}\ket{\{n^{m}_{\alpha}\}} \\
=&\sum_{\gamma \neq\beta\kk} j^{\nu}_{\beta\gamma \kk}\times \\
&\begin{cases} 
(-1)^{\sum_{i \ \mathrm{between} \ \beta ,\gamma }n_{i\kk}}   & \text{if } \ket{\{n^{m}_{\alpha}\}} = \ket{...,n_{\gamma \kk}=1, ...,n_{\beta \kk}=0,...},\ket{\{n^{\ell}_{\alpha}\}} = \ket{...,n_{\gamma \kk}=0, ...,n_{\beta \kk}=1,...} \\
0 & \text{otherwise}.
\end{cases}
\end{aligned}
\end{equation}
For $\bra{n}j^{\mu}\ket{m}$ and $ \bra{\ell} j^\lambda \ket{n}$, we need to be more careful and consider four possibilities. 
If $\ket{n}$ is orthogonal to both $\ket{m}$ and $\ket{\ell}$, then there are two possibilities for the product of matrix elements to be nonzero. 
The first is if $n^{n}_{\gamma\kk} = n^{n}_{\beta\kk}=1 $ and there exists another empty state $n^{n}_{\alpha \kk}=0$ which is filled in $\ket{m},\ket{\ell}.$ The second is if $n^{n}_{\gamma\kk} = n^{n}_{\beta\kk}=0 $ and there exists another filled state $n^{n}_{\alpha \kk}=0$ which is empty for $\ket{m},\ket{\ell}.$ The first case contributes to $S^{\mu \lambda \nu}(\omega_1,\omega_2)$ a term
\begin{equation} \label{part 1 in chi''}
\begin{aligned}
&- \sum_{\alpha,\beta,\gamma \ \mathrm{all \ different}\kk} \left[f_{\gamma\kk}(1-f_{\beta \kk})f_{\alpha \kk} - (1-f_{\gamma\kk})f_{\beta \kk}f_{\alpha \kk}\right] \bra{\psi_{\gamma \kk}}j^\mu\ket{\psi_{\alpha \kk}}
\bra{\psi_{\beta \kk}}j^\nu\ket{\psi_{\gamma \kk}}
\bra{\psi_{\alpha \kk}}j^\lambda\ket{\psi_{\beta \kk}}\times\\
&\delta(\omega_1+E_{\gamma \kk}-E_{\alpha \kk}) \delta(\omega_2+E_{\gamma \kk}-E_{\beta \kk})\\
=&\sum_{\alpha \neq\beta, \alpha \neq\gamma,\kk} (f_{\beta \kk}-f_{\gamma \kk})f_{\alpha \kk} j^\lambda_{\alpha\beta\kk} j^\nu_{\beta \gamma\kk} j^\mu_{\gamma \alpha\kk} \delta(\omega_1 +\Delta E_{\gamma \alpha\kk})\delta(\omega_2 +\Delta E_{\gamma \beta\kk}).
\end{aligned}
\end{equation}
Similarly, the second case contributes the term  
\begin{equation} \label{part 2 in chi''}
-\sum_{\alpha \neq\beta, \alpha \neq\gamma,\kk}(f_{\beta \kk}-f_{\gamma \kk})(1-f_{\alpha \kk}) j^\mu_{\alpha\beta\kk} j^\nu_{\beta \gamma\kk} j^\lambda_{\gamma \alpha\kk} \delta(\omega_1 +\Delta E_{\alpha \beta\kk})\delta(\omega_2 +\Delta E_{\gamma \beta\kk}).
\end{equation}
Note that, compared to the second case, the first case has an extra minus sign from the inner product of many-body states due to the fermionic anti-commutation relation.
Now, let us consider the remaining two possibilities, i.e. $\ket{n} = \ket{m}$ or $\ket{n} = \ket{\ell}$. 

If $\ket{n} = \ket{m}$, then we have 
\begin{equation} \label{part 3 in chi''}
\begin{aligned}
-& \sum_{\beta \gamma \kk}(f_{\beta \kk}-f_{\gamma \kk}) j^\mu_{\beta \beta\kk}j^\nu_{\beta \gamma\kk} j^\lambda_{\gamma \beta \kk} \delta(\omega_1)\delta(\omega_2+\Delta E_{\gamma \beta \kk})\\
-&\sum_{\beta \gamma \kk}(f_{\beta \kk}-f_{\gamma \kk})\left[\sum_{(\alpha\kk')\neq(\beta \kk),(\alpha\kk')\neq(\gamma \kk)}j^\mu_{\alpha\alpha\kk'}f_{\alpha \kk'}\right]  j^\nu_{\beta \gamma\kk} j^\lambda_{\gamma \beta \kk}\delta(\omega_1)\delta(\omega_2+\Delta E_{\gamma \beta \kk}).
\end{aligned}
\end{equation}

If $\ket{n}=\ket{\ell}$, we have 
\begin{equation} \label{part 4 in chi''}
\begin{aligned}
-& \sum_{\beta \gamma \kk}(f_{\beta \kk}-f_{\gamma \kk}) j^\mu_{\gamma \beta\kk}j^\nu_{\beta \gamma\kk} j^\lambda_{\gamma \gamma \kk} \delta(\omega_1+\Delta E_{\gamma \beta \kk})\delta(\omega_2+\Delta E_{\gamma \beta \kk})\\
-&\sum_{\beta \gamma \kk}(f_{\beta \kk}-f_{\gamma \kk}) j^\mu_{\gamma \beta\kk}j^\nu_{\beta \gamma\kk} \left[\sum_{(\alpha\kk')\neq(\beta \kk),(\alpha\kk')\neq(\gamma \kk)}j^\lambda_{\alpha\alpha\kk'}f_{\alpha \kk'}\right] \delta(\omega_1+\Delta E_{\gamma \beta \kk})\delta(\omega_2+\Delta E_{\gamma \beta \kk}).
\end{aligned}
\end{equation}
Therefore, the correlation function $ S^{\mu \lambda \nu}(\omega_1,\omega_2)$ is the sum of Eqs.~\eqref{part 1 in chi''}, \eqref{part 2 in chi''}, \eqref{part 3 in chi''}, and \eqref{part 4 in chi''}.

Using the four equations above, we can also compute $S^{\lambda \mu \nu} (\omega_2-\omega_1,\omega_2) $. 
There are four nonzero contributions:
\begin{equation} \label{part 5 in chi'' except minus sign}
\begin{aligned}
\sum_{\alpha \neq\beta, \alpha \neq\gamma,\kk} (f_{\beta \kk}-f_{\gamma \kk})f_{\alpha \kk} j^\mu_{\alpha\beta\kk} j^\nu_{\beta \gamma\kk} j^\lambda_{\gamma \alpha\kk} \delta(\omega_1 +\Delta E_{\alpha \beta\kk})\delta(\omega_2 +\Delta E_{\gamma \beta\kk}),
\end{aligned}
\end{equation}
\begin{equation} \label{part 6 in chi'' except minus sign}
-\sum_{\alpha \neq\beta, \alpha \neq\gamma,\kk}(f_{\beta \kk}-f_{\gamma \kk})(1-f_{\alpha \kk}) j^\lambda_{\alpha\beta\kk} j^\nu_{\beta \gamma\kk} j^\mu_{\gamma \alpha\kk} \delta(\omega_1 +\Delta E_{\gamma \alpha\kk})\delta(\omega_2 +\Delta E_{\gamma \beta\kk}),
\end{equation}
\begin{equation}  \label{part 7 in chi'' except minus sign}
\begin{aligned}
-& \sum_{\beta \gamma \kk}(f_{\beta \kk}-f_{\gamma \kk}) j^\lambda_{\beta \beta\kk}j^\nu_{\beta \gamma\kk} j^\mu_{\gamma \beta \kk} \delta(\omega_1+\Delta E_{\gamma \beta \kk})\delta(\omega_2+\Delta E_{\gamma \beta \kk})\\
-&\sum_{\beta \gamma \kk}(f_{\beta \kk}-f_{\gamma \kk})\left[\sum_{(\alpha\kk')\neq(\beta \kk),(\alpha\kk')\neq(\gamma \kk)}j^\lambda_{\alpha\alpha\kk'}f_{\alpha \kk'}\right]  j^\nu_{\beta \gamma\kk} j^\mu_{\gamma \beta \kk}\delta(\omega_1+\Delta E_{\gamma \beta \kk})\delta(\omega_2+\Delta E_{\gamma \beta \kk}),
\end{aligned}
\end{equation}
\begin{equation} \label{part 8 in chi'' except minus sign}
\begin{aligned}
-& \sum_{\beta \gamma \kk}(f_{\beta \kk}-f_{\gamma \kk}) j^\lambda_{\gamma \beta\kk}j^\nu_{\beta \gamma\kk} j^\mu_{\gamma \gamma \kk} \delta(\omega_1)\delta(\omega_2+\Delta E_{\gamma \beta \kk})\\
-&\sum_{\beta \gamma \kk}(f_{\beta \kk}-f_{\gamma \kk}) j^\lambda_{\gamma \beta\kk}j^\nu_{\beta \gamma\kk} \left[\sum_{(\alpha\kk')\neq(\beta \kk),(\alpha\kk')\neq(\gamma \kk)}j^\mu_{\alpha\alpha\kk'}f_{\alpha \kk'}\right] \delta(\omega_1)\delta(\omega_2+\Delta E_{\gamma \beta \kk}),
\end{aligned}
\end{equation}
and $S^{\lambda \mu \nu} (\omega_2-\omega_1,\omega_2) $ is the sum of Eqs.~\eqref{part 5 in chi'' except minus sign}, \eqref{part 6 in chi'' except minus sign}, \eqref{part 7 in chi'' except minus sign}, and \eqref{part 8 in chi'' except minus sign}.

Now we can see that, since $[\chi'']^{\mu\lambda \nu}(\omega_1,\omega_2)=-\pi^2 \left[S^{\mu \lambda \nu}(\omega_1,\omega_2) -S^{\lambda \mu \nu} (\omega_2-\omega_1,\omega_2) \right]$, Eq.~\eqref{part 1 in chi''} and the term in Eq.~\eqref{part 6 in chi'' except minus sign} that is quadratic in $f$ cancel, and the same thing happens between Eq.~\eqref{part 2 in chi''} and Eq.~\eqref{part 5 in chi'' except minus sign}. 
Furthermore, terms quadratic in $f$ cancel between Eq.~\eqref{part 4 in chi''} and Eq.~\eqref{part 7 in chi'' except minus sign}, as well as  between Eq.~\eqref{part 3 in chi''} and Eq.~\eqref{part 8 in chi'' except minus sign}. 
Therefore, $[\chi'']^{\mu\lambda \nu}(\omega_1,\omega_2)$ is given by 
\begin{equation} \label{eq: free-fermion result of spectral density for current}
\begin{aligned}
[\chi'']^{\mu\lambda \nu}(\omega_1,\omega_2)=&+\pi^2\sum_{\alpha \beta \gamma\kk} (f_{\beta \kk}-f_{\gamma \kk}) j^\mu_{\alpha\beta\kk} j^\nu_{\beta \gamma\kk} j^\lambda_{\gamma \alpha\kk} \delta(\omega_1 +\Delta E_{\alpha \beta\kk})\delta(\omega_2 +\Delta E_{\gamma \beta\kk})\\
&-\pi^2\sum_{\alpha \beta \gamma\kk} (f_{\beta \kk}-f_{\gamma \kk}) j^\lambda_{\alpha\beta\kk} j^\nu_{\beta \gamma\kk} j^\mu_{\gamma \alpha\kk} \delta(\omega_1 +\Delta E_{\gamma \alpha\kk})\delta(\omega_2 +\Delta E_{\gamma \beta\kk}).
\end{aligned}
\end{equation}

We can use these results and the analogous logic to compute the quasistatic Bures connection contributions $\Gamma^{\mathrm{f}}$ and $\Gamma^{\mathrm{in}}$, since $\Gamma^{\mathrm{f}}$ depends on $[\chi'']^{\mu \lambda \nu}(\omega_1,\omega_2)$, and $\Gamma^{\mathrm{in}}$ depends on $S^{\mu \lambda \nu}(\omega_1,\omega_2)$.

\subsection{Computing the Bures Connection for Noninteracting Systems at Zero Temperature}
\label{Computing the Bures connection for free-fermion crystalline systems at zero temperature}
In this section we will compute the Bures connection for systems of noninteracting fermions due to a quasistatic vector potential perturbation. 
For simplicity, we will take the zero-temperature limit which can exclude the thermal information and only capture the quantum part of the Bures connection. 
Note that although $\Gamma^{\mathrm{f}}_{\mathrm{q.s.}}$ in Eq.~\eqref{eq: star 1} depends on the spectral density in Eq.~\eqref{eq: free-fermion result of spectral density for current}, the $\omega_1 = \omega_2$ terms in it have no contributions to Eq.~\eqref{eq: star 1} due to the $\tanh$ function and the principle value. 
Therefore, we only need to consider the case with $\omega_1 \neq \omega_2$. 
Combining this constraint with the exclusion $\beta \neq \gamma$ of diagonal matrix elements due to the occupation factors, we can write the part of $[\chi'']^{\mu \lambda \nu}(\omega_1,\omega_2)$ which has nonzero contributions to $\Gamma^{\mathrm{f}}_{\mathrm{q.s.}}$ as
\begin{equation} \label{nonzero contribution}
\begin{aligned}
[\chi'']^{\mu \lambda \nu}(\omega_1,\omega_2)\Big|_{\omega_1\neq \omega_2}=&+\pi^2\sum_{\alpha \neq \beta , \alpha \neq  \gamma,\kk} \Delta f_{\beta \gamma\kk } j^\mu_{\alpha\beta\kk} j^\nu_{\beta \gamma\kk} j^\lambda_{\gamma \alpha\kk} \delta(\omega_1 +\Delta E_{\alpha \beta\kk})\delta(\omega_2 +\Delta E_{\gamma \beta\kk})\\
&-\pi^2\sum_{\alpha \neq \beta , \alpha \neq  \gamma,\kk}\Delta f_{\beta \gamma\kk } j^\lambda_{\alpha\beta\kk} j^\nu_{\beta \gamma\kk} j^\mu_{\gamma \alpha\kk} \delta(\omega_1 +\Delta E_{\gamma \alpha\kk})\delta(\omega_2 +\Delta E_{\gamma \beta\kk})\\
&+\pi^2\sum_{ \beta \gamma\kk} \Delta f_{\beta \gamma\kk } j^\mu_{\beta \beta\kk} j^\nu_{\beta \gamma\kk} j^\lambda_{\gamma \beta\kk} \delta(\omega_1)\delta(\omega_2 +\Delta E_{\gamma \beta\kk})\\
&-\pi^2\sum_{\beta \gamma\kk} \Delta f_{\beta \gamma\kk } j^\lambda_{\gamma\beta\kk} j^\nu_{\beta \gamma\kk} j^\mu_{\gamma \gamma\kk} \delta(\omega_1)\delta(\omega_2 +\Delta E_{\gamma \beta\kk}).
\end{aligned}
\end{equation}
Using the diagonal and off-diagonal matrix elements of the current operator
\begin{equation}
\begin{aligned}\label{eq:current_matrix_elements}
j^\mu_{\alpha \beta\kk}= -i \left[X^\mu,H_0\right]_{\alpha \beta\kk}=
\begin{cases}
\partial^\mu E_{\alpha \kk} \quad & \mathrm{if} \ \alpha = \beta \\
ir^{\mu}_{\alpha \beta \kk}\Delta E_{\alpha \beta\kk} & \mathrm{if} \  \alpha \neq \beta,
\end{cases}
\end{aligned}
\end{equation}
Eq.~\eqref{nonzero contribution} becomes
\begin{equation}
\begin{aligned}
[\chi'']^{\mu \lambda \nu}(\omega_1,\omega_2)\Big|_{\omega_1\neq \omega_2}=&+i\pi^2\omega_1\omega_2 (\omega_1-\omega_2)\sum_{\alpha \neq \beta , \alpha \neq  \gamma,\kk} \Delta f_{\beta \gamma\kk } r^\mu_{\alpha\beta\kk} r^\nu_{\beta \gamma\kk} r^\lambda_{\gamma \alpha\kk} \delta(\omega_1 +\Delta E_{\alpha \beta\kk})\delta(\omega_2 +\Delta E_{\gamma \beta\kk})\\
&-i\pi^2\omega_1\omega_2 (\omega_1-\omega_2)\sum_{\alpha \neq \beta , \alpha \neq  \gamma,\kk}\Delta f_{\beta \gamma\kk } r^\lambda_{\alpha\beta\kk} r^\nu_{\beta \gamma\kk} r^\mu_{\gamma \alpha\kk} \delta(\omega_1 +\Delta E_{\gamma \alpha\kk})\delta(\omega_2 +\Delta E_{\gamma \beta\kk})\\
&+\pi^2\omega_2^2\sum_{ \beta \gamma\kk} \Delta f_{\beta \gamma\kk } \partial^\mu \Delta E_{\beta \gamma\kk} r^\nu_{\beta \gamma\kk} r^\lambda_{\gamma \beta\kk} \delta(\omega_1)\delta(\omega_2 +\Delta E_{\gamma \beta\kk}).
\end{aligned}
\end{equation}
Using this in Eq.~\eqref{eq: star 1} [and taking into account the global minus sign from the current perturbation in Eq.~\eqref{eq:vec_potential_ham}] we have
\begin{equation}
\begin{aligned}
\Gamma^{\mathrm{para}}_{\mathrm{q.s.}} =\frac{1}{\pi^2} A_\mu A_\nu A_\lambda \mathrm{P} & \mathrm{Re} \left[ \int  d\omega_1 \int  d\omega_2 \tanh \frac{\beta (\omega_1 -\omega_2)}{2} \frac{ [\chi'']^{\mu \lambda \nu} (\omega_1,\omega_2)}{(\omega_1-\omega_2)^2\omega_2}\right]\\
= A_\mu A_\nu A_\lambda \mathrm{Re} \biggl[&i\sum_{\substack{\alpha \neq \beta \\ \alpha \neq  \gamma,\kk}} \Delta f_{\beta \gamma\kk } \frac{\Delta E_{\beta \alpha\kk}}{\Delta E_{\gamma \alpha\kk}}\mathrm{sign}(\Delta E_{\gamma \alpha\kk}){}r^\mu_{\alpha\beta\kk} r^\nu_{\beta \gamma\kk} r^\lambda_{\gamma \alpha\kk} -i\sum_{\substack{\alpha \neq \beta\\ \alpha \neq  \gamma,\kk}}\Delta f_{\beta \gamma\kk }\frac{\Delta E_{\gamma \alpha\kk}}{\Delta E_{\beta \alpha\kk}} \mathrm{sign}(\Delta E_{\alpha \beta\kk})r^\lambda_{\alpha\beta\kk} r^\nu_{\beta \gamma\kk} r^\mu_{\gamma \alpha\kk} \\
&+\sum_{ \beta \gamma\kk} \Delta f_{\beta \gamma\kk } \frac{\partial^\mu \Delta E_{\beta \gamma\kk}}{\Delta E_{\beta \gamma\kk}}\mathrm{sign}({\Delta E_{\gamma \beta\kk}})   r^\nu_{\beta \gamma\kk} r^\lambda_{\gamma \beta \kk} \biggr].
\end{aligned}
\end{equation}
We can see that the second term in $\mathrm{Re}[...]$ is the complex conjugate of the first term by taking complex conjugate and exchanging dummy indices $\beta$ and $\gamma$. 
Additionally, the third term in $\mathrm{Re}[...]$ is real, thus we have
\begin{equation}
\begin{aligned}
\Gamma^{\mathrm{para}}_{\mathrm{q.s.}} = A_\mu A_\nu A_\lambda 
\mathrm{Re} \biggl[2i\sum_{\substack{\alpha \neq \beta \\ \alpha \neq  \gamma,\kk}} \Delta f_{\beta \gamma\kk } \frac{\Delta E_{\beta \alpha\kk}}{\Delta E_{\gamma \alpha\kk}}\mathrm{sign}(\Delta E_{\gamma \alpha\kk})r^\mu_{\alpha\beta\kk} r^\nu_{\beta \gamma\kk} r^\lambda_{\gamma \alpha\kk}-\sum_{\alpha \beta \kk} \Delta f_{\beta \alpha\kk } \frac{\partial^\mu \Delta E_{\beta \alpha\kk}}{\Delta E_{\beta \alpha\kk}}\mathrm{sign}({\Delta E_{ \beta \alpha\kk}})   r^\nu_{\beta \alpha\kk} r^\lambda_{\alpha \beta\kk} 
\biggr],
\end{aligned}
\end{equation}
which is Eq.~\eqref{eq: computed response part of Bures connection for current example} in the main text.

Now consider $\Gamma^{\mathrm{dia}}_{\mathrm{q.s.}}$ , which includes the first-order spectral density between $j^{\mu \nu}$ and $j^\lambda$, i.e.
\begin{equation}
\begin{aligned}
\chi''_{j^{\mu \nu} j^\lambda} (\omega) = \pi \sum_{\alpha \beta\kk} \Delta f_{\beta \alpha\kk} j^{\mu \nu}_{\alpha \beta\kk} j^{\lambda}_{\beta \alpha\kk} \delta(\omega-\Delta E_{\beta \alpha\kk}),
\end{aligned}
\end{equation}
where $\alpha \neq \beta$. 
Inserting the current matrix elements from Eq.~\eqref{eq:current_matrix_elements}, we find
\begin{equation}
\begin{aligned}
j^{\mu \nu}_{\alpha \beta\kk } =& i \left[ X^\mu,j^\nu \right]_{\alpha \beta\kk} = i \left[ i\partial^\mu j^\nu_{\alpha \beta\kk} +r^\mu_{\alpha \gamma\kk}j^\nu_{\gamma \beta\kk}-j^\nu_{\alpha \gamma\kk}r^\mu_{\gamma \beta\kk} \right]\\
=& -\left[i\partial^\mu r^\nu_{\alpha \beta\kk}+r^\nu_{\alpha \beta\kk} (r^\mu_{\alpha \alpha\kk}-r^\mu_{\beta \beta\kk})\right] \Delta E_{\alpha \beta\kk} - \left[ i r^\nu_{\alpha \beta\kk} \partial^\mu\Delta E_{\alpha \beta\kk} + i r^\mu_{\alpha \beta\kk} \partial^\nu\Delta E_{\alpha \beta\kk}  \right]\\
&- \sum_{ \gamma\neq\alpha,  \gamma\neq \beta } \left[ r^\mu_{\alpha \gamma\kk} r^\nu_{\gamma \beta\kk} \Delta E_{\gamma \beta\kk} -r^\nu_{\alpha \gamma\kk} r^\mu_{\gamma \beta\kk} \Delta E_{\alpha \gamma\kk} \right].
\end{aligned}
\end{equation}
Inserting these expressions into Eq.~\eqref{eq: star 2} gives
\begin{equation}
\begin{aligned}
\Gamma^{\mathrm{dia}}_\mathrm{q.s.} =\frac{1}{2} A_\mu A_\nu A_\lambda  \mathrm{Re}  \Bigg[&+\sum_{\alpha\beta \kk}\Delta f_{ \beta \alpha \kk} (\partial^\mu r^\nu_{\alpha \beta \kk}) r^\lambda_{ \beta \alpha \kk} \mathrm{sign}(\Delta E_{ \beta \alpha \kk}) -i \Delta f_{ \beta \alpha \kk}r^\nu_{\alpha \beta \kk }(r^\mu_{\alpha \alpha \kk}-r^\mu_{\beta \beta \kk}) r^\lambda_{ \beta \alpha \kk} \mathrm{sign}(\Delta E_{ \beta \alpha \kk})\\
&+\sum_{\alpha\beta \kk}\Delta f_{ \beta \alpha \kk} \left(  r^\nu_{\alpha \beta \kk}r^\lambda_{ \beta \alpha \kk} \frac{\partial^\mu\Delta E_{ \beta \alpha \kk}}{\Delta E_{ \beta \alpha \kk}} +  r^\mu_{\alpha \beta \kk}r^\lambda_{ \beta \alpha \kk} \frac{\partial^\nu\Delta E_{ \beta \alpha \kk}}{\Delta E_{ \beta \alpha \kk}} \right) \mathrm{sign}(\Delta E_{ \beta \alpha \kk})\\
&+i\sum_{ \gamma\neq\alpha,  \gamma\neq \beta,\kk} \Delta f_{ \beta \alpha \kk}\left( r^\mu_{\alpha \gamma \kk} r^\nu_{\gamma \beta \kk} r^\lambda_{ \beta \alpha \kk} \frac{\Delta E_{\gamma \beta \kk}}{\Delta E_{ \beta \alpha \kk}} -r^\nu_{\alpha \gamma \kk} r^\mu_{\gamma \beta \kk}r^\lambda_{ \beta \alpha \kk} \frac{\Delta E_{\alpha \gamma \kk}}{\Delta E_{ \beta \alpha \kk}} \right) \mathrm{sign}(\Delta E_{ \beta \alpha \kk})\Bigg].
\end{aligned}
\end{equation}
Next, we note that the first two terms are complex conjugates of each other, as are the two terms inside the final summation. 
Exploiting this and relabeling $\mu \leftrightarrow \nu$ in the second part of the third term, we have
\begin{equation}\label{eq:appendx_gamma_dia}
\begin{aligned}
\Gamma^{\mathrm{dia}}_\mathrm{q.s.} = A_\mu A_\nu A_\lambda  \mathrm{Re} \Biggl[\sum_{\alpha\beta \kk}\Delta f_{ \beta \alpha \kk}\mathrm{sign}(\Delta E_{ \beta \alpha \kk})
\bigg(&\frac{1}{2} (\partial^\mu r^\nu_{\alpha \beta \kk}) r^\lambda_{ \beta \alpha \kk} +  r^\nu_{\alpha \beta \kk}r^\lambda_{ \beta \alpha \kk} \frac{\partial^\mu\Delta E_{ \beta \alpha \kk}}{\Delta E_{ \beta \alpha \kk}}  \\
&-i r^\nu_{\alpha \beta \kk }r^\mu_{\alpha \alpha \kk} r^\lambda_{ \beta \alpha \kk} +i\sum_{\substack{\gamma\neq \alpha  \\ \gamma\neq\beta  }} r^\mu_{\alpha \gamma \kk} r^\nu_{\gamma \beta \kk} r^\lambda_{ \beta \alpha \kk} \frac{\Delta E_{\gamma \beta \kk}}{\Delta E_{ \beta \alpha \kk}}  \bigg) \Biggr],
\end{aligned}
\end{equation}
which is Eq.~\eqref{eq: computed diamagnetic part of Bures connection for current example} in the main text.

Now consider $\Gamma^{\mathrm{in}}_\mathrm{q.s.}$, which according to Eq.~\eqref{eq: star 3} depends on $S^{\mu \lambda \nu}(\omega_1,\omega_2)$ and is nonvanishing only when $\omega_1, \omega_2, \omega_1-\omega_2 \neq 0$. 
Using the symmetry under $\mu \leftrightarrow \lambda$, $\omega_1 \rightarrow \omega_2-\omega_1,\omega_2 \rightarrow \omega_2$, we can rewrite $\Gamma^{\mathrm{in}}_\mathrm{q.s.}$ as [taking into account the global minus sign from the current perturbation in Eq.~\eqref{eq:vec_potential_ham}]
\begin{equation} 
\begin{aligned}
\Gamma^{\mathrm{in}}_\mathrm{q.s.} = -\frac{1}{4} \mathrm{Re} \left[ A_\mu A_\nu A_\lambda \int  d\omega_1 d\omega_2 \tanh \frac{\beta \omega_1}{2} \tanh \frac{\beta (\omega_1-\omega_2)}{2} \mathrm{P} \frac{S^{\mu \lambda \nu}(\omega_1,\omega_2)+S^{ \lambda\mu \nu}(\omega_2-\omega_1,\omega_2)}{\omega_1 \omega_2 (\omega_1-\omega_2)}  \right].
\end{aligned}
\end{equation}

The constraints $\omega_1, \omega_2, \omega_1 -\omega_2 \neq 0 $ which arise from the $\tanh$ function and the principle value, single out the sum of Eqs.~\eqref{part 1 in chi''}, \eqref{part 2 in chi''}, \eqref{part 5 in chi'' except minus sign}, and \eqref{part 6 in chi'' except minus sign}, so the terms we need to consider in $S^{\mu \lambda \nu}(\omega_1,\omega_2)+S^{ \lambda\mu \nu}(\omega_2-\omega_1,\omega_2)$ are
\begin{equation}
\begin{aligned}
&[S^{\mu \lambda \nu}(\omega_1,\omega_2)+S^{ \lambda\mu \nu}(\omega_2-\omega_1,\omega_2)]\Big|_{\omega_ 1\neq \omega_2, \omega_1\neq 0}\\
=&+\sum_{\alpha \neq\beta , \alpha\neq\gamma,\kk} \Delta f_{\beta \gamma \kk}[f_{\alpha \kk}-(1-f_{\alpha \kk}) ] j^\lambda_{\alpha\beta \kk} j^\nu_{\beta \gamma \kk} j^\mu_{\gamma \alpha \kk} \delta(\omega_1 +\Delta E_{\gamma \alpha \kk})\delta(\omega_2 +\Delta E_{\gamma \beta \kk})\\
&+\sum_{\alpha \neq\beta , \alpha\neq\gamma,\kk} \Delta f_{\beta \gamma \kk }[f_{\alpha \kk}-(1-f_{\alpha \kk}) ] j^\mu_{\alpha\beta \kk} j^\nu_{\beta \gamma \kk} j^\lambda_{\gamma \alpha \kk} \delta(\omega_1 +\Delta E_{\alpha \beta \kk})\delta(\omega_2 +\Delta E_{\gamma \beta \kk})\\
=&+i\omega_1 \omega_2 (\omega_1-\omega_2)\sum_{\alpha \neq\beta , \alpha\neq\gamma,\kk} \Delta f_{\beta \gamma \kk}[f_{\alpha \kk}-(1-f_{\alpha \kk}) ]r^\lambda_{\alpha\beta \kk} r^\nu_{\beta \gamma \kk} r^\mu_{\gamma \alpha \kk} \delta(\omega_1 +\Delta E_{\gamma \alpha \kk})\delta(\omega_2 +\Delta E_{\gamma \beta \kk})\\
&+i\omega_1 \omega_2 (\omega_1-\omega_2)\sum_{\alpha \neq\beta , \alpha\neq\gamma,\kk} \Delta f_{\beta \gamma \kk}[f_{\alpha \kk}-(1-f_{\alpha \kk}) ]r^\mu_{\alpha\beta \kk} r^\nu_{\beta \gamma \kk} r^\lambda_{\gamma \alpha \kk} \delta(\omega_1 +\Delta E_{\alpha \beta \kk})\delta(\omega_2 +\Delta E_{\gamma \beta \kk}),
\end{aligned}
\end{equation}
which gives 
\begin{equation}
\begin{aligned}
\Gamma^{\mathrm{in}}_\mathrm{q.s.} = -\frac{1}{4} A_\mu A_\nu A_\lambda \mathrm{Re} \Bigg[ i \sum_{\substack{\alpha \neq \beta \\ \alpha \neq \gamma,\mathbf{k}}} \mathrm{sign}(\Delta E_{ \beta \alpha \kk}) \mathrm{sign}(\Delta E_{\gamma \alpha \kk}) \Delta f_{\beta \gamma \kk} \big[f_{\alpha \kk}-(1-f_{\alpha \kk})  \big] (r^\lambda_{\alpha\beta \kk} r^\nu_{\beta \gamma \kk} r^\mu_{\gamma \alpha \kk}+r^\mu_{\alpha\beta \kk} r^\nu_{\beta \gamma \kk} r^\lambda_{\gamma \alpha \kk})
\Bigg].
\end{aligned}
\end{equation}

We can see that the second term is the complex conjugate of the first term (even without exchanging indices among $\mu,\nu,\lambda$), so
\begin{equation}
\Gamma^{\mathrm{in}}_\mathrm{q.s.} =  A_\mu A_\nu A_\lambda \mathrm{Re} \Bigg[ \frac{1}{2} i\sum_{\substack{\alpha \neq \beta \\ \alpha \neq \gamma,\mathbf{k}}} \mathrm{sign}(\Delta E_{ \beta \alpha \kk}) \mathrm{sign}(\Delta E_{\gamma \alpha \kk}) \Delta f_{ \gamma \beta \kk} \big[f_{\alpha \kk}-(1-f_{\alpha \kk})  \big] r^\lambda_{\alpha\beta \kk} r^\nu_{\beta \gamma \kk} r^\mu_{\gamma \alpha \kk} \Bigg],
\end{equation}
which is Eq.~\eqref{eq: computed intrinsic part of Bures connection for current example} in the main text.

\subsection{Cancellation of Terms in the Bures Connection}
\label{Cancellation of terms in Bures connection}
From Eqs.~\eqref{eq: computed response part of Bures connection for current example}--\eqref{eq: computed intrinsic part of Bures connection for current example}, we have
\begin{equation} \label{eq: computed total Bures connection for current example}
\begin{aligned}
&\Gamma^{\mathrm{para}}_\mathrm{q.s.}+\Gamma^{\mathrm{dia}}_\mathrm{q.s.}+\Gamma^{\mathrm{in}}_\mathrm{q.s.} \\
=&   +A_\mu A_\nu A_\lambda \mathrm{Re}  \sum_{\alpha \beta \kk}\Bigg[\Delta f_{\beta \alpha \kk}\mathrm{sign}(\Delta E_{\beta \alpha \kk})\Big[\frac{1}{2} (\partial^\mu r^\nu_{\alpha \beta \kk}) r^\lambda_{\beta \alpha\kk}-i r^\nu_{\alpha \beta \kk}r^\mu_{\alpha \alpha\kk} r^\lambda_{\beta \alpha\kk}\Big] \Bigg]\\
&+  A_\mu A_\nu A_\lambda \mathrm{Re} \sum_{\substack{\alpha \neq \beta \\ \alpha \neq  \gamma,\kk}}\Bigg[+2i \Delta f_{\beta \gamma \kk} \frac{\Delta E_{\beta \alpha\kk}}{\Delta E_{\gamma \alpha\kk}}\mathrm{sign}(\Delta E_{\gamma \alpha\kk})r^\mu_{\alpha\beta\kk} r^\nu_{\beta \gamma\kk} r^\lambda_{\gamma \alpha\kk}+i\Delta f_{\beta \gamma\kk}\mathrm{sign}(\Delta E_{\beta \gamma\kk}) \frac{\Delta E_{\alpha \beta\kk}}{\Delta E_{\beta \gamma\kk}} r^\mu_{\alpha\beta\kk} r^\nu_{\beta \gamma\kk} r^\lambda_{\gamma \alpha\kk} \\
& \quad \quad \quad \ \quad \quad \quad \quad \quad \quad +\frac{i}{2}  \mathrm{sign}(\Delta E_{\beta \alpha\kk}) \mathrm{sign}(\Delta E_{\gamma \alpha\kk}) \Delta f_{ \gamma \beta\kk} \big[f_{\alpha \kk}-(1-f_{\alpha\kk})  \big] r^\mu_{\alpha\beta\kk} r^\nu_{\beta \gamma\kk} r^\lambda_{\gamma \alpha\kk} \Bigg].
\end{aligned}
\end{equation}
Let us first focus on the second term in Eq.~\eqref{eq: computed total Bures connection for current example}, which is
\begin{equation} \label{eq: intermediate step for cancellation}
\begin{aligned}
 -A_\mu A_\nu A_\lambda \mathrm{Im} \sum_{\substack{\alpha \neq \beta \\ \alpha \neq  \gamma,\kk}} \Bigg\{ r^\mu_{\alpha\beta \kk} r^\nu_{\beta \gamma \kk} r^\lambda_{\gamma \alpha \kk}\times \Big[
&+2\Delta f_{\beta \gamma \kk } \frac{\Delta E_{ \beta \alpha \kk}}{\Delta E_{\gamma \alpha \kk}}\mathrm{sign}(\Delta E_{\gamma \alpha \kk})+\Delta f_{\beta \gamma \kk}\mathrm{sign}(\Delta E_{\beta \gamma \kk}) \frac{\Delta E_{\alpha \beta \kk}}{\Delta E_{\beta \gamma \kk}}  \\
&+\frac{1}{2}  \mathrm{sign}(\Delta E_{ \beta \alpha \kk}) \mathrm{sign}(\Delta E_{\gamma \alpha \kk}) \Delta f_{ \gamma \beta \kk} \big[f_{\alpha \kk}-(1-f_{\alpha\kk})  \big]  \Big] \Bigg\},
\end{aligned}
\end{equation}
then after separating terms in the sums into single-occupancy and double-occupancy cases, and relabeling dummy Greek indices so as to unify the summation into $\alpha, \beta \in \mathrm{occ}, \gamma \in \mathrm{unocc}$ for all double-occupancy cases and $\alpha    \in \mathrm{occ} , \beta, \gamma \in \mathrm{unocc}$ for all single-occupancy cases, Eq.~\eqref{eq: intermediate step for cancellation} simplifies to
\begin{equation}
\begin{aligned}
 - \mathrm{Im} \Bigg\{& +A_\mu A_\nu A_\lambda \sum_{\substack{\alpha,  \beta   \in \mathrm{occ},\kk \\ \gamma \in \mathrm{unocc}}}r^\mu_{\alpha\beta \kk} r^\nu_{\beta \gamma \kk} r^\lambda_{\gamma \alpha \kk}\times
\left[ 1+\Delta E_{ \beta \alpha \kk}\left(\frac{1}{\Delta E_{\gamma \alpha \kk}}-\frac{1}{\Delta E_{\gamma \beta \kk}}\right) +\frac{\Delta E_{ \alpha \gamma \kk}+\Delta E_{\beta \gamma \kk}}{\Delta E_{ \alpha \beta \kk}}\mathrm{sign}(\Delta E_{\alpha \beta \kk})\right]\\
&+A_\mu A_\nu A_\lambda \sum_{\substack{\alpha    \in \mathrm{occ},\kk \\ \beta, \gamma \in \mathrm{unocc}}}r^\mu_{\alpha\beta \kk} r^\nu_{\beta \gamma \kk} r^\lambda_{\gamma \alpha \kk}\times\left[1+\Delta E_{\gamma \beta \kk}\left(\frac{1}{\Delta E_{ \alpha \beta \kk}}-\frac{1}{\Delta E_{ \alpha \gamma \kk}}\right)+\frac{\Delta E_{ \alpha \gamma \kk}+\Delta E_{\alpha \beta \kk }}{\Delta E_{\beta \gamma \kk}}\mathrm{sign}(\Delta E_{\beta \gamma \kk })\right]  \Bigg\}.
\end{aligned}
\end{equation}
We can observe that the term inside $-\mathrm{Im} \{...\}$ is real, which means the whole term is zero. 
The remaining contribution to the Bures connection is 
\begin{equation} \label{eq: inter 2}
\begin{aligned}
-  \mathrm{Re} \left\{\sum_{\substack{\alpha    \in \mathrm{occ},\kk \\  \beta \in \mathrm{unocc}}}+\sum_{\substack{\alpha    \in \mathrm{unocc} \\  \beta \in \mathrm{occ},\kk}}  A_\mu A_\nu A_\lambda\left[ \frac{1}{2} (\partial^\mu r^\nu_{\alpha \beta \kk}) r^\lambda_{ \beta \alpha \kk}-i r^\nu_{\alpha \beta \kk }r^\mu_{\alpha \alpha \kk} r^\lambda_{ \beta \alpha \kk} \right] \right\},
\end{aligned}
\end{equation}
however, the second term in Eq.~\eqref{eq: inter 2} is purely imaginary and it also has no contribution. 
Therefore, the Bures connection reduces to 
\begin{equation} 
\Gamma^{\mathrm{para}}_\mathrm{q.s.}+\Gamma^{\mathrm{dia}}_\mathrm{q.s.}+\Gamma^{\mathrm{in}}_\mathrm{q.s.}=- \frac{1}{2}A_\mu A_\nu A_\lambda \mathrm{Re} \left[\sum_{\substack{\alpha    \in \mathrm{occ},\kk \\  \beta \in \mathrm{unocc}}}+\sum_{\substack{\alpha    \in \mathrm{unocc} \\  \beta \in \mathrm{occ},\kk}}    (\partial^\mu r^\nu_{\alpha \beta \kk}) r^\lambda_{ \beta \alpha \kk} \right] ,
\end{equation}
which we can further simplify as
\begin{equation}
\begin{aligned}
\Gamma^{\mathrm{para}}_\mathrm{q.s.}+\Gamma^{\mathrm{dia}}_\mathrm{q.s.}+\Gamma^{\mathrm{in}}_\mathrm{q.s.}=&- \frac{1}{2}A_\mu A_\nu A_\lambda \mathrm{Re} \left[\sum_{\substack{\alpha    \in \mathrm{occ},\kk \\  \beta \in \mathrm{unocc}}}+\sum_{\substack{\alpha    \in \mathrm{unocc} \\  \beta \in \mathrm{occ},\kk}}    (\partial^\mu r^\nu_{\alpha \beta \kk}) r^\lambda_{ \beta \alpha \kk} \right] \\
=&- \frac{1}{2}A_\mu A_\nu A_\lambda \left[\sum_{\substack{\alpha    \in \mathrm{occ},\kk \\  \beta \in \mathrm{unocc}}}+\sum_{\substack{\alpha    \in \mathrm{unocc} \\  \beta \in \mathrm{occ},\kk}}    (\partial^\mu r^\nu_{\alpha \beta \kk}) r^\lambda_{ \beta \alpha \kk} \right]\\
=&-\frac{1}{2}  A_\mu A_\nu A_\lambda \biggl[ \sum_{\substack{\alpha    \in \mathrm{occ},\kk \\  \beta \in \mathrm{unocc}}}(\partial^\mu r^\nu_{\alpha \beta \kk}) r^\lambda_{ \beta \alpha \kk} -\sum_{\substack{\alpha    \in \mathrm{unocc} \\  \beta \in \mathrm{occ},\kk}}r^\nu_{\alpha \beta \kk} (\partial^\mu r^\lambda_{ \beta \alpha \kk}) +\sum_{\substack{\alpha    \in \mathrm{unocc} \\  \beta \in \mathrm{occ},\kk}}\partial^\mu (r^\nu_{\alpha \beta \kk} r^\lambda_{ \beta \alpha \kk}) \biggr]\\
=&-\frac{1}{2}  A_\mu A_\nu A_\lambda \sum_{\substack{\alpha    \in \mathrm{unocc} \\  \beta \in \mathrm{occ},\kk}} \partial^\mu (r^\nu_{\alpha \beta \kk} r^\lambda_{ \beta \alpha \kk}),
\end{aligned}
\end{equation}
where from the first line to the second line we used that the term is purely real, from the second line to the third line we integrated the second term by parts, and from the third line to the fourth line we exchanged indices for the last term by $\nu \leftrightarrow \lambda$, $\alpha \leftrightarrow \beta$ such that the last and first terms cancel. 
Thus the total derivative term is the only nonvanishing contribution. 
We note that the only nonvanishing contribution to the Bures connection comes from the first term in $\Gamma^\mathrm{dia}_\mathrm{q.s.}$ in Eq.~\eqref{eq:appendx_gamma_dia}. 

\twocolumngrid
\bibliography{refs-fisher}
\end{document}